\documentclass[preprint]{aastex61}
\usepackage{amsmath}
\usepackage{amssymb}
\newcommand{\vtd}{{(\tau_0)}}
\newcommand{\e}{\mathbf{e}}

\renewcommand{\bm}[1]{\mbox{\boldmath{$#1$}}}

\def\a{{\bm{a}}}
\def\r{{\bm{r}}}
\def\b{{\bm{b}}}
\def\v{{\bm{v}}}

\def\x{{\bm{x}}}
\def\y{{\bm{y}}}
\def\B{{\bm{B}}}
\def\hatbB{{\hat{\bm{B}}}}

\def\be{\begin{eqnarray}}
\def\ee{\end{eqnarray}}

\usepackage{graphicx}
\usepackage{needspace}
\usepackage{xspace}
\def\qsl{{\fontfamily{pcr}\selectfont QSL} {\fontfamily{pcr}\selectfont Squasher}\xspace}

\shorttitle{Coiling and Squeezing}

\begin{document}

\title{Coiling and Squeezing: Properties of the Local Transverse Deviations of Magnetic Field Lines}

\author{Svetlin Tassev}
\affiliation{Harvard-Smithsonian Center for Astrophysics, 60 Garden Street,
Cambridge, MA 02138, USA}
\affiliation{Braintree High School, 128 Town Street, Braintree, MA 02184, USA}
\author{Antonia Savcheva}
\affiliation{Harvard-Smithsonian Center for Astrophysics, 60 Garden Street,	Cambridge, MA 02138, USA}
\correspondingauthor{Svetlin Tassev}
\email{svetlin.tassev@cfa.harvard.edu}
\begin{abstract}
We study the properties of the local transverse deviations of magnetic field lines at a fixed moment in time. Those deviations ``evolve'' smoothly in a plane normal to the field-line direction as one moves that plane along the field line. Since the evolution can be described by a planar flow in the normal plane, we derive most of our results in the context of a toy model for planar fluid flow. We then generalize our results to include the effects of field-line curvature. We show that the type of flow is determined by the two non-zero eigenvalues of the gradient of the normalized magnetic field. The eigenvalue difference quantifies the local rate of squeezing or coiling of neighboring field lines, which we relate to standard notions of fluid  vorticity and shear. The resulting squeezing rate can be used in the detection of null points, hyperbolic flux tubes and current sheets. Once integrated along field lines, that rate gives a squeeze factor, which is an approximation to the squashing factor, which is usually employed in locating quasi-separatrix layers (QSLs), which are possible sites for magnetic reconnection. Unlike the squeeze factor, the squashing factor can miss QSLs for which field lines are squeezed and then unsqueezed. In that regard, the squeeze factor is a better proxy for locating QSLs than the squashing factor. In another application of our analysis, we construct an  approximation to the local rate of twist of neighboring field lines, which we refer to as the coiling rate. That rate can be integrated along a field line to give a coiling number, $\mathrm{N_c}$. We show that unlike the standard local twist number, $\mathrm{N_c}$ gives an unbiased approximation to the number of twists neighboring field lines make around one another. $\mathrm{N_c}$ can be useful for the study of flux rope instabilities, such as the kink instability, and can be used in the detection of flux ropes.
\end{abstract}


\tableofcontents

\section{Introduction\label{sec:Intro}}

The structure of the coronal magnetic field and its restructuring determines the energetics of solar flares and coronal mass ejections. Partitioning the magnetic field into flux domains is often the first step in studying the structure of magnetic fields. Such partitioning can be accomplished by reconstructing the magnetic skeleton, which is formed by features such as null points, separatrix surfaces and separators. That type of partitioning has been studied extensively in the past \citep[e.g.][]{Gorbachev88,1996SoPh..169...91L,Parnell10}. A broader type of partitioning of magnetic fields has been relatively recently employed using sheets of large, yet continuous, variation of the magnetic field, called Quasi-Separatrix Layers (QSLs) \citep[e.g.][]{1994ApJ...437..851L,Priest95}. 
QSLs have been used successfully in studying solar eruptions  \citep[e.g.][]{Savcheva12a,Janvier13,Janvier14,Liu14,Savcheva15} as well as laboratory magnetoplasma \citep[e.g.][]{PhysRevLett.103.105002}, and have been associated with locations with exponentially large values of the dimensionless squashing factor value, $Q$ \citep[e.g.][]{Titov02,Titov07}, where large current build-up can develop \citep[e.g.][]{1994ApJ...437..851L}. 

Two magnetic features that are of significant theoretical and observational interest are  hyperbolic flux tubes \citep[HFTs;][]{Titov07,Savcheva12a,Savcheva12b} -- formed at the intersection of QSLs -- and flux ropes. The latter consist of magnetic field lines wrapping around a common flux rope axis. Flux ropes play a fundamental role in solar eruptions \citep[e.g.][]{2016ApJ...818..148L} by exhibiting  instabilities, such as  torus or kink instability \cite[e.g.][]{Aulanier10,Torok04}. Meanwhile, HFTs  can generically form under flux ropes when the QSL wrapping around a flux rope, separating it from the overlying magnetic arcade, intersects with itself. Observable features associated with HFTs are flare ribbons whose locations and motions are reproduced by the photospheric traces of the corresponding HFT \citep{Liu14,Savcheva15}. 

Without prior knowledge, locating  magnetic features of interest -- even as basic as flux ropes -- in realistic magnetic field models is a non-trivial task since the domain decomposition obtained using QSLs can be extremely complex. As not all QSLs are associated with producing current sheets (indeed, QSLs are exhibited even in potential fields, carrying no current; see e.g. \cite{Titov07}), quite often one has no other option but to resort to picking by hand the QSLs of interest through ad-hoc thresholding in current density or magnetic field \citep[e.g.][]{Savcheva15}. Non-QSL-based  methods for detecting flux ropes have been proposed \citep[e.g.][]{2009ApJ...699.1024Y,2016A&A...594A..98Y,2017ApJ...846..106L}, yet again they rely on  ad-hoc thresholds: in the case of the method described by \citet{2017ApJ...846..106L}, that threshold is even time-dependent. 

An important property of flux ropes used in their stability analyses is the twist number \citep[e.g.][]{berger_field_1984}, measuring the number of twists a field line in the flux rope makes around the flux rope core. As the twist number is a non-local quantity, requiring a proper choice of flux rope axis \citep[e.g.][]{2017ApJ...840...40G}, one usually resorts to computing the \textit{local} twist number\footnote{We avoid using the symbol $\mathcal{T}_w$ to denote the local twist number since \cite{2006JPhA...39.8321B} use that symbol to denote \textit{both} the local and non-local twist (see their equations~(12) and (16)). Meanwhile, \cite{2016ApJ...818..148L} use $\mathcal{T}_w$ to denote only the local approximation to the twist, calling the non-local twist $\mathcal{T}_g$; see their equations~(7) and (13). Since we focus on the local deviations of field-lines, in most of this paper, we are concerned only with the \textit{local} twist number and how it compares with the coiling number we introduce. However,  in Section~\ref{sec:simpleFR}, we do compare those local quantities with the non-local twist number for simple flux-rope configurations.}, $\mathrm{N_t}$, instead, which is proportional to the component of the current parallel to the magnetic field \citep[see][eq. 16]{2006JPhA...39.8321B}. For non-diverging current distributions, $\mathrm{N_t}$ approaches the exact twist number at the core of a flux rope \citep[e.g.][]{2016ApJ...818..148L}. As twist is a dimensionless quantity, it is tempting to use the natural inequality,  $\mathrm{N_t}\gtrsim 1$, as a  robust flux-rope detection threshold. That threshold would seem to imply that nearby field lines exhibit at least $\mathcal{O}(1)$ twists around one another to qualify as being part of a flux rope. However, in this paper we analytically show  that the twist number gives a biased estimate of the number of turns infinitesimally separated field lines make around one another. Indeed, in a follow-up paper \citep{Tassev2019}, we apply the above threshold to realistic magnetic fields, and we find that along with flux ropes, it picks out many sheared, untwisted structures, which  further corroborates that $\mathrm{N_t}$ gives a biased estimate of the local field-line twisting.

In this paper,  we construct an unbiased local equivalent of the twist number, which we call the coiling number, $\mathrm{N_c}$. In \citep{Tassev2019}, we show that the natural inequality $\mathrm{N_c}\gtrsim  1$ can be used successfully as a detection threshold for flux ropes. For each field line, $\mathrm{N_c}$ is given as an integral of a local coiling rate ($\omega_c$) over the field line length. 

We also introduce a quantity, complementary to the coiling rate, which we dub the squeezing rate ($\rho_{\mathcal{Z}}$), measuring the local logarithmic rate of squeeze of neighboring field lines.  The field-line integral over $\rho_{\mathcal{Z}}$ gives $\ln(\mathcal{Z})$, with $\mathcal{Z}$ defined as the squeeze factor. We show that  $\mathcal{Z}\sim Q$ in a sense made precise in the paper; and therefore, $\rho_{\mathcal{Z}}$ gives an estimate of the local contributions to $Q$ -- with the latter having no exact corresponding local squashing rate. The squeezing rate is large in the vicinity of HFTs and null points, and can be used for their detection.

\begin{figure}[t!]\epsscale{0.6}
	\plotone{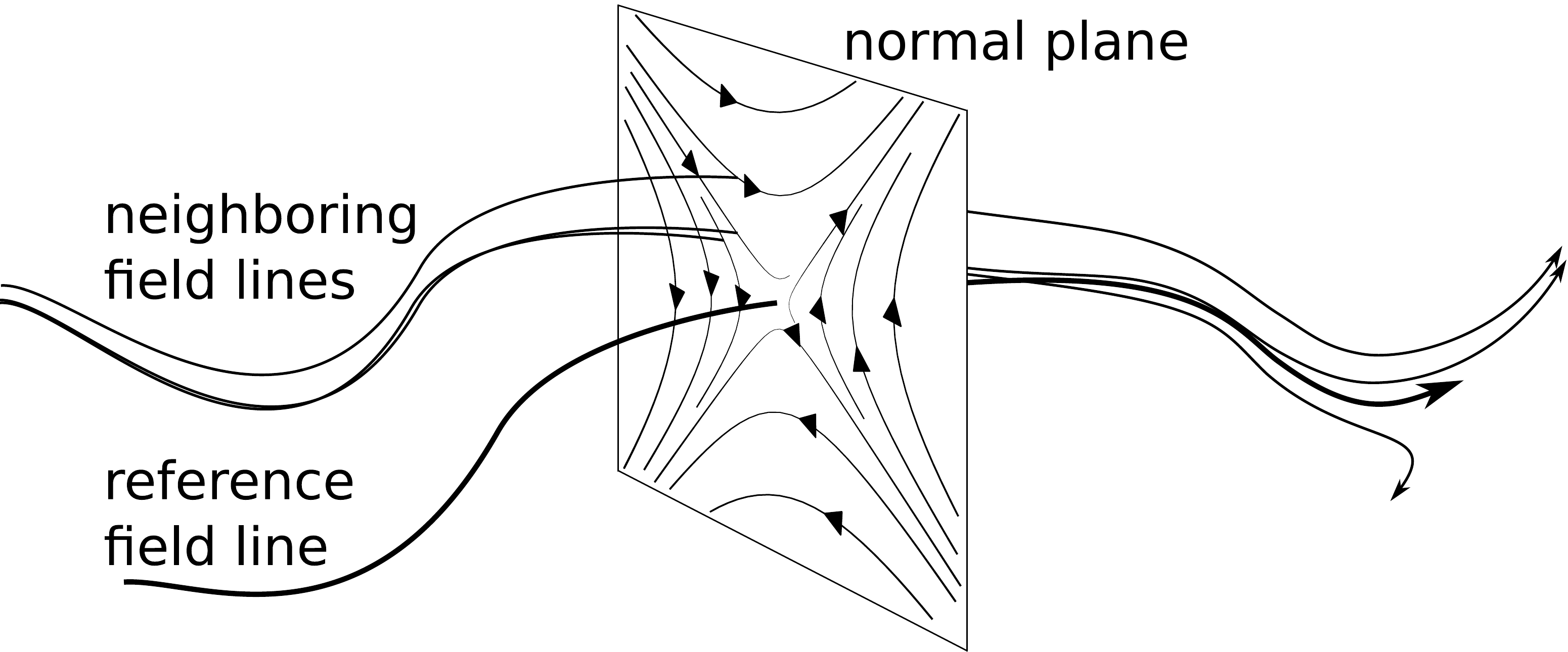}\caption{
		The figure illustrates the behavior of neighboring field lines in 3D. The field lines intersect the plane normal to one of them (the \textit{reference} field line). As that normal plane is moved along the length of the reference field line, the intersection points of its neighboring field lines with the plane follow a flow in the normal plane. In the illustration above, that flow corresponds to a saddle. See Fig.~\ref{fig:types} for other types of flows. Complications in the study of the planar flow arise from the curvature of the reference field line, which causes the normal plane to change orientation as it is moved along that field line. Another complication arises from the fact that the transverse flow type in general depends on rotations of the normal plane basis around the reference field line. We fix the basis by requiring that the rate of rotation of individual neighboring field lines around the reference field line is independent of whether that rate is calculated using the standard non-local twist rate (cf. eq.~(\ref{omegaInit})), or in the basis spanning the normal plane. See the text for further discussion. }\label{fig:transverse}
\end{figure}

\begin{figure}[t!]\epsscale{0.5}
	\plotone{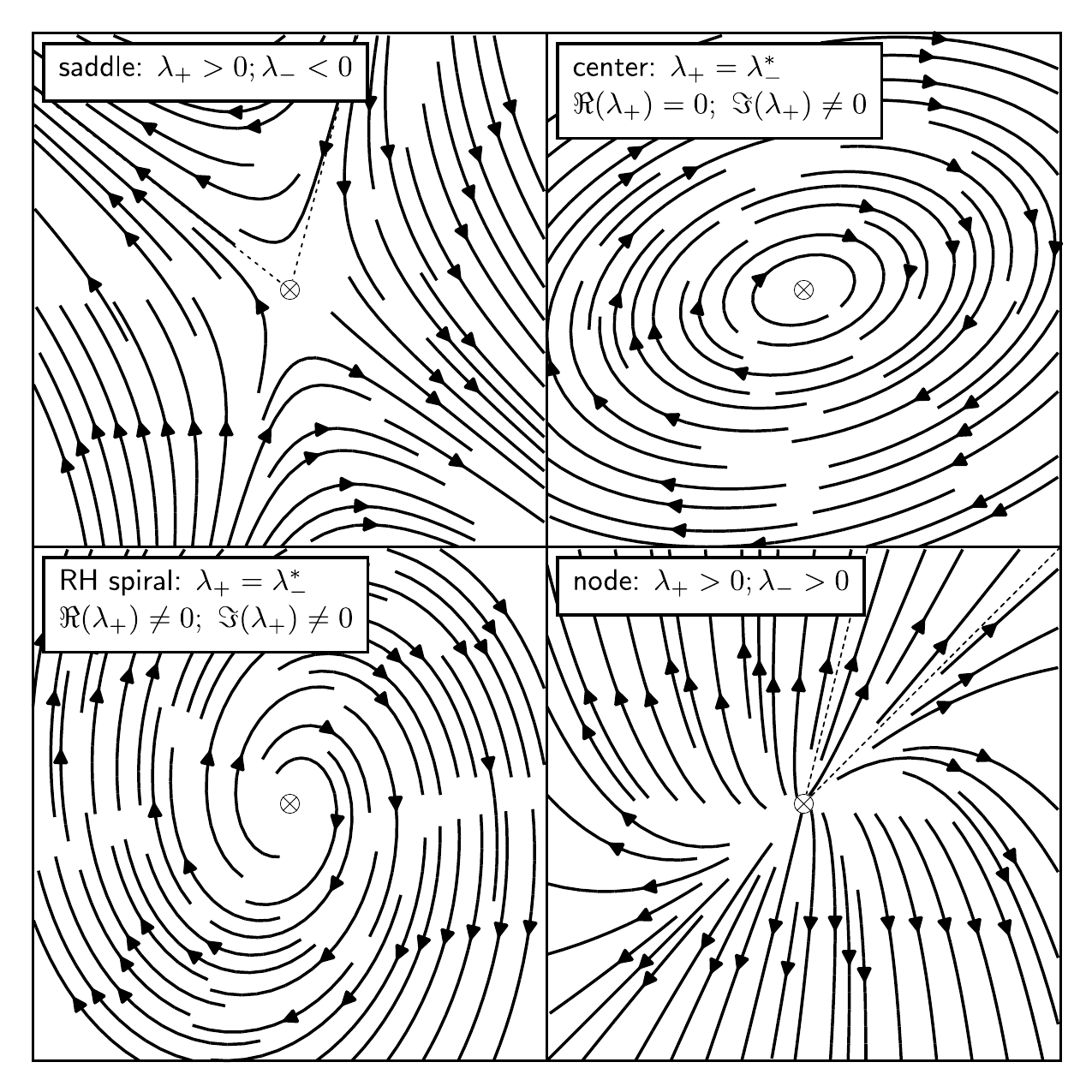}\caption{
		The figure illustrates streamlines traced in different types of planar fluid flow around a critical point. In the context of magnetic fields, one should think of these integral lines as the lines traced by the intersection of field lines in the infinitesimal vicinity of a reference field line,  with a plane normal to that reference field line. As the normal plane is moved along the length of the reference field line, the location of the intersection of the neighboring field lines with that plane is changing; thus, exhibiting a flow pattern similar to that of a planar fluid flow  (see Fig.~\ref{fig:transverse}).
		The reference field line is denoted with $\otimes$ at the center of each panel to indicate that it is going into the page. The dashed lines in the top-left and bottom-right panels indicate the eigenvector directions of the flow, which correspond to the asymptotes of the flow. The node flow (bottom-right) can be decomposed into a saddle flow (top-left) plus uniform expansion. Similarly, the spiral flow (bottom-left) can be decomposed into an elliptical flow (top-right) plus uniform expansion. That is the basis of the decomposition of the velocity gradient into a trace part (the rate-of-expansion tensor: $\Theta$) and a traceless part ($\Psi$). See the text for further discussion.}\label{fig:types}
\end{figure}

To construct the  quantities described above at a particular location, we pick the field line passing through that location -- we call that field line the \textit{reference} field line below -- and explore the kinematics of the  displacements of nearby magnetic field lines relative to that reference field line. We focus on the transverse part of those deviations, calculated in a plane normal to the reference field line (see Fig.~\ref{fig:transverse}). As the normal plane is moved along the length of the reference field line, the intersection points between that plane and the nearby field lines can be described by a planar ``flow''. By construction, that flow exhibits a critical point at the location where the reference field line intersects the normal plane. As we focus on the kinematics of \textit{nearby} field lines, that transverse flow is approximately linear. Therefore, the types of transverse field-line flow  are given by the  standard equilibrium solutions of linear systems, such as a saddle, cycle, spiral or node (see Fig.~\ref{fig:types}). The type of solution is
unique for the normal plane that is only \textit{minimally rotated} around its normal when moved along the reference field line (see Section~\ref{sec:uniq}). That solution  is determined solely by the two eigenvalues of the gradient of the velocity of the nearby field lines in the normal plane. Here, velocity is defined as the time derivative of the transverse deviations, where ``time'' corresponds to the field-line length parameter along the reference field line. Even at this point, one can conjecture that critical points corresponding to saddles and nodes exhibiting large eigenvalue difference, can be used in locating HFTs and the vicinity of null points; whereas spirals and centers are the typical flow patterns in flux ropes. Thus, we construct our $\rho_{\mathcal{Z}}$ and $\omega_c$  using the properties of the transverse field-line flow. 


Studying the transverse flow of field lines has  the advantage of simplicity as one can apply  intuition from the study of planar fluid flows. We use that to our advantage as in Section~\ref{sec:toy} we introduce most of our results in the context of a toy model of a steady-state linear planar fluid flow. Only then (Section~\ref{sec:fldev}) do we address issues arising from the fact that the field-line flow is ``time''-dependent (i.e. the flow changes as one moves along the reference field line); that the normal plane is constantly changing orientation due to the curvature of the reference field line (see Fig~\ref{fig:transverse}); and that the type of flow in the normal plane depends on overall rotations of the basis spanning the normal plane around the reference field line. We specify that basis uniquely by the requirement  that the rate of rotation of individual neighboring field lines around the reference field line is independent of whether that rate is calculated using the standard non-local twist rate (cf. eq.~(\ref{omegaInit})), or in the basis spanning the normal plane. In the end, we demonstrate that for that unique normal-plane basis, the transverse flow type is entirely determined by the non-zero eigenvalues of the gradient of the normalized magnetic field, $\bm{\nabla}\hatbB$, which match those of the gradient of the transverse velocity of the flow. The real part of the difference of the non-zero eigenvalues of $\bm{\nabla}\hatbB$ gives $\rho_{\mathcal{Z}}$ (Section~\ref{sec:localQ}), while the imaginary part equals $2\omega_c$ (Section~\ref{sec:coiling}). In Section~\ref{sec:simpleFR} we write our results for generic, axially-symmetric, force-free flux ropes, and show numerical solutions for example flux rope configurations. In Section~\ref{sec:summary} we present a summary of our result. In Appendix~\ref{sec:app} we write our results in curvilinear coordinates.

\section{A toy model: linear steady-state planar flow}\label{sec:toy}

In this section we write our coiling and squeezing rates for a toy model of a linear steady-state 2-dimensional (2D) flow. We use this model to show the relationship between the quantities we introduce in this paper and standard fluid notions, such as vorticity and shear. Along the way, we relate the flow vorticity to the coiling and squeezing rates using properties of the global geometric deformation of the flow.

\subsection{Flow decomposition}\label{sec:decompstd}
Let us consider the steady-state 2D linear flow with a velocity field ($\v$) given below as a function of position, $\r$:
\be\label{2dflow}
v_i(\r(\tau))=\frac{d r_i(\tau)}{d\tau}=m_{ij}r_j(\tau)\ .
\ee
The constant matrix $m$ above equals the velocity gradient, $m_{ij}=\nabla_j v_i$.
Clearly $\r=0$ is a critical point, which depending on the eigenvalues $\lambda_{\pm}$ of $m$, can be a right-handed (RH) or left-handed (LH) stable/unstable spiral (two complex conjugate eigenvalues), a node (real eigenvalues of the same sign), a saddle (real eigenvalues of opposite signs); or in certain special cases: a center (purely imaginary eigenvalues) and an (im)proper node (see Fig.~\ref{fig:types} for example illustrations).  Even if a flow does not exhibit a critical point, the equation above can always be written for the (linearized) relative displacement vector ($\r$) between two neighboring streamlines as a function of time ($\tau$). Then it is easy to see how eq.~(\ref{2dflow}) can be later reinterpreted  as giving the transverse deviation rate between two neighboring magnetic field lines (separated by a transverse displacement $\r$) as a function of the field line length parameter $\tau$. In that case, we show that  one must replace $m$ with the transverse gradient of the normalized magnetic field, $\hatbB$ (cf. eq.~(\ref{dr})).

The velocity gradient ($m$) is usually decomposed as follows \citep[e.g.][]{2017mcpo.book.....T}:
\be\label{stddecomprates1}
m=\Theta+\Psi=\Theta+\Sigma+\Omega\ ,
\ee
with:
\be\label{stddecomprates}
\Theta\equiv \mathrm{I}\,\frac{\mathrm{tr}(m)}{2},\quad  \Psi\equiv m-\Theta,\quad  \Sigma\equiv \frac{\Psi+\Psi^{\mathrm{T}}}{2},\quad  \Omega\equiv \frac{m-m^{\mathrm{T}}}{2}=\frac{\Psi-\Psi^{\mathrm{T}}}{2}\ ,
\ee
where  $\mathrm{I}$ is the identity matrix.
Clearly, $\Theta$ is the diagonal, trace part of $m$, usually called the rate-of-expansion tensor. Then, $\Psi$ is the traceless part of $m$. The latter is in turn usually decomposed into a symmetric traceless part (the rate-of-shear tensor; $\Sigma$) and an antisymmetric part (the rate-of-rotation tensor; $\Omega$). This split is manifestly possible for all types of critical points. The rate of expansion corresponds to uniform contraction/expansion of the area (in 3D, that would be the volume) of a fluid element. Once that uniform expansion flow is subtracted from $m$, that renders the critical point of the remaining traceless part of the velocity gradient ($\Psi$) into either a saddle (for real eigenvalues of $m$) or a center (for complex eigenvalues of $m$). 

The standard decomposition of $m$ above, however, is not unique. To see that, note that for infinitesimal $\delta\tau$, eq.~(\ref{2dflow}) can be written in finite difference form as:
\be\label{2dflowFD}
r_i(\tau+\delta\tau)=(\mathrm{I}+\delta\tau m)_{ij}r_j(\tau)\ .
\ee
The infinitesimal transformation $(\mathrm{I}+\delta\tau m)$ entering above, can be an infinitesimal squeeze ($\delta Z$), shear ($\delta S$), scaling ($\delta C$) or rotation ($\delta R$) map, or a composition of those. With an appropriate choice of basis, those infinitesimal maps can be written as (for the corresponding finite transformations, see eq.~(\ref{maps}) below):
\be\label{infmaps}
\delta Z(z)&=&\begin{pmatrix}
	1+z\delta \tau  & 0\cr
	0 &1-z\delta \tau 
\end{pmatrix}, \ 
\delta S(s)=\begin{pmatrix}
1 &\quad  s\delta \tau\cr
0 &\quad 1
\end{pmatrix} ,\nonumber\\
\delta C(c)&=&\begin{pmatrix}
	1+c\delta \tau  & 0\cr
	0 &1+c\delta \tau 
\end{pmatrix},\ 
\delta R(\varpi)=\begin{pmatrix}
	1 &\  -\varpi\delta\tau \cr
	\varpi\delta \tau  &1
\end{pmatrix}\ .
\ee
Those four transformations depend on four parameters: $z$, $s$, $c$ and $\varpi$, which are defined through the equations above. 

Clearly, the rate-of-expansion tensor represents the rate of scale change, and thus can be written in terms of $\delta C$ as $(\mathrm{I}+\delta\tau \Theta)=\delta C$, which is proportional to the identity matrix. Since it commutes with the rest of the maps, we need to focus only on the traceless part, $\Psi$, of the velocity gradient. Note that $\Psi$ depends on three degrees of freedom since $m$ has four elements, but we imposed the condition  $\mathrm{tr}(\Psi)=0$. One of those degrees of freedom amounts to a trivial choice of orthonormal basis, which is defined up to an overall rotation. Thus, $\Psi$ has only two non-trivial degrees of freedom, which means that it can be decomposed in various ways using the infinitesimal maps $\delta Z$, $\delta S$ and $\delta R$, given in eq.~(\ref{infmaps}). Let us explore those decompositions below.

\begin{figure}[h!]\epsscale{0.65}
\plotone{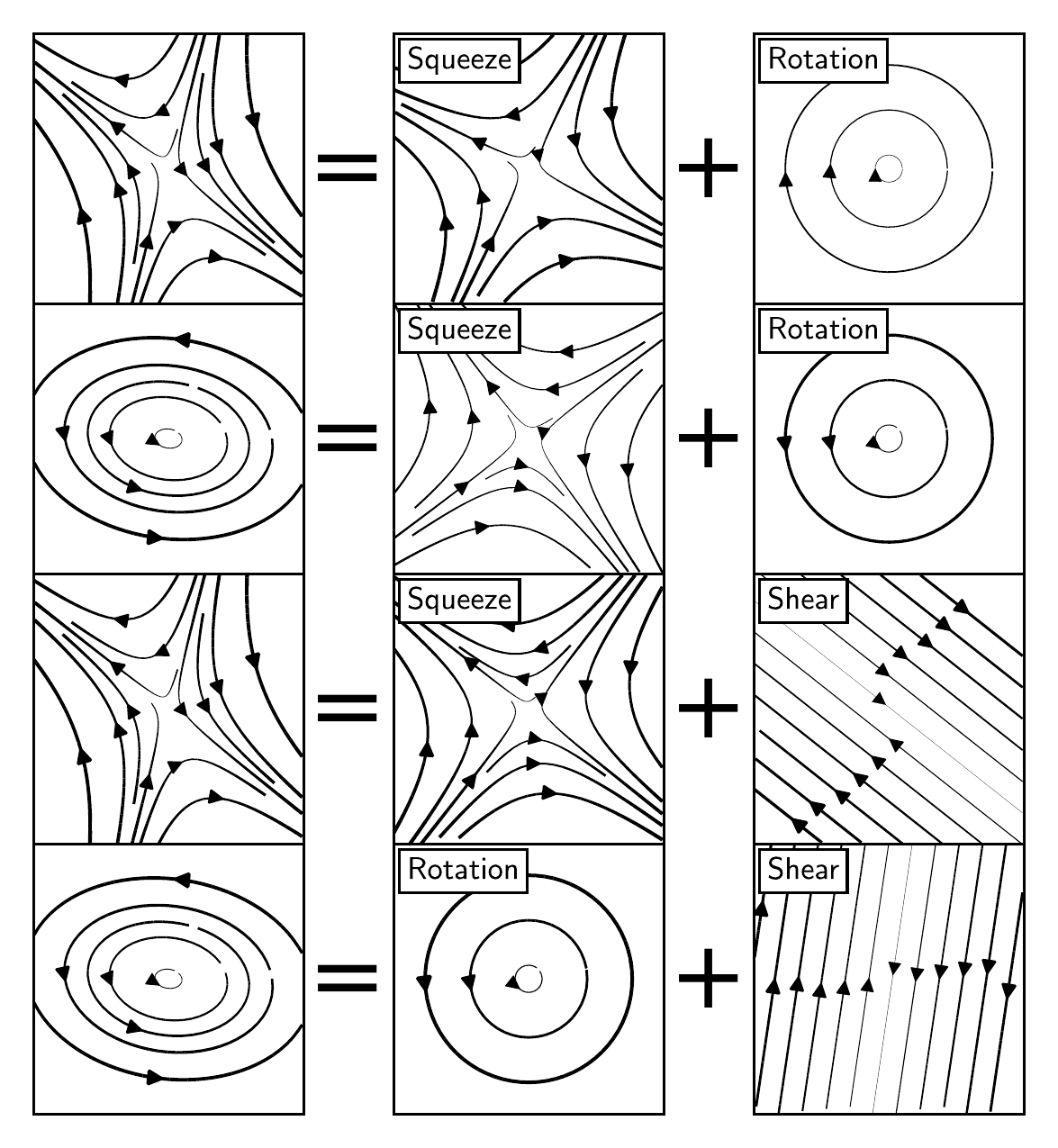}\caption{
In this figure we illustrate different decompositions of the two generic types of flows (saddle and center) that can be produced by the traceless part ($\Psi$) of the velocity gradient of a planar fluid flow. In a finite difference sense, the evolution of the flow  due to $\Psi$ between any two successive time-steps can be decomposed into a composition of two infinitesimal maps (chosen out of the following list: squeeze, rotation, shear). The velocity field produced by $\Psi$ is a linear superposition of the velocity fields produced by the two infinitesimal maps into which it is decomposed. In the figure, we show streamlines traced into the flow of those velocity fields (with the magnitude of the velocity indicated by the thickness, not by the density, of streamlines). The standard decomposition of $\Psi$ into a rate-of-shear and a rate-of-rotation tensors corresponds to a decomposition into an infinitesimal squeeze and rotation maps. That standard flow decomposition is shown in the first two rows of the figure above. That decomposition is not unique as the bottom two rows illustrate, where we show alternative decompositions of the flow. 
See the text for further discussion.}\label{fig:decomp}
\end{figure}

Since the rate of shear, $\Sigma$, is symmetric, it has two orthogonal eigenvectors, and because it is traceless, its eigenvalues are opposite in sign. Thus,  $(\mathrm{I}+\Sigma\delta\tau)$ can be treated as an infinitesimal squeeze map, $\delta Z$, along the eigenvectors of $\Sigma$. As $\Omega$ is antisymmetric, $(\mathrm{I}+\Omega\delta\tau)$ is an infinitesimal rotation map, $\delta R$, independent of the choice of orthonormal basis vectors. Thus, (up to an overall rotation aligning one of the basis vectors with one the principle axes of $\delta Z$) the decomposition  $\Psi=\Sigma+\Omega$, can be written as: 
\be\label{linearSup}
\mathrm{I}+ \Psi\delta\tau\approx(\mathrm{I}+\Sigma\delta\tau )(\mathrm{I}+ \Omega\delta\tau)=\delta Z\delta R\approx\delta R\delta Z\ .
\ee
As eq.~(\ref{2dflowFD}) is linear in $\delta\tau$, we kept only terms to linear order in $\delta\tau$ in the above equation. Thus, the velocity fields generated by each of the infinitesimal maps above can be simply added (as in linear superposition) to obtain the velocity field of the composite infinitesimal map $\delta R\delta Z$. This can be seen by using $\Psi$ from 
eq.~(\ref{linearSup}) in combining equations~(\ref{2dflow}) and (\ref{stddecomprates1}). Thus, in the deformation map language, the standard decomposition of $\Psi$ into  rate-of-shear and rate-of-rotation tensors corresponds to the superposition of an infinitesimal squeeze flow and an infinitesimal rotation flow. An illustration of that superposition is shown in the first two rows of  Fig.~\ref{fig:decomp} for the two possible flows generated by $\Psi$: a saddle and a center.

As noted earlier, one can use other (non-standard) decompositions of $\Psi$, however. For saddles, one can write $(\mathrm{I}+\Psi\delta\tau )$ using the Schur decomposition  as a rotated upper triangular matrix, which can in turn be decomposed \footnote{In the context of fluid flows, see \citep{2017arXiv171002760K} for an example usage of the Schur decomposition; and see \citep{KOLAR2007638} for an example of using, what they call, shear-plus-residual decomposition.}  as $\delta Z\delta S$. That  decomposition can be achieved by choosing the (repeated) eigenvectors of $\delta S$ to coincide with one of the eigenvectors of $\delta Z$ as well as with one of the eigenvectors of $\Psi$. That shear-squeeze decomposition is illustrated in the third row of  Fig.~\ref{fig:decomp}.  For centers, one can decompose  $(\mathrm{I}+\Psi\delta\tau )$ as  $\delta R\delta S$, with the shear aligned with one of the semi-axes of the concentric elliptical trajectories around the center\footnote{\label{ft:eig}A closed-form expression  which allows one to determine the direction $\r_{\mathrm{sm}}$ of the semi-major axis of elliptical flow around a center generated by a matrix $\Psi$ is not readily available in the literature. Thus, we include that here for completeness. First, one needs to find the complex conjugate eigenvectors $\bm e^{\pm}$ of $\Psi$. Then, after a bit of algebra, one can show that  the following relationship holds between the $x$ and $y$ components of $\r_{\mathrm{sm}}$ and $\bm e^{\pm}$: $$\tan^{-1}\left(\frac{e^{\pm}_y}{e^{\pm}_x}\right)=\tan^{-1}\left(\frac{r_{\mathrm{sm},y}}{r_{\mathrm{sm},x}}\right)\pm i\tanh^{-1}\left(\frac{S_i}{S_a}\right),$$ where $i=\sqrt{-1}$, and $S_i/S_a$ is the (fixed) aspect ratio (the ratio of the semi-minor ($S_i$) to semi-major ($S_a$) axis) of one of the concentric ellipses around the center. Taking the real and imaginary parts of that expression allows one to find the direction of $\r_{\mathrm{sm}}$, as well as the aspect ratio of the elliptical flow.}.  That non-standard decomposition is illustrated in the fourth row of  Fig.~\ref{fig:decomp}.

In the rotation-squeeze decomposition of a center flow, one can show that the rotation rate is different than the one obtained in the rotation-shear decomposition; and furthermore, the latter is not unique as it depends on whether the shear is aligned with the semi-minor or semi-major axis of the elliptical flow. Similarly, the squeeze rate of a saddle flow depends on whether the flow is decomposed into squeeze plus shear or into squeeze plus rotation flows (for the latter, see Section~\ref{sec:sym}). Thus, defining local squeeze and rotation rates is prone to pitfalls if one is not careful in stating precisely what one is trying to quantify. Below we include an example which is a clear illustration of that.

One standard quantifier of flow rotation is vorticity, $\bm\omega\equiv \bm{\nabla}\times\v$, where $\v$ is the velocity field of the flow. A planar flow (in a plane spanned by the basis $\hat{\bm x}$, $\hat{\bm y}$) has a vorticity vector which is normal (along the third basis vector $\hat{\bm z}$) to the flow and is given by 
\be\label{omegaz}
\omega_{z}=\nabla_xv_y-\nabla_yv_x=m_{yx}-m_{xy}\ , 
\ee
which has a magnitude equal to $\sqrt{-2\mathrm{tr}(\Omega^2)}$ and is, therefore, a rotation invariant. Thus, the vorticity equals twice the rotational rate ($\varpi$; see eq.~(\ref{infmaps})) one infers from the rotation-squeeze decomposition of the velocity gradient tensor. As the squeeze map is symmetric, it does not produce vorticity for that decomposition. Meanwhile, in the rotation-shear decomposition of a center flow, both the infinitesimal shear and rotation produce vorticity, and therefore, the rotational rate associated with the infinitesimal rotation map in that decomposition is not as straightforward to relate to vorticity. 

A standard intuitive way \citep[e.g.][]{2017mcpo.book.....T} to understand vorticity is to imagine placing a vane with orthogonal fins in the flow. Such a vane will spin at an angular velocity given by half the vorticity as can be seen by averaging the angular velocity obtained from eq.~(\ref{2dflow}) for any two orthogonal vectors $\r$. Such a vane would also rotate in sheared flows, as well as in saddle flows with non-orthogonal eigenvectors. Yet, in the case of shear or saddle flow, fluid elements do not revolve around  $\r=0$; and thus, the number of turns they make around $\r=0$ is always less than one, despite the fact that the imaginary vane makes $\mathrm{N_t}$ turns for a time interval $\Delta\tau$:
\be\label{twist_fluids}
\mathrm{N_t}=\frac{1}{2\pi}\left(\frac{\omega_z}{2} \Delta \tau\right)\ .
\ee
Moreover, even for elliptical flows, below we  show that  the rate at which a vane would rotate fails to match the average rate at which fluid elements revolve around $\r=0$  (see Section~\ref{sec:ellSq}). Thus, clearly vorticity is not the quantity we should be using to quantify the number of turns a fluid element makes around $\r=0$ over $\Delta\tau$.  

At this point, the reader should compare the above equation with the equation for the standard local magnetic twist number (\cite{2006JPhA...39.8321B}; see also our eq.~(\ref{twistN})), which we reproduce here for convenience:
\be\label{twistN1}
\mathrm{N_t}=\frac{1}{2\pi}\int_L d\tau \frac{\alpha(\tau)}{2}\ .
\ee
Here, the integral is  over a field line $L$; while $\alpha$ is the generalization of the force-free parameter  (which we define for any magnetic field $\B$ as $\alpha\equiv(\bm \nabla\times\bm B)\cdot\bm B/B^2$). For magnetic fields, $\alpha$ plays the role of the fluid vorticity above (compare eq.~(\ref{omegaz}) with eq.~(\ref{alpha_tmp})).  In Section~\ref{sec:coiling}, we show that $\alpha$ equals twice the average of the angular rate of motion of neighboring field lines (whether that angular motion is due to rotation or shear) around a reference field line (cf. eq.~(\ref{omegaAve})). The average is taken over neighboring field lines, assuming they are  uniformly distributed in angle, similar to the way the orthogonal fins of the vane, floating in the fluid flow described above, sample streamlines equally spaced in angle. Thus, similar to the fluids case, the standard local twist number fails to give the correct number of turns magnetic field lines make around a reference field line when they exhibit saddle-like transverse flow. $\mathrm{N_t}$ fails to return the correct number of turns even for elliptical transverse field-line flows (see Sections~\ref{sec:ellSq} and \ref{sec:w0alpha}). We  address that issue in the next section.

\subsection{Coiling and squeezing rates, and the geometric interpretation of vorticity and the force-free parameter}

In the previous section, we wrote the vorticity in the standard way, in terms of the rotation rate tensor, which we related to an infinitesimal rotation map. In this section we  reinterpret the vorticity of a 2D flow in terms of finite (as opposed to infinitesimal) deformation maps of the flow. Along the way, we will be able to construct  local measures of the rotation and squeeze rates of neighboring field lines.

Let us start by writing the finite squeeze ($Z$), shear ($S$), scaling ($C$) and rotation ($R$)  maps, obtained by composing (infinitely many of) their corresponding  infinitesimal maps (eq.~(\ref{infmaps})): 
\be\label{maps}
Z(z_f)&=&\begin{pmatrix}
	z_f  &\quad  0\cr
	0 &\quad 1/z_f
\end{pmatrix}, \ 
S(s_f)=\begin{pmatrix}
	1 &\quad  s_f\cr
	0 &\quad 1
\end{pmatrix} ,\nonumber\\
C(c_f)&=&\begin{pmatrix}
	c_f  &\quad  0\cr
	0 &\quad c_f
\end{pmatrix},\ 
R(\theta)=\begin{pmatrix}
	\cos(\theta) & \quad -\sin(\theta) \cr
	\sin(\theta)  &\quad \cos(\theta)
\end{pmatrix}\ ,
\ee
with the finite map factors $z_f$, $s_f$, $c_f$, $\theta$ defined through the equations above. Any rate of expansion ($\Theta$) is uniquely captured by a scale map ($C$) which, being proportional to the identity matrix, commutes with all other maps. Therefore, similar to the previous section, here we focus solely on the traceless part,  $\Psi$, of the velocity gradient. 

In this section, we decompose $\Psi$ into a set of finite transformations (given by eq.~(\ref{maps})) and one infinitesimal transformation (eq.~(\ref{infmaps})). The role of the finite maps is  to undo certain deformations to the flow, so that we can use only \textit{one} infinitesimal map to describe its evolution. When combining infinitesimal maps, we showed that the velocity field produced by them can be treated as a linear superposition of the  velocity field produced by each individual infinitesimal map (see the discussion around eq.~(\ref{linearSup})). However, under the decomposition we are about to write down, not all maps are  infinitesimal, and therefore one can no longer use linear superposition to study the map combinations. Yet, the interpretation of the results in the end will be as intuitive.

Let us focus on  transformations of the type:
\be\label{finmapdecomp}
\mathrm{I}+\Psi\delta\tau=R\, A\,  \delta B\,  A^{-1}\, R^{-1}\ ,
\ee 
with $A$ and $\delta B$ being  a finite and an infinitesimal transformation map, respectively. The finite rotation maps above are used to rotate the flow to the principle axes of the transformation $A$. The rotation maps do not affect the invariants of $\Psi$, and we will therefore omit them from most of the discussion below. The  decomposition shown in the equation above has three parameters (one for each of the maps), matching the degrees of freedom of $\Psi$.

Note that the structure of the decomposition in eq.~(\ref{finmapdecomp}) is chosen to be such that the flow, $\tilde \r(\tau)$, generated by the traceless part of (the constant) $m$ in eq.~(\ref{2dflowFD})  can be solved to give the position of a fluid element after a finite interval $\Delta\tau=n\delta\tau$ (with $n\to\infty$)  as:
\be\label{2dflowFDfinite}
\tilde r_i(\tau+n\delta\tau)&=&(\mathrm{I}+\delta\tau \Psi)_{ij}\tilde r_j\bigg(\tau+(n-1)\delta\tau\bigg)=(\mathrm{I}+\delta\tau \Psi)_{ij}^n\tilde r_j(\tau)\nonumber\\
&=&\big(R\, A\,  \delta B\,  A^{-1}\, R^{-1}\big)^n_{ij}\tilde r_j(\tau)=\big(R\, A\,  \left(\delta B\right)^n\,  A^{-1}\, R^{-1}\big)_{ij}\tilde r_j(\tau)\nonumber\\
&=&\big(R\, A\,  B\,  A^{-1}\, R^{-1}\big)_{ij}\tilde r_j(\tau)\ .
\ee
In the last equality we used the fact that the composition of infinitesimal maps $\delta B$ results in a finite map $B$. Thus, the flow produced by $\Psi$ after some finite $\Delta \tau$ matches the expression for the infinitesimal flow (eq.~(\ref{finmapdecomp})) with the infinitesimal $\delta B$ replaced by a finite $B$. Thus, one can think of this decomposition as  a global deformation of the flow done by $A$, with the $\tau$-evolution  captured by $B$ (see e.g. Fig~\ref{fig:decompFin}, discussed below). This will allow us to interpret our results more easily below.

Transformation $A$ can be either a shear or a squeeze (as any rotation can be absorbed into $R$). Therefore, the non-trivial combinations of the pairs  $(A,\delta B)$ are: (squeeze, shear), (shear, squeeze), (shear, rotation), (squeeze, rotation), where we eliminated repeated pairs as the composition of $A\,\delta A$ is still a transformation of the same type, $A$, which is not sufficient to  parametrize generic elliptic and saddle flows. 

Of the pairs listed above, one can check explicitly that the (squeeze, shear) pair gives a composition $Z\,  \delta S\,  Z^{-1}$, which is again an infinitesimal shear map, and therefore, cannot describe generic elliptic and saddle flows. The pair (shear, squeeze) has two real eigenvalues and therefore can describe saddles. The (shear, rotation) and (squeeze, rotation) pairs have two complex conjugate eigenvalues and therefore can describe centers. We find the latter pair more intuitive when discussing elliptical flows, and therefore we  focus on it  below. We  return to saddle flows in Section~\ref{sec:saddleSq}.

\subsubsection{Elliptical flows as Squeeze -- Infinitesimal rotation -- Un-squeeze maps}\label{sec:ellSq}

\begin{figure}[t!]\epsscale{0.9}
\plotone{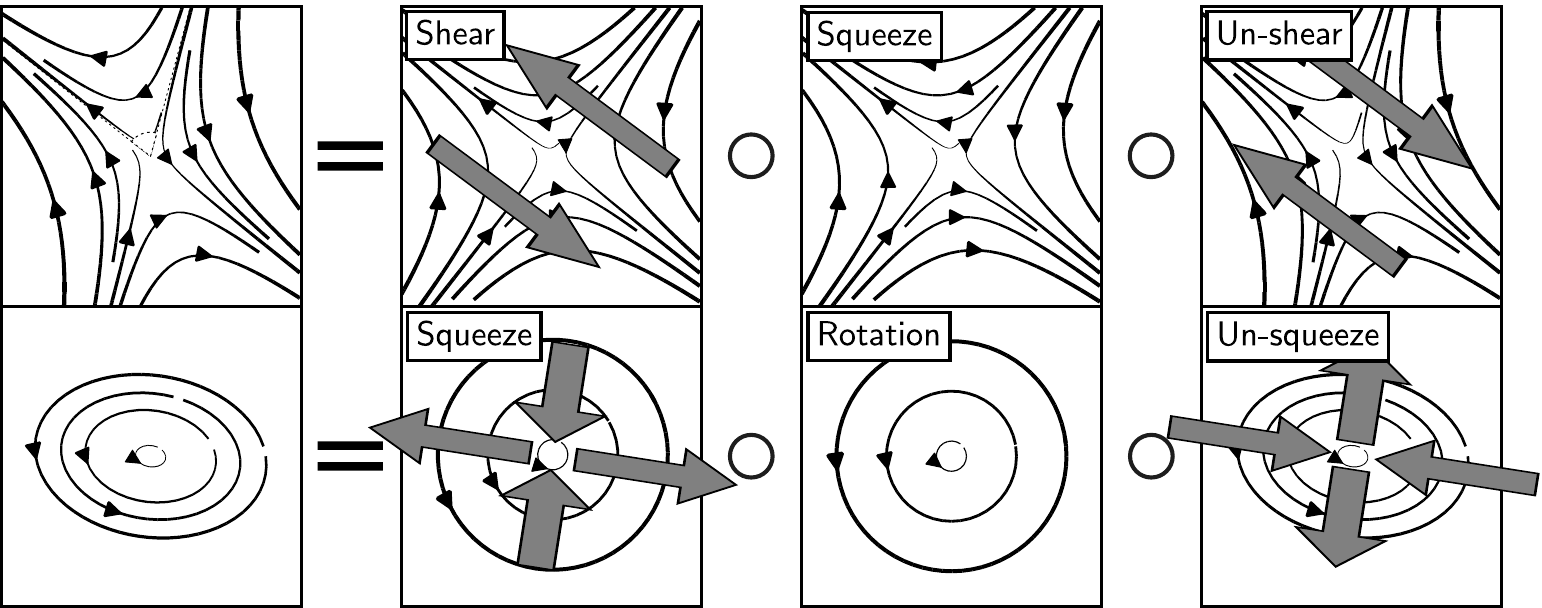}\caption{
Same as in Fig.~\ref{fig:decomp}. However, here the flow produced by $\Psi$ is decomposed into two mutually inverse finite maps bracketing an infinitesimal map. As not all maps in this decomposition are infinitesimal, the velocity field produced by $\Psi$ is no longer a linear superposition of the velocity fields produced by the individual maps. Instead, one should think of the maps as: first undoing a finite deformation of the flow (forth column), then applying an infinitesimal map (third column), and then redoing the finite deformation of the flow (second column). Thus, an elliptical flow (second row) can be regarded as a deformed uniform circular flow: at each time-step, one has to un-squeeze the elliptical flow into a circular flow, then rotate the fluid infinitesimally, and then squeeze it back into an elliptical flow. One can perform a similar procedure for saddle flow with non-orthogonal asymptotes: at each time-step, one can un-shear the flow to make the asymptotes orthogonal, then infinitesimally squeeze the flow, then shear it back again. Due to the structure of the decomposition, the middle step can be infinitesimal or finite (see the text). We show that the flow vorticity (in the context of magnetic fields, vorticity corresponds to the generalized force-free parameter, $\alpha$) depends on the overall geometric distortion of the flow. For saddle flows, vorticity depends on the angle, $\vartheta$, between the two asymptotes of the flow (shown as dashed lines in the top-left panel). For elliptical flows,  vorticity depends on the aspect ratio of the ellipses traced out by the fluid elements. The coiling rate ($\omega_c$) defined in this paper corresponds to the rate of rotation in the middle step of the Squeeze -- Rotation -- Un-squeeze decomposition of elliptical flows (bottom row). The squeezing rate ($\rho_{\mathcal{Z}}$) defined in this paper, corresponds to the rate of squeezing in the middle step of the Shear -- Squeeze -- Un-shear decomposition of saddle flows (top row). See the text for further discussion.}\label{fig:decompFin}
\end{figure}

Let us first focus on describing centers using the pair ($A,\delta B$)=(squeeze, rotation) in eq.~(\ref{finmapdecomp}). The rotation matrices in that equation can be used to rotate  the basis vectors by an angle $\theta$ so that they are aligned with the semi-axes of the elliptical flow around the center generated by $\Psi$. Then $Z^{-1}(z_f)$ can be used to un-squeeze the elliptical flow and render it into a uniform circular flow, rotating with a constant angular frequency $\omega_c$, which we will call the \textit{coiling rate}. At each time-step ($\delta\tau$), that rotation is generated by $\delta R(\omega_c)$. After that, we need to squeeze  the circular flow back into an elliptical flow. For an elliptical flow described by concentric ellipses with a fixed aspect ratio $S_i/S_a$ (where $S_i$ and $S_a$ are the semi-minor and semi-major axes of one of those ellipses, respectively), the squeeze factor must equal $z_f=\sqrt{S_a/S_i}$ as $Z(z_f)$ (eq.~(\ref{maps})) rescales the component $r_x$ of $\r$ by $z_f$ and $r_y$  by $1/z_f$. Thus, we can write:
\be\label{sqrot}
\mathrm{I}+\Psi\,\delta\tau=R(\theta)\, Z\left(\sqrt{\frac{S_a}{S_i}}\right)\,  \delta R(\omega_c)\,  Z^{-1}\left(\sqrt{\frac{S_a}{S_i}}\right)\, R^{-1}(\theta)\ .
\ee
That decomposition is illustrated in the second row of Fig.~\ref{fig:decompFin}.

Using equations~(\ref{infmaps})  and (\ref{maps}), we can write out the right-hand-side of eq.~(\ref{sqrot}) explicitly as a matrix and then calculate its invariants. We find that the eigenvalues\footnote{In the context of fluid flows, a vortex detection criterion using the imaginary part of the complex eigenvalues (referred to as the \textit{swirling strength} of a vortex) of the velocity gradient was used by \cite{zhou_1999}. Even though that criterion can be directly ported to  magnetism by using the gradient of the magnetic field, our method differs in several significant ways. (1) As we show in Section~\ref{sec:coiling}, we use the gradient of the \textit{normalized} magnetic vector field to calculate $\omega_c$ (which is why we choose to call it by a different name: the \textit{coiling rate}).  For fluids, normalizing the velocity field would break Galilean invariance for the vortex detection method; however, in the case of magnetic fields in the solar corona, we do have a preferred reference frame. In Section~\ref{sec:fldev}, we show that  the non-zero eigenvalues of the  gradient of the normalized magnetic field in 3D match those of the 2D gradient of the rate-of-deviation of neighboring field lines in the plane normal to a reference field line. This allows for a simple interpretation of our results. (2) \cite{zhou_1999} use a vortex detection criterion by choosing an ad hoc threshold for the swirling strength. Our flux rope detection criterion is a threshold on the integral of $\omega_c$ over the field line length, which in \citep{Tassev2019} we set at $\mathrm{N_c}\gtrsim  1$ (see eq.~\ref{coilingFinal1}). In other words, our criterion is that neighboring field lines should make at least one winding around one another to classify as part of a flux rope.} of $\Psi$ in this decomposition are $(\pm i\omega_c)$. As $\Theta$ does not affect the difference of the eigenvalues of $m$, we can write that the difference between the eigenvalues ($\lambda_{\pm}$) of $m$ equals that of $\Psi$:
\be\label{omega0lambdafluids}
\lambda_{+}-\lambda_-=2 i\omega_c\ ,
\ee
and therefore, the period with which fluid streamlines revolve around $\r=0$ equals $2\pi/\omega_c$. This is not surprising in light of eq.~(\ref{2dflowFDfinite}), from which one can see that for the (squeeze, rotation) decomposition, the periodicity of the flow generated by $\Psi$ matches that of $B$, which in our case is the composition of infinitesimal $\delta R(\omega_c)$ maps. Thus, we can define a coiling number for elliptical flows as:
\be\label{coiling_fluids}
\mathrm{N_c}\equiv\frac{1}{2\pi}\omega_c\Delta\tau=\frac{1}{4\pi}\Im\{\lambda_{+}-\lambda_-\}\Delta\tau\ ,
\ee
with the coiling rate, $\omega_c$, defined as the rotational rate of the flow in the decomposition of eq.~(\ref{sqrot}) (bottom row of Fig.~\ref{fig:decompFin}).

It is important to note that the same result above (eq.~(\ref{omega0lambdafluids})) is obtained in the (shear, rotation) decomposition, highlighting the robustness of $\omega_c$. However, we find the result for vorticity obtained in the (squeeze, rotation) decomposition to be more intuitive (see below), which is why we focus on that decomposition.

Since the flow is elliptical, the angular velocity of the streamlines is not constant. Therefore, the exact number of turns a particular fluid element makes around $\r=0$  in a finite  $\Delta\tau$ depends on the initial position of the fluid element as well as on $\Delta\tau$ (cf. Section~\ref{sec:coiling}, as well as \citep{2006JPhA...39.8321B}). However, since the flow is periodic, when $\mathrm{N_c}$ is an integer, all fluid elements would have made exactly $\mathrm{N_c}$ turns around the origin, independent of their initial positions. Another way of thinking about $\mathrm{N_c}$ is that the un-squeezed flow (middle step in the decomposition shown in the second row of Fig.~\ref{fig:decompFin}) is uniformly rotating with $\omega_c$; and therefore, we can treat its rotation rate as an average of the actual fluid flow rotation rate in the sense that the periodicities of the two flows match.

Similar to the above analysis that lead to eq.~(\ref{omega0lambdafluids}), we can use eq.~(\ref{sqrot}) to calculate the  vorticity of the flow (eq.~(\ref{omegaz})):
\be\label{omegazcenter}
\omega_z=m_{yx}-m_{xy}=\Psi_{yx}-\Psi_{xy}=\left(\frac{S_i}{S_a}+\frac{S_a}{S_i}\right)\omega_c\ , \mbox{ implying:  }|\omega_z|\geq 2|\omega_c|\ ,
\ee
with the sign of $\omega_c$ chosen to match that of $\omega_z$.
Thus, for center flows, we can see that  vorticity is proportional to the coiling rate as well as to a geometric factor, related to the aspect-ratio of the elliptical flow. Thus, the local twist number, eq.~(\ref{twist_fluids}), is boosted relative to the coiling number, eq.~(\ref{coiling_fluids}), by terms dependent on that aspect ratio. 
From the discussion above, we can see that  the coiling number, $\mathrm{N_c}$, is a much better quantifier  than the local twist number for the average number of turns fluid elements make around the origin in a finite $\Delta\tau$. At the very least,  as discussed above, when $\mathrm{N_c}$ is an integer, it matches the actual number of turns the fluid elements make, while $\mathrm{N_t}$ always overestimates that number for non-circular flows.

The above result is easy to understand intuitively. Going back to our submerged vane example, a vane floating at the origin would be predominantly affected by fluid elements at the co-vertex of their streamlines (the points of closest approach to the origin, lying on the tips of the minor axis). For ellipses with large aspect ratio, one can see that at the co-vertex, fluid elements sweep large angles (relative to the origin) over short amount of time even for fluid flows with large period. Thus, our imaginary vane would rotate at an angular speed which is boosted relative to the average angular speed of the fluid elements around the origin.

In the context of transverse magnetic field-line deviations, $\alpha$  is the magnetic equivalent of $\omega_z$ (see the discussion around eq.~(\ref{twistN1})). In Section~\ref{sec:coiling}, we  find that we can write eq.~(\ref{omegazcenter}) for $\alpha$ as (cf. eq.~(\ref{finalCoiling})):
\be\label{finalCoiling1}
	\alpha=\left(\frac{S_i}{S_a}+\frac{S_a}{S_i}\right)\omega_c\ ,
\ee
with $\omega_c$ given by eq.~(\ref{omega0lambdafluids}), where the eigenvalues correspond to to the non-zero eigenvalues of the 3D gradient of the normalized magnetic field (compare equations~(\ref{omega0lambdafluids}) and (\ref{omega0Final})). Thus, for magnetic fields we can define a coiling number as:
\be\label{coilingFinal1}
	\mathrm{N_{c}}\equiv\frac{1}{2\pi}\int_L d\tau\  \omega_c(\tau)\ .
\ee
Similar to the coiling number in the fluid context, $\mathrm{N_c}$ is a much more robust estimate of the number of turns neighboring magnetic field lines make around a reference field line, getting that number exactly right for constant $\omega_c$ when $\mathrm{N_c}$  is an integer. That is again in contrast with the biased magnetic local twist number, which is boosted by the aspect ratio of the transverse elliptical field-line flow. 

\subsubsection{Saddle flows as Shear -- Infinitesimal squeeze -- Un-shear maps}\label{sec:saddleSq}

Now let us focus on representing saddle flows using the pair ($A,\delta B$)=(shear, squeeze) in eq.~(\ref{finmapdecomp}). Similar to the discussion of elliptical flows above, we can use the rotation matrices in that equation to rotate  the basis vectors by an angle $\theta$ so that one of the basis vectors is aligned with 
one of the eigenvectors of $\Psi$. Then $S^{-1}(s_f)$ can be used to un-shear the flow and render the second eigenvector of $\Psi$ orthogonal to the first. The flow would then be a saddle flow with orthogonal eigenvectors which, at each time-step ($\delta\tau$), can be generated by an infinitesimal squeeze map $\delta Z(\rho_\mathcal{Z}/2)$ with a squeeze rate $\rho_{\mathcal{Z}}$, where the factor of $1/2$ is introduced for convenience. After the infinitesimal squeeze, we need to shear back the flow with $S(s_f)$ so that the directions of its asymptotes again match the eigenvectors of $\Psi$. If we denote the angle between the eigenvectors of $\Psi$ as $\vartheta$ (see top-left panel of Fig.~\ref{fig:decompFin}), then the required shear factor needs to equal $s_f=\cot(\vartheta)$, which one can check explicitly using eq.~(\ref{maps}).
Thus, we can write this decomposition as:
\be\label{sqsh}
\mathrm{I}+\Psi\delta\tau=R(\theta)\, S\big(\cot(\vartheta)\big)\,  \delta Z\left(\frac{\rho_{\mathcal{Z}}}{2}\right)\,  S^{-1}\big(\cot(\vartheta)\big)\, R^{-1}(\theta)\ .
\ee
That decomposition is illustrated in the first row of Fig.~\ref{fig:decompFin}.

Following the analysis in the previous section, using equations~(\ref{infmaps})  and (\ref{maps}), we can write out the right-hand-side of eq.~(\ref{sqsh}) explicitly as a matrix and then calculate its invariants. We find that the eigenvalues of $\Psi$ in this decomposition are $(\pm \rho_{\mathcal{Z}}/2)$. As $\Theta$ does not affect the difference in eigenvalues of $m$, we can conclude that the difference between the eigenvalues of $m$ equals that of $\Psi$. Therefore,  the (relative logarithmic) squeeze rate between the directions corresponding to the eigenvectors of $\Psi$, and hence $m$ (as $\Theta$ does not affect the eigenvectors of $m$ either), is given by $\rho_{\mathcal{Z}}$. Indeed, if $\hat{\v}_{\pm}$ are the two unit eigenvectors of $m$ corresponding to $\lambda_{\pm}$, then for two fluid parcels lying on the asymptotes of the flow (in the eigenvector directions relative to the origin), we can write $\r_+(\tau)=r_+(\tau)\hat{\v}_+$ and $\r_-(\tau)=r_-(\tau)\hat{\v}_-$. Then, from eq.~(\ref{2dflow}), one can check explicitly that $d\ln(r_+(\tau)/r_-(\tau))/d\tau=\lambda_{+}-\lambda_-$, and therefore, we can summarize:
\be\label{squeezrate}
\frac{d}{d\tau}\ln\left(\frac{r_+(\tau)}{r_-(\tau)}\right)=\lambda_{+}-\lambda_-=\rho_{\mathcal{Z}}\ .
\ee

Using the matrix entering on the right-hand-side of eq.~(\ref{sqsh}), we can express the  vorticity of the flow (eq.~(\ref{omegaz})) as:
\be\label{omegazsaddle}
\omega_z=m_{yx}-m_{xy}=\Psi_{yx}-\Psi_{xy}=\rho_{\mathcal{Z}}\cot(\vartheta)\ .
\ee
Thus, we can see that similar to the analysis of elliptical flows in the previous section, the vorticity (and hence, the force-free parameter $\alpha$ for magnetic fields) for saddle flows  has a clear dependence on the geometry of the flow. Through its dependence on the angle ($\vartheta$) between the flow eigenvectors, the vorticity quantifies the shear necessary to make the asymptotes of the flow orthogonal. For flows with orthogonal asymptotes (produced by an infinitesimal squeeze), we have $\vartheta=\pi/2$, which means that  there is no overall shear in this decomposition, and no vorticity. 

In the context of transverse magnetic field-line deviations, one can use the squashing factor ($Q$) \citep{Titov07} to quantify the divergence of neighboring field lines, separated by $\r$ in the plane normal to one of them. Let us see how  $\rho_{\mathcal{Z}}$ relates to $Q$. We review the calculation of $Q$ in Section~\ref{sec:localQintro}. Here we highlight just the parts necessary for this section. The squashing factor is a function (see eq.~(\ref{QJ})) of the transformation matrix $J$ relating $\r$ evaluated at two different locations along a field line: at $\tau$ and $\tau+\Delta\tau$ for some finite $\Delta\tau$:
\be
r_i(\tau+\Delta\tau)=J_{ij}r_i(\tau)=(\mathrm{I}+\delta\tau m)^{\Delta\tau/\delta\tau}_{ij}r_i(\tau)\ ,
\ee
where we used eq.~(\ref{2dflowFD}). One can decompose $m$ into a vector times the Pauli matrices vector, then use the exponentiation of a Pauli vector formula to obtain in the limit of $\delta\tau\to0$:
\be
J=\exp\left(\frac{\mathrm{tr}(m)}{2}\Delta\tau\right)\times\left[\cosh\left(\frac{\rho_{\mathcal{Z}}}{2}\Delta\tau\right)\mathrm{I}+\Psi\,\frac{2\sinh\left(\frac{\rho_{\mathcal{Z}}}{2}\Delta\tau\right)}{\rho_{\mathcal{Z}}}\right]\ ,
\ee
where we used equations~(\ref{stddecomprates}) and (\ref{squeezrate}). This can in turn be used to calculate the squashing factor  (using eq.~(\ref{QJ})). After some algebra, we obtain:
\be\label{Q2dflowExact}
Q=2\left[\csc^2(\vartheta)\,\cosh\big(\rho_{\mathcal{Z}}\Delta\tau\big)-\cot^2(\vartheta)\right]\ ,
\ee
with $\vartheta$ given by eq.~(\ref{omegazsaddle}).
The equation above is valid for any steady-state planar flow.  

We are interested in large $Q\gg 1$, which holds\footnote{In the limit $\vartheta\to 0$, $m$ has repeated eigenvectors, and despite the eigenvalues being the same (implying $\rho_{\mathcal{Z}}\to 0$), $Q$ can still grow large, as can be seen from eq.~(\ref{Q2dflowExact}). However, one can check that $Q$ in that case grows only as $\Delta\tau^2$, and not exponentially, and therefore we disregard those special cases.} for large \textit{real} arguments of $\cosh()$, in which case the logarithm of $Q$ approaches $\ln(Q)\sim \rho_{\mathcal{Z}}\Delta\tau$. Therefore, we can  approximate $\ln(Q)$ as an integral over the local squeezing rate, given by $\rho_{\mathcal{Z}}$ in eq.~(\ref{squeezrate}). We denote that approximation of $Q$ with $\mathcal{Z}$, and we will refer to the latter as the \textit{squeeze factor}. Therefore, the squeeze factor is the solution to the equation (compare with eq.~(\ref{rhoQreal}) and (\ref{qhoQFinal})):
\be\label{rhoQreal1}
\frac{d \ln Q}{d\tau} \sim\frac{d \ln \mathcal{Z}}{d\tau}= \rho_{\mathcal{Z}}= \lambda_{+}-\lambda_{-}\quad \mbox{for real eigenvalues of $m$},
\ee
with the squeezing rate, $\rho_{\mathcal{Z}}$, defined as the squeezing rate of the flow in the decomposition\footnote{Note that $\lambda_+-\lambda_-$ also equals the squeezing rate one would obtain from the infinitesimal squeeze plus infinitesimal shear decomposition of the flow (third row of Fig.~\ref{fig:decomp}). In that decomposition, the vorticity of the flow equals the negative of the shear rate entering in $\delta S$. In contrast, the decomposition used in this section allows one to interpret vorticity through geometric distortions of the flow (eq.~(\ref{omegazsaddle})), bringing it in parallel with the discussion presented in Section~\ref{sec:ellSq} for elliptical flows.} of eq.~(\ref{sqsh}) (top row of Fig.~\ref{fig:decompFin}). 

From the above equation, we can see that we can treat the squeezing rate as a local approximation for the logarithmic rate of squashing for constant $m$.
In Section~\ref{sec:localQden} we generalize
the above equation (using somewhat different arguments) to field lines in 3D with non-zero curvature and show that in that case, the eigenvalues entering above correspond to the two non-zero eigenvalues of the 3D gradient of the normalized magnetic field.

\subsubsection{Squeezing in elliptical flows}\label{sec:sqIm}

Our result (eq.~(\ref{rhoQreal1})) from the previous section applies to  squeezing under saddle flows. In the context of magnetic fields, under elliptical transverse flows, a flux tube is periodically squashed and un-squashed; thus, on average, $Q$ remains the same. This can be seen from eq.~(\ref{Q2dflowExact}), which is valid even for complex eigenvalues of $m$, in which case $\rho_{\mathcal{Z}}=2i\omega_c$ (compare equations~(\ref{omega0lambdafluids}) and (\ref{squeezrate})). Note that $Q$ can periodically  still grow large, however. It is maximized at $\Delta\tau=\pi/(2\omega_c)$, and every $\Delta\tau=\pi/\omega_c$ later along the field-line, in which case one can show that $Q$ equals $Q_{\mathrm{max.}}=S_a^2/S_i^2+S_i^2/S_a^2\geq 2$, where we used eq.~(\ref{omegazcenter}) and (\ref{omegazsaddle}) to eliminate $\vartheta$ from eq.~(\ref{Q2dflowExact}). Comparing with eq.~(\ref{Qaspect}) which we write down in our review Section~\ref{sec:localQintro}, a flux tube, with an initially circular cross-section, within a quarter of a period (by ``period'' we loosely refer to the field-line length corresponding to $2\pi/\omega_c$) is transversely squashed into an ellipse with an aspect ratio equal to the square of the aspect ratio of the integral lines of the transverse flow. In the next quarter of a period, the flux tube gets completely un-squashed back into a circle, and so on. Thus, one can define an average squashing rate (localized) within each quarter of a period: starting with an initial $Q=2$, within a quarter of a period, $\ln(Q)$ changes by:
\be\label{possibleImRhoZtmp}
\left|\frac{\Delta\ln{Q}}{\Delta\tau}\right|&=&\frac{2|\omega_c|}{\pi}\ln\left(\frac{Q_{\mathrm{max.}}}{2}\right)=\frac{2|\omega_c|}{\pi}\ln\left[\frac{1}{2}\left(\frac{S_a^2}{S_i^2}+\frac{S_i^2}{S_a^2}\right)\right]\ .
\ee

So, at this point we have two different possibilities ($\rho_{\mathcal{Z},1|2}$) for the logarithmic squeezing rate under elliptical flows. If one is not interested in periodic localized flux-tube squashing, one can use:
\be\label{possibleImRhoZ1}
\mbox{Option 1:}\quad \frac{d \ln \mathcal{Z}_1}{d\tau}\equiv\rho_{\mathcal{Z},1}= 0\quad \mbox{for complex eigenvalues of $m$}
\ee
since $Q$ periodically returns to 2 as discussed above.
If localized periodic squashing is of interest, however, 
one can take that into account by defining the squeeze factor as:
\be\label{possibleImRhoZ2}
\mbox{Option 2:}\quad \frac{d \ln \mathcal{Z}_2}{d\tau}\equiv\rho_{\mathcal{Z},2}=\frac{2|\omega_c|}{\pi}\ln\left(\frac{\alpha^2}{2\omega_c^2}-1\right)\quad \mbox{for complex eigenvalues of $m$.}
\ee
In the equation above, we used eq.~(\ref{finalCoiling1}) to express eq.~(\ref{possibleImRhoZtmp}) using the generalized force-free parameter. In Section~\ref{sec:coiling}, we  express $\alpha$ and $\omega_c$ as scalars constructed using the gradient of the normalized magnetic field.

The ambiguity in the definition of $\mathcal{Z}$ above for elliptical flows is a consequence of the fact that the choice of squeezing rate is, in the end,  application specific. In other words, our definition for $\mathcal{Z}$ offers fine-grained control over which effects causing squeezing are included and which -- not. Note, however, that in defining the coiling number in eq.~(\ref{coiling_fluids}) we ignored any partial coiling caused by transverse saddle flows. So, under Option 1 above (eq.~(\ref{possibleImRhoZ1})), $\mathcal{Z}$ carries complementary information to that of $\mathrm{N_c}$: the former would depend only on the real part of the eigenvalue difference of $m$, while the latter depends only on the imaginary part\footnote{Indeed, an infinitesimal rotation map can be diagonalized to give an infinitesimal  squeeze map with an imaginary squeeze factor; see eq.~(\ref{infmaps}).}.

\section{Field-line deviations}\label{sec:fldev}

In this section, we write an equation for the relative displacement of neighboring field-lines in the 2D subspace spanned by the plane normal to those field lines. Those transverse field-line deviation vectors are changing as one moves the normal plane down the field lines. Thus, the flow of the transverse  deviations with the field-line length parameter ($\tau$) is planar, and takes the form of eq.~(\ref{2dflow}), which was the starting point of our analysis in the previous section. However, in order to write that equation in terms of the magnetic field $\B$, one has to address  issues arising from the fact that generic magnetic field lines have curvature, and therefore the normal plane does not have a constant orientation in 3D. We address those complications in this section. 

To simplify the analysis, we write our equations in a 3D Cartesian basis. We generalize the results to curvilinear coordinates in Appendix~\ref{sec:app}.

\subsection{Evolution of the field-line deviation vector in 3D}
Let us pick two neighboring field lines with radius vectors $\x(\tau)$ and $\y(\tau)$ with $\tau$ being the field-line length parameter. We pick an initial condition  such that $\tau=\tau_0$ when both field lines intersect the plane normal to the first field line at $\x(\tau_0)$. The magnetic field lines $\x(\tau)$ and $\y(\tau)$ are calculated as the integral curves of the unit magnetic field vectors, $\hatbB$. Thus, in Cartesian coordinates, we have:
\be\label{fl}
\frac{d\x(\tau)}{d\tau}=\hatbB(\x(\tau))\ ,
\ee
and similarly for $\y(\tau)$.

Then we can construct the field-line deviation vector $\delta\x(\tau)\equiv \y(\tau)-\x(\tau)$. Using (\ref{fl}), the deviation vector $\delta \x(\tau)$ is a solution to the linearized equation:
\be\label{dev}
\frac{d\delta \x(\tau)}{d\tau}=\hatbB\Big(\y(\tau)\Big)-\hatbB\Big(\x(\tau)\Big)\approx \bigg(\delta \x(\tau)\cdot\bm{\nabla}_{\x} \bigg)\hatbB\Big(\x(\tau)\Big)\ .
\ee
Let us write this equation in component form, suppressing the $\tau$ dependence on the right hand side  (we use the Einstein summation convention):
\be\label{ddx}
\frac{d\delta x_i(\tau)}{d\tau}=\left(\nabla_j \hat B_i\right) \delta x_j\equiv M_{ij}\delta x_j\ ,
\ee
where the last equality defines the (generally) non-symmetric matrix $M$. Index placement in this section is unimportant as we work in a Cartesian basis.

 Let us additionally define the deviation vector $\r(\tau)\equiv \y(\tau'(\tau))-\x(\tau)$, with $\tau'$ chosen such that $\r(\tau)$ is orthogonal to the field line at $\x(\tau)$: $$\bm{r}(\tau)\cdot\hatbB(\x(\tau))=0\ .$$Above we used the fact that $\hatbB(\x)$ is the tangent vector for the field line passing through $\x$. Choosing the field lines $\x(\tau)$ and $\y(\tau)$ infinitesimally close to each other, we can write $\tau'(\tau)=\tau+\varepsilon(\tau)$ with $\varepsilon$ being small. Thus, we can linearize: $$\r(\tau)\approx \y(\tau)-\x(\tau)+\varepsilon \frac{d\y(\tau)}{d\tau}=\delta\x(\tau)+\varepsilon\hatbB(\y(\tau))\approx\delta\x(\tau)+\varepsilon\hatbB(\x(\tau))\ ,$$ where we used eq.~(\ref{fl}). Setting $\bm{r}(\tau)\cdot\hatbB(\x(\tau))=0$, we find $\varepsilon(\tau)=-\delta\x(\tau)\cdot\hatbB(\x(\tau))$, which in turn implies\footnote{ The same result was obtained by \cite{2017ApJ...848..117S}.} (in matrix notation):
\be\label{r}
\bm{r}(\tau)= {P_\perp}\delta\x(\tau)\ ,
\ee
where the projection operator is given in component form by: $P_{\perp,ij}(\x(\tau))=\mathrm{I}_{ij}-\hat B_i(\x(\tau))\hat B_j(\x(\tau))$. From eq.~(\ref{r}) we can see that $\r$ equals the projection of $\delta \x$ onto the plane normal to the reference field line at location $\x(\tau)$.

We can write an evolution equation for $\r(\tau)$ similar to eq.~(\ref{dev}) for $\delta\x(\tau)$. 
To do that, we need to find the derivative of the projected deviation, $\r$, which is given by eq.~(\ref{r}). Thus, we need to take the derivatives of $\hatbB$ and then $P_{\perp}$ with respect to $\tau$:
\be
\frac{d\hat B_i(\x(\tau))}{d\tau}=\frac{dx_j(\tau)}{d\tau}\nabla_j \hat B_i(\x(\tau))=\hat B_j(\x(\tau))\nabla_j \hat B_i(\x(\tau))=M_{ij}\hat B_j\ ,\nonumber\\
\frac{dP_{\perp, ij}(\x(\tau))}{d\tau}=-\frac{d\left[\hat B_i(\x(\tau))\hat B_j(\x(\tau))\right]}{d\tau}=-\left(M(\mathrm{I}-P_{\perp})+(\mathrm{I}-P_{\perp})M^{\mathrm{T}}\right)_{ij}\label{dP}\ ,
\ee
where  $\mathrm{T}$ denotes matrix transpose.
Let us also write down the following useful identities:
\be
P_{\perp}^n=P_{\perp}\ ,&\quad& P_{\perp}(\mathrm{I}-P_\perp)=0 \ ,\nonumber\\
\hat B_i M_{ij}=0\ ,  \quad
P_{\perp}M&=&M\ , \quad  M^\mathrm{T}P_{\perp}=M^\mathrm{T} \ .
\label{usefulEq}
\ee
The second line above follows from $2\hat B_i \nabla_j \hat B_i=\nabla_j \hat B^2=\nabla_j  1=0$. These properties of $\hatbB$ and its derivatives will lead to important simplifications below, which is the reason we wrote the field-line equation (eq.~(\ref{fl})) using the normalized magnetic field, and not the magnetic field itself.

Combining equations~(\ref{ddx}), (\ref{r}), (\ref{dP}) and (\ref{usefulEq}), we can write:
\be\label{dr3}
\frac{dr_i}{d\tau}=\left(MP_{\perp}-(\mathrm{I}-P_{\perp})M^{\mathrm{T}}\right)_{ij}r_j\equiv{\tilde m}_{ij} r_j\ ,
\ee
where we used the fact that $P_\perp^2=P_\perp$. The term $(\mathrm{I}-P_{\perp})M^{\mathrm{T}}$ above takes into account the rotation of the plane orthogonal to $\hatbB$ due to the changing direction of $\hatbB$, which in turn is due to the curvature of the field lines (cf. eq.~(\ref{curv}) and the discussion around it).

\subsection{Evolution of the field-line deviation vector in the normal plane}\label{sec:2d}

Let us remind the reader what the end goal is for this paper. It is obtaining \textit{local} approximations to the  squashing factor rate as well as to the field-line coiling rate. In order for those scalars to be locally defined around some field-line location $\x(\tau)$, they can depend only on the first few derivatives of $\delta \x(\tau)$, and $\r(\tau)$ in particular. If all derivatives mattered, that  would allow one to construct a Taylor expansion of $\delta \x(\tau)$ (for analytic $\delta \x(\tau)$) along a non-infinitesimal field-line length, which will render the quantities non-local. As we will see later on, we  in fact need only the first derivative of $\delta \x(\tau)$ to be able to construct those quantities. Therefore, as the coiling and squashing rates are scalar quantities, we can conclude that  they can only depend  on invariant quantities constructed from the appropriate index contractions  of the matrix  $M_{ij}$ (entering in eq.~(\ref{ddx})) with itself and with $\hat B_i$. That already includes possible contractions with $\tilde m$ (eq.~(\ref{dr3})) as it is constructed out of $M$ and $\hatbB$. Thus, in what follows we are going to focus exclusively on the first derivative of the transverse deviation vector, $\r(\tau)$.

That derivative does not necessarily lie in the plane normal to $\hatbB(\tau)$. To see that, one can multiply both sides of eq.~(\ref{dr3}) by $P_\perp$ to find that generally: 
\be\label{ineqdr}
P_\perp \frac{d\r}{d\tau}\neq \frac{d\r}{d\tau}\ .
\ee
The reason for the inequality is that, by construction, $\r(\tau)$ always lies in the plane normal to $\hatbB$, and that plane changes orientation with $\hatbB$. 

However, in this paper we are only interested in transverse field-line deviations, disregarding the curvature of the field lines (cf. eq.~(\ref{curv})), which arises from the $\tau$-dependent $\hatbB$ direction.  So, to study those transverse deviations, we can pick a $\tau$-dependent 2D basis (subject to certain constraints; see below) which spans the changing normal plane, and write $\r$ in that 2D transverse basis. Let us denote the components of $\r$ in one of those (non-unique; see below) $\tau$-dependent 2D transverse orthonormal bases as: $r^{\diamondsuit}_i=(r_x,r_y)$. We can trivially extend that basis to 3D by including $\hatbB$ as the third orthogonal unit vector, so that the trivial extension of $r^{\diamondsuit}_i$ to 3D is $(r_x,r_y,0)$ and the components of $\hatbB$ in that basis are $(0,0,1)$. 

By construction, the components $r^{\diamondsuit}_i$ must be related to the components in the  fixed Cartesian basis by a $\tau$-dependent rotation matrix, $R_{ij}(\tau)$. Extracting only the first two non-zero components, we can write (with the sum written out explicitly to highlight the upper limit):
\be\label{rsuit}
r^{\diamondsuit}_i=\sum\limits_{j=1}^3 R_{ij}(\tau) r_j(\tau) , \ \mbox{with $i=1,2$ .}
\ee
The matrix $R$ is not unique since we can always make additional rotations in the normal plane, while leaving the components of $\hatbB$ equal to (0,0,1) in the rotated frame. We  address that issue in Section~\ref{sec:uniq}.

In order to make progress, we find it simpler to  undo the rotation on the right-hand-side above up to some $\tau_0$ using the constant $R^\mathrm{T}(\tau_0)$ matrix (which equals $R^{-1}(\tau_0)$ as $R$ is orthogonal). Let us denote the components of vectors and operators expressed in that rotated coordinate system with a superscript $\vtd$. Then the components, $r^{\vtd}_{i}$, of $\r$  in that 3D basis  are given by:
\be\label{rvtd}
r^{\vtd}_{i}(\tau)=R^{\mathrm{T}}_{ij}(\tau_0)R_{jk}(\tau)r_k(\tau)\ .
\ee

As we discussed in the beginning of this section, we will be focusing only on the first $\tau$ derivative of $\r(\tau)$ around some $\tau_0$. For infinitesimal $\delta \tau$, we can write that derivative in finite difference form, which implies that we need to focus only on $\r(\tau_0)$ and $\r(\tau_0+\delta\tau)$. 
The former can be easily obtained in component form from the equation above: $r^\vtd_i(\tau_0)=r_i(\tau_0)$; while the latter can be written out  as:
\be
r^\vtd_i(\tau_0+\delta\tau)=R^{\mathrm{T}}_{ij}(\tau_0)R_{jk}(\tau_0+\delta\tau)r_k(\tau_0+\delta\tau)\equiv\delta R_{ij}(\delta\tau,\tau_0)r_j(\tau_0+\delta\tau)\ ,
\ee
where the last equation defines the infinitesimal rotation matrix $\delta R$ (assuming $R(\tau)$ is picked to vary smoothly). In what follows, we write an equation for $\delta R$.

The rotation matrix $\delta R$ infinitesimally rotates the basis vectors for the quantities evaluated at $(\tau_0+\delta\tau)$, but leaves those evaluated at $\tau_0$ unchanged since $\delta R(\delta\tau=0,\tau_0)=\mathrm{I}$ as can be seen from the definition above.   We loosely refer to this mixture of basis vectors at the two different locations, $\tau_0$ and $(\tau_0+\delta\tau)$, along the field line  as the \textit{bent} basis, as it ``bends'' from $\tau_0$ to $\tau_0+\delta\tau$ in such a way so as to leave the  components of $\hatbB$ unchanged (see the discussion above). 
In other words, $\delta R$ is chosen such that $\hat B_i^\vtd(\tau_0+\delta\tau)=\delta R_{ij}\hat B_j(\tau_0+\delta\tau)=\hat B_i(\tau_0)$.

Using the Rodrigues' rotation formula\footnote{One can use the Rodrigues' rotation formula to find the rotation matrix ($\tilde R$) in the plane defined by two unit vectors, $\bm{a}$ and \bm{b}, that takes the components $b_i$ to the components $a_i$. That matrix is given by: $$\tilde R_{ij}= \mathrm{I}_{ij}+2a_ib_j-\frac{(a_i+b_i)(a_j+b_j)}{1+\a\cdot\b}\ .$$ It is straightforward to check that indeed $\tilde R\tilde R^\mathrm{T}=\mathrm{I}$, and $\tilde R_{ij}b_j=a_i$. We use this result to obtain eq.~(\ref{rotRod}).}, after a bit of algebra, one can show that the sought for rotation matrix transforming vector components to the bent frame is given by:
\be\label{rotRod}
\delta R=\mathrm{I}+\delta\tau \left[(\mathrm{I}-P_\perp) M^{\mathrm{T}}-M(\mathrm{I}-P_\perp)\right]\ ,
\ee
with all operators evaluated at $\tau_0$.
Indeed, checking that $\delta R\delta R^{\mathrm{T}}=\mathrm{I}$ to linear order in $\delta \tau$ is a straightforward exercise, which guarantees that $\delta R$ is a rotation matrix. Then, using eq.~(\ref{dP}), one can check that to linear order in $\delta\tau$:
\be
\hat B_i^\vtd(\tau_0+\delta\tau)=\delta R_{ij}\hat B_j(\tau_0+\delta\tau)=\delta R_{ij}(\delta \tau,\tau_0)\left(\hat B_j(\tau_0)+\delta \tau M_{jk} \hat B_k(\tau_0)\right)=\hat B_i(\tau_0)\ ,
\ee 
where we used eq.~(\ref{usefulEq}).

Note that similar to $R(\tau)$, $\delta R(\delta\tau,\tau_0)$ is not unique as we can additionally make a rotation in the plane normal to $\hatbB$. We  address that ambiguity in Section~\ref{sec:uniq}.
However, let us first use the rotation matrix $\delta R$ to calculate $r^\vtd_i(\tau_0+\delta\tau)$. First, note that eq.~(\ref{dr3}) can be written with finite differences as:
\be\label{tmp1}
\bm r(\tau+\delta\tau)=\left[\mathrm{I}+\delta\tau  \tilde m\right]\bm{r}(\tau)\ .
\ee
From that, we obtain:
\be
r^\vtd_i(\tau_0+\delta\tau)&=&\delta R_{ij}r_j(\tau_0+\delta\tau)=\delta R_{ij}\left[\mathrm{I}+\delta\tau\tilde m\right]_{jk}r_{k}(\tau_0)\nonumber\\
&=&\left[\mathrm{I}+\delta\tau P_\perp MP_\perp\right]_{ij}r_j(\tau_0)\ ,
\ee
where we used equations (\ref{rotRod}) and (\ref{usefulEq}).
Therefore, in the bent basis, we can write the first derivative of the components of $\r(\tau)$ at $\tau_0$ as:
\be\label{m3d}
\left.\frac{dr^\vtd_i(\tau)}{d\tau}\right|_{\tau=\tau_0}= \frac{r^\vtd_i(\tau_0+\delta\tau)-r_i(\tau_0)}{\delta\tau}=(P_\perp MP_\perp)_{ij}r^\vtd_j\equiv m^\vtd_{ij}r^\vtd_j\ ,
\ee
where we used the fact that $r^\vtd_i(\tau_0)=r_i(\tau_0)$ since in the bent frame, quantities evaluated at $\tau_0$ are not rotated by $\delta R$.
We can see that $m^\vtd$, defined in the last equality above, is simply the transverse part of $M$, which is reassuringly reasonable given the non-trivial way we arrived at it.

Now we are ready to write an equation for the first derivative of the components of $\r(\tau)$ at $\tau_0$ in the 2D transverse basis. Let us apply the constant rotation matrix $R(\tau_0)$, which is  independent of both $\tau$ and $\delta\tau$, to both sides of eq.~(\ref{m3d}):
\be\label{drbox}
\left.\frac{d\left(R_{ij}(\tau_0)r^\vtd_j(\tau)\right)}{d\tau}\right|_{\tau=\tau_0}=\left[R_{ij}(\tau_0)m^\vtd_{jk}\left(R^\mathrm{T}(\tau_0)\right)_{kl}\right]\left[R_{lm}(\tau_0)r^\vtd_m\right]\ .
\ee
From equations~(\ref{rsuit}) and (\ref{rvtd}), one can  write the  vector components $R_{ij}(\tau_0)r_j^\vtd$, entering on both sides above, as:
\be
r_i^\diamondsuit(\tau)=\sum\limits_{j=1}^3 R_{ij}(\tau_0)r_j^\vtd(\tau)  , \ \mbox{with $i=1,2$ ,}
\ee
with the $i=3$ component vanishing. Moreover, since $m^\vtd$ is the transverse part of $M$ (eq.~(\ref{m3d})), that means that $R_{ij}(\tau_0)m^\vtd_{jk}\left(R^\mathrm{T}(\tau_0)\right)_{kl}$ entering in eq.~(\ref{drbox}) has non-zero elements only in the upper-left $2\times 2$ block since $\hatbB$ has components (0,0,1) in the normal-plane basis extended to 3D (see discussion above). We  denote that non-zero block as the 2-by-2 matrix $m$:
\be\label{extm}
m_{il}(\tau_0)=\sum\limits_{j,k=1}^3 R_{ij}(\tau_0)m^\vtd_{jk}\left(R^\mathrm{T}(\tau_0)\right)_{kl} , \ \mbox{with $i=1,2$ and $l=1,2$ .}
\ee
In other words, the trivial extension  of $m$  (by padding it with zeros) to 3D equals $m^\vtd$ up to a 3D rotation rotating the components of $\hatbB(\tau)$ to (0,0,1).

Furthermore, it is easy to see from eq.~(\ref{m3d})  that  the $\tau$ derivative of the components of $\r(\tau)$ evaluated at $\tau_0$ in the bent basis is entirely in the normal plane as well:
\be\label{drboxP}
\left.\frac{dr^\vtd_i(\tau)}{d\tau}\right|_{\tau=\tau_0}=\left.P_{\perp,ij}\frac{dr^\vtd_i(\tau)}{d\tau}\right|_{\tau=\tau_0}\ ,
\ee
where we used eq.~(\ref{usefulEq}). The above result is in contrast to what we obtained in the fixed Cartesian basis, eq.~(\ref{ineqdr}). Thus, once we rotate that derivative with $R(\tau_0)$ (left-hand-side of eq.~(\ref{drbox})), that derivative will have a vanishing third component, and its first two components can be written as $dr_i^\diamondsuit/d\tau$ with $i=1,2$.

To conclude, after rotating with $R(\tau_0)$, the rotated $m^\vtd$, $r_i^{\vtd}$ and $dr_i^\vtd/d\tau$ are explicitly in the 2D subspace  of the  normal plane. From now on, equations involving quantities evaluated in that 2D transverse subspace will be denoted with a $\diamondsuit$ at the start of each line, instead of modifying the symbol for each such quantity. Collecting the results of the discussion above, in the transverse 2D subspace, eq.~(\ref{drbox}) can be written as:
\be\label{dr}
\boxed{
\diamondsuit \quad \left.\frac{dr_i(\tau)}{d\tau}\right|_{\tau=\tau_0}=\sum\limits_{j=1}^2m_{ij}(\tau_0)r_j(\tau_0) , \ \mbox{with $i=1,2$}
}\ ,
\ee
with $m$ given by eq.~(\ref{extm}). The above equation gives the first derivative of the components of $\r$ entirely in the 2D transverse subspace, which was what we  set out to find.

As a side note, earlier we argued that the we are not interested in the changes in the orientation of the normal plane, which intuitively must be due to the curvature ($\kappa$) of the field line $\x(\tau)$. Let us check that explicitly. The part of $M$ which is not captured by $m^\vtd$ is $M(\mathrm{I}-P_\perp)$ since from eq.~(\ref{usefulEq}) we have $m^\vtd=P_\perp MP_\perp=MP_\perp$. We can write that as:
\be\label{curv}
M_{ik}-m^\vtd_{ik}=M_{ij}(\mathrm{I}-P_\perp)_{jk}=M_{ij}\hat B_j \hat B_k= \hat B_k (\hatbB\cdot\bm{\nabla})\hat B_{i}=\hat B_k\frac{d\hat B_{i}(\x(\tau))}{d\tau}=\kappa \hat N_i \hat B_k\ ,
\ee
where we used eq.~(\ref{fl}) and the definition of $M$ (eq.~(\ref{ddx})), as well as the Frenet-Serret formulas from differential geometry. In the above equation $\hat {\bm N}$ is the normal unit vector to the field line, which along with the tangent vector (in our case, that is $\hatbB$, which runs contrary to standard differential geometry notation) and the binormal unit vector, form the basis for the Frenet-Serret frame of a curve. The torsion of the curve does enter\footnote{See equation (3.140) of the manuscript ``Fundamentals of Plasma Physics'' by J.~D.~Callen, archived at  \url{https://web.archive.org/web/20170622161238/http://homepages.cae.wisc.edu/~callen/chap3.pdf}} in the antisymmetric part of $m^\vtd$, and therefore is part of the force-free parameter $\alpha$ as well (cf. eq.~(\ref{tmp2}) and surrounding discussion). But it will not be of special significance to our analysis, since it is defined in terms of the changes to the field-line tangent and normal vectors, which we explicitly disregard once we focus on the field line motions in the normal plane (but see footnote~\ref{ft:kappa}). And indeed,  torsion and curvature are properties of a single field line, while in this paper we explore the transverse behavior of an ensemble of nearby field lines.

\subsection{Specifying the bent basis uniquely}\label{sec:uniq}

Before we move on, there is one more issue to address. As mentioned above, the bent basis (and hence $\delta R$) is not unique as one can make an infinitesimal rotation in the normal plane and still keep the difference between the components $r_i^\vtd(\tau_0+\delta\tau)$ and $r_i(\tau_0+\delta\tau)$ small. To fix the bent basis uniquely, we require that the infinitesimal angle ($\delta\phi$) of rotation of $\hat \r$ in the normal plane between $\tau$ and $(\tau+\delta\tau)$ is the same, whether it is calculated using the components of $\hat \r$ in the fixed 3D Cartesian basis or using the components of $\hat \r$ in the 2D transverse basis (and therefore in the bent basis as well) if we  pretended that basis was held fixed (see below). In the 3D Cartesian basis, that angle can be found using (cf. eq.~(\ref{omegaInit})):
\be\label{dphi1}
\delta \phi&=&\hatbB(\tau_0)\cdot \big(\hat \r (\tau_0)\times \hat \r(\tau_0+\delta\tau)\big)=\epsilon_{ijk} \hat B_{i}(\tau_0)\hat r_j(\tau_0)\hat r_k(\tau_0+\delta\tau)\nonumber\\
&=&\delta\tau\epsilon_{ijk} \hat B_{i}(\tau_0)\hat r_j(\tau_0)\tilde  m_{kl}\hat r_l(\tau_0)\ ,
\ee
where $\epsilon_{ijk}$ is the Levi-Civita symbol. In the last equality above, we linearized $\hat r_k(\tau_0+\delta\tau)$ and applied the antisymmetry of $\epsilon_{ijk}$, combined with:
\be\label{drhat}
\frac{d\hat{r}_i}{d\tau}=\frac{1}{r}\left(\mathrm{I}_{ij}-\hat r_i\hat r_j\right)\frac{d{r}_j}{d\tau}=\left(\mathrm{I}_{ij}-\hat r_i\hat r_j\right) \tilde m_{jk}\hat r_k\ ,
\ee
where in the last step we used eq.~(\ref{dr3}).
Plugging in  the expression for $\tilde  m$ (eq.~(\ref{dr3})) in  eq.~(\ref{dphi1}) and using eq.~(\ref{usefulEq}), we find (cf. eq.~(\ref{omegaFinal})):
\be\label{dphi2}
\delta \phi=\delta\tau\epsilon_{ijk} \hat B_{i}(\tau_0)\hat r_j(\tau_0)M_{kl}\hat r_l(\tau_0)\ .
\ee

We want the above expression to match the ``naive'' angle, $\delta \phi^{2d}$, that one would calculate from the components of $\r$ in the 2D transverse basis assuming the basis vectors were held constant:
\be
\diamondsuit \quad\delta \phi^{2d}&\equiv&\left.\delta\tau\frac{d}{d\tau}\left[\tan^{-1}\left(\frac{r_y}{r_x}\right)\right]\right|_{\tau=\tau_0}\nonumber
\\&=&\delta\tau\left.\left[\frac{r_x}{\sqrt{r_x^2+r_y^2}}\frac{d}{d\tau}\left( \frac{r_y}{\sqrt{r_x^2+r_y^2}}\right)-\frac{r_y}{\sqrt{r_x^2+r_y^2}}\frac{d}{d\tau}\left( \frac{r_x}{\sqrt{r_x^2+r_y^2}}\right)\right]\right|_{\tau=\tau_0}\ .
\ee
Since (as discussed above) the 2D basis spans a plane in the bent basis orthogonal to $\hatbB(\tau)$, one can rewrite the equation above in the bent basis:
\be\label{phivtd}
\delta \phi^{2d}=\delta\tau\epsilon_{ijk} \hat B_{i}(\tau_0)\hat r_j(\tau_0)\left.\frac{d\hat r_k^\vtd(\tau)}{d\tau}\right|_{\tau=\tau_0}=\delta\tau\epsilon_{ijk} \hat B_{i}(\tau_0)\hat r_j(\tau_0)m_{kl}^\vtd\hat r_l(\tau_0)\ ,
\ee
where we used the fact that the bent basis only changes the components of quantities evaluated at ($\tau_0+\delta\tau$). The equation above can be easily checked by writing it in a  basis such that the components of $\hatbB$ equal (0,0,1) (which is what the rotation matrix $R(\tau_0)$ achieves in eq.~(\ref{drbox})). Plugging the expression for $m^\vtd$ (eq.~(\ref{m3d})) above, using $P_\perp \r=\r$, and comparing with eq.~(\ref{dphi2}), we find that indeed:
\be\label{minimR}
\delta \phi^{2d}=\delta \phi\ .
\ee
Thus, the rotation matrix $\delta R$ is the unique infinitesimal rotation matrix that does not introduce any spurious rotations of the components of $r_i$ in the normal plane. Thus, the normal-plane basis is uniquely specified by the requirement  that the rate of rotation of individual neighboring field lines around the reference field line is independent of whether that rate is calculated using the standard non-local twist rate (cf. equations~(\ref{dphi1}) and (\ref{omegaInit})), or in the basis spanning the normal plane. We will refer to that property (eq.~(\ref{minimR}), combined with equations~(\ref{dphi2}) and (\ref{phivtd})) of the bent basis (defined by $\delta R$) as the basis  being \textit{minimally rotated}.

\subsection{Invariants\label{sec:inv}}

As we discussed in the beginning of Section~\ref{sec:2d}, we are interested in the local invariants one can build out of the matrices introduced so far. So, let us show that the following useful relationship between the invariants of $M$,  $\tilde m$,  $m^\vtd$ and $m$ (defined in equations (\ref{ddx}), (\ref{dr3}), (\ref{m3d}) and (\ref{dr}), respectively) holds\footnote{A standard result in linear algebra states that the eigenvalues, and hence the determinant, of a square matrix can be written as a function of the traces of the various powers of that matrix.}:
\be\label{invs}
\mathrm{tr}\left[M^n\right]=\mathrm{tr}\left[\tilde {m}^n\right]&=&\mathrm{tr}\left[\left( m^\vtd\right)^n\right]=\mathrm{tr}\left[m^n\right]
\ee
for integer powers $n\geq 1$. The last equality follows trivially from the fact that $m^\vtd$ is the 3D extension of $m$ (see the discussion around eq.~(\ref{extm})).

Below we show the first two equalities in eq.~(\ref{invs}) for $n=1,2$. The result for the rest of the powers follows analogously. Indeed, we can use eq.~(\ref{usefulEq}) and the cyclic property of the trace to write:
\be
\mathrm{tr}\left[\tilde {m}\right]&=&\mathrm{tr}\left[MP_{\perp}-(\mathrm{I}-P_{\perp})M^{\mathrm{T}}P_{\perp})\right]=\mathrm{tr}\left[P_{\perp}MP_{\perp}\right]
=\mathrm{tr}\left[P_{\perp}M\right]=\mathrm{tr}\left[M\right]\nonumber\\
\mathrm{tr}\left[\tilde {m}^2\right]&=&\mathrm{tr}\left[MP_{\perp}MP_{\perp}\right]=\mathrm{tr}\left[(P_{\perp}MP_{\perp})^2\right]=\mathrm{tr}\left[M^2\right]\ .
\ee

From eq.~(\ref{invs}), we can conclude that two of the eigenvalues of $M$,  $\tilde m$,  $m^\vtd$ and $m$ match, while the third eigenvalue of $M$,  $\tilde m$ and  $m^\vtd$ is zero. Indeed, from the definition of the projection operator, it is easy to see that $m^\vtd= P_{\perp}MP_{\perp}$ has an eigenvalue of zero, corresponding to an eigenvector equal to $\hatbB$. Similarly for $\tilde m$:
\be\label{zeroEig}
\tilde m\hatbB=\left(MP_{\perp}-(\mathrm{I}-P_{\perp})M^{\mathrm{T}}\right)\hatbB=\left(MP_{\perp}-(\mathrm{I}-P_{\perp})M^{\mathrm{T}}\right)P_{\perp}\hatbB=0\ ,
\ee
where we used eq.~(\ref{usefulEq}). Note, however, that even though $M$ must have a right eigenvector with a corresponding zero eigenvalue, that does not necessarily need to be $\hatbB$ (which is, however, a left eigenvector of $M$ as can be seen from (\ref{usefulEq}): $\hatbB^\mathrm{T} M=0$). 

\section{The squeeze factor}\label{sec:localQ}

In this section we  obtain an approximation to the squashing factor as an integral over purely local quantities. We call that approximation the \textit{squeeze factor} in reference to the decomposition we used in Section~\ref{sec:saddleSq}.
Indeed, the results of this section parallel those for the simple 2D planar flow presented in Section~\ref{sec:saddleSq}. Yet, our approach in this section will be markedly different; thus, offering an alternative way to understand our results.

To derive an expression for the squeezing  factor, we start by  writing down the general equations for the squashing factor, $Q$. In doing so, we schematically follow \cite{Titov07}; however, we simplify the analysis by restricting it only to orthonormal basis vectors (see below).

\subsection{Review of the standard squashing factor calculation}\label{sec:localQintro}

\begin{figure}[t!]\epsscale{0.6}
	\plotone{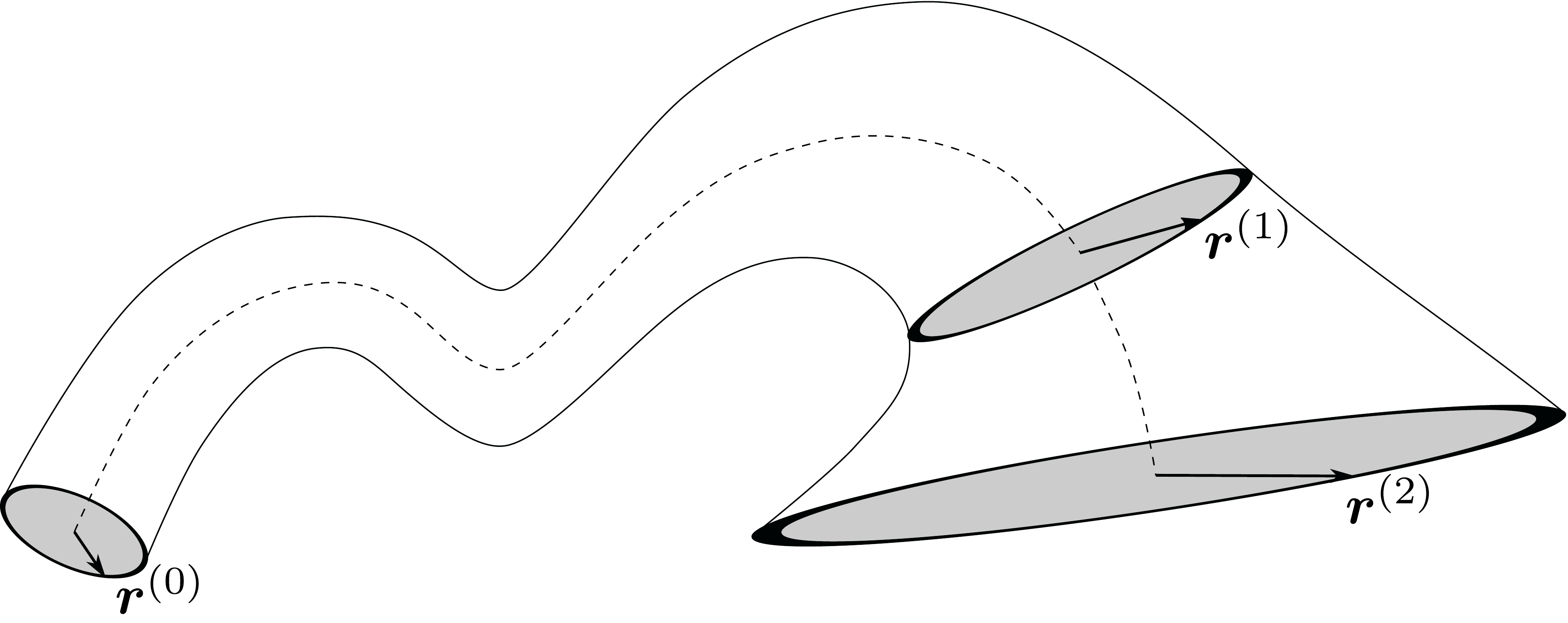}\caption{Here we illustrate a field line (dashed line) and a set of neighboring field lines forming a flux tube around it. The field lines in the flux tube are such that at $\tau_0$ they intersect the plane normal to the reference field line in a circle (gray disk on the left). That circle is generally squashed into an ellipse as one moves the normal plane further down the flux tube. Such cross-sectional ellipses are shown at $\tau_1$ and $\tau_2$ (gray disks on the right), which in the text are assumed to be infinitesimally apart. The standard squashing factor $Q$ quantifies the degree of squashing of the cross-sectional ellipse between two footpoints, which are chosen to be $\tau_0$ and $\tau_1$ in the text. By definition, $Q$ is constant along a field line as long as the footpoints are fixed. However, $Q$ does change as one moves the footpoint locations, for instance, when moving the right footpoint from $\tau_1$ to $\tau_2$. We show that the additional logarithmic squashing, produced by moving one of the footpoints, can be approximated by a purely local property of the magnetic field: the squeezing rate, $\rho_\mathcal{Z}$. The transverse field-line deviation vector $\r^{(i)}\equiv \r(\tau_i)$ ($i=0,1,2$) is the radius-vector of one flux-tube field line relative to the reference field line measured in the different normal planes. 
	}\label{fig:Tube}
\end{figure}

We start with a reference field line (dashed line in Fig.~\ref{fig:Tube}) for which we want to calculate $Q$. That quantity is constant along any field line once the footpoints of the field line have been chosen. Let us pick three points along the length of the reference field line at values of the field-line length parameter ($\tau$) equal to $\tau_0$, $\tau_1$ and $\tau_2$, with points $\tau_1$ and $\tau_2$ infinitesimally close to each other. In this section, we  focus only on points $\tau_0$ and $\tau_1$, which we will treat as the footpoints of the reference field line.

We can launch a set of neighboring field lines, intersecting in a circle the plane normal to the reference field line at $\tau_0$. The cross-section of that  flux tube of neighboring field lines is generally deformed into an ellipse as one follows the field lines (see Fig.~\ref{fig:Tube}). The amount of that deformation is quantified by $Q$ as discussed below.

Let us introduce the following notation: $\bm{r}^{(i)}\equiv \bm{r}(\tau_i)$ (with $i=0,1,2$). 
From eq.~(\ref{r}), we know that each of those vectors lies in the plane orthogonal to the reference field line at the respective $\x(\tau_i)$ (see Fig.~\ref{fig:Tube}). We will refer to each one of those normal planes as plane ($i$) from now on. 

Let the map between the components of $\bm{r}^{(1)}$ in an orthonormal basis spanning plane (1) and the components of $\bm{r}^{(0)}$ in an orthonormal basis spanning plane (0) be given by:
\be\label{Jdef}
\diamondsuit \quad r_{i}^{(1)}= J_{ij}r_{j}^{(0)}\ ,
\ee
with $J$ being a $2\times 2$ matrix. As we did for eq.~(\ref{dr}), we place a $\diamondsuit$ next to all equations written in the varying 2D basis of the normal planes as a reminder that we are not working in the fixed 3D Cartesian basis.

The squashing factor \citep{Titov07,Pariat12}, $Q$, between $\tau_0$ and $\tau_1$ can then be written as:
\be\label{QJ}
\diamondsuit \quad Q=\frac{\mathrm{tr}\big(JJ^\mathrm{T}\big)}{|\mathrm{det}\big(J\big)|}\ .
\ee 
The squashing factor is invariant under rotations in either plane (0) or (1) \citep{Titov07}. For an easy way to see this, note that under a rotation in plane (0), we have $r'^{(0)}=R_{(0)}r^{(0)}$ and under a rotation in plane (1)  we have $r'^{(1)}=R_{(1)}r^{(1)}$, with $R_{(0,1)}$ being rotation matrices. Substituting those expressions in $r^{(1)}=Jr^{(0)}$, we can write $r'^{(1)}=J'r'^{(0)}$ with $J'=R_{(1)}JR_{(0)}^{-1}$, which is the transformation equation for $J$ under rotations in both planes (0) and (1). Plugging $J'$ in eq.~(\ref{QJ}), we obtain: 
\be
\diamondsuit \quad Q'=\frac{\mathrm{tr}\left(J'J'^{\mathrm{T}}\right)}{|\mathrm{det}\left(J'\right)|}=\frac{\mathrm{tr}\left(R_{(1)}JR_{(0)}^{T}R_{(0)}J^{\mathrm{T}}R_{(1)}^{\mathrm{T}}\right)}{\left|\mathrm{det}\left(R_{(1)}JR_{(0)}^{\mathrm{T}}\right)\right|}=\frac{\mathrm{tr}\left(JJ^{\mathrm{T}}\right)}{|\mathrm{det}\left(J\right)|}=Q\ , 
\ee
where we used  the cyclic property of the trace, as well as the fact that the rotation matrices are orthogonal. Thus, we can see that indeed Q is independent of both $R_{(0)}$ and $R_{(1)}$.

We find it useful to introduce yet another $2\times 2$ matrix: 
\be\label{Mdef}
\diamondsuit \quad E\equiv \left( JJ^\mathrm{T} \right) ^{-1}\ .
\ee
It is the matrix of the cross-sectional ellipse, which is formed at the intersection of the flux tube of neighboring field lines with plane (1). To see that, note that (up to a choice of units) one can write $\left(\bm{r}^{(0)}\right)^2=1$ for all tube field lines, since, by construction, we picked a flux tube that intersects plane (0) in a circle. Using eq.~(\ref{Jdef}), we can then write: 
\be\label{ellipseM}
\diamondsuit \quad 1=r^{(0)}_ir^{(0)}_i=\left(J^{-1}\right)_{ij}r^{(1)}_j\left(J^{-1}\right)_{ik}r^{(1)}_k=r^{(1)}_iE_{ij} r^{(1)}_j\ ,
\ee
from which one can see that $E$ is indeed the matrix of the cross-sectional ellipse in plane (1). Note that the fact that the quadratic form (in $\r^{(1)}$) specified by eq.~(\ref{ellipseM}) is  an ellipse can be seen from the inequality $\mathrm{det}\left(E\right)=1/\mathrm{det}\left(J\right)^2>0$. Comparing the above equation with the implicit equation for an ellipse written in canonical form, and using the rotation invariance of the trace and determinant of $E$ (which can be confirmed following the proof of the invariance of $Q$ above), one can write:
\be\label{Mai}
\diamondsuit \quad \mathrm{tr}\big(E\big)=\frac{1}{s_a^2}+\frac{1}{s_i^2}\ , \quad 
\mathrm{det}\big(E\big)=\left(s_as_i\right)^{-2}\ ,
\ee
where $s_{a}$ and $s_i$ are the semi-major and semi-minor axes of the cross-sectional  ellipse, respectively.

Using the definition of $E$, eq.~(\ref{Mdef}), we can express $Q$ from eq.~(\ref{QJ}) as:
\be\label{QM}
\diamondsuit \quad Q=\frac{\mathrm{tr}\big(E\big)}{\sqrt{\mathrm{det}\big(E\big)}} \ .
\ee
In writing the above equation, we used the fact that since $E$ is a $2\times 2 $ matrix, we have $\mathrm{tr}(E^{-1})=\mathrm{tr}(E)/\mathrm{det}(E)$. Combining equations~(\ref{Mai}) and (\ref{QM}), we can write the squashing factor as the sum:
\be\label{Qaspect}
Q=\frac{s_a}{s_i}+\frac{s_i}{s_a}\geq 2\ ,
\ee
where $s_a/s_i\geq 1$ is the aspect ratio of the cross-sectional  ellipse in plane (1). That result makes the invariance of $Q$ under rotations in plane (1) explicit \citep{Titov07}.

\subsection{The derivative of the squashing factor with respect to the footpoint locations}

In this section, we will take the derivative of the squashing factor with respect to the location of one of the footpoints of the reference field line for which $Q$ is calculated. To do that, let us move one of the footpoints an infinitesimal distance, $\delta\tau$, along the reference field line from $\tau_1$ to $\tau_2=\tau_1+\delta\tau$ (see Fig.~\ref{fig:Tube}) and use the results from the previous section.

Plugging eq.~(\ref{Jdef}) in eq.~(\ref{dr}) evaluated at $\tau_1$, and using the fact that $\bm{r}^{(0)}$ was arbitrarily chosen, we can write:
\be\label{dJ}
\diamondsuit \quad \frac{dJ}{d\tau}=mJ\ .
\ee
Taking the derivative of $E$  in eq.~(\ref{Mdef}) with respect to $\tau$,  and combining with eq.~(\ref{dJ}), we find 
\be\label{dM}
\diamondsuit \quad \frac{dE}{d\tau}=-\left(m^{\mathrm{T}}E+Em\right)\ .
\ee

Let us now take the derivative of $Q$ with respect to $\tau$.  Using equations (\ref{QM}) and (\ref{dM}), we find:
\be\label{dQ}
\diamondsuit \quad \frac{d \ln (Q)}{d\tau}=\mathrm{tr}\left(m\right)-2\frac{\mathrm{tr}\left(Em\right)}{\mathrm{tr}\left(E\right)}\ .
\ee
To obtain the above equation, we used the fact that $E$ is a symmetric matrix (as can be seen from eq.~({\ref{Mdef})), along with the identity $\mathrm{det}\left(E\right)=\exp\left[\mathrm{tr}\left(\ln(E)\right)\right]$, which allows us to write the familiar equation for the derivative of a determinant: $d(\mathrm{det}\left(E\right))/d\tau=\mathrm{det}\left(E\right)\mathrm{tr}\left(E^{-1}dE/d\tau\right)$.
	
Note that in eq.~(\ref{dQ}) we can replace $m$ with its symmetric part as any anti-symmetric part of $m$ has no contribution as $E$ is symmetric. However, eq.~(\ref{dQ}) can still depend implicitly on the anti-symmetric part of $m$ through $E$, given by eq.~(\ref{dM}). 

The system of equations (\ref{dM}) and (\ref{dQ}) specifies the evolution of $Q(\tau)$ as long as $m(\tau)$ is known, and as long as we specify initial conditions for $E$ and $Q$. Initially, we can set $\tau=\tau_0$, for which the tube of neighboring field lines has a circular cross-section by construction. Thus, after setting $\tau_1=\tau_0$ in eq.~(\ref{ellipseM}), we can read off the initial condition for $E$: $E_{ij}(\tau_0)=\mathrm{I}_{ij}$. From eq.~(\ref{QM}), the initial condition for $Q$ is, therefore, given by $Q(\tau_0)=2$. To close the system of equations, we would need an expression for $m(\tau)$. However, we will avoid writing an explicit expression for the $2\times 2$ matrix $m$ and instead in the next section we will use the results of Section~\ref{sec:inv} for its invariants.

The right-hand side of eq.~(\ref{dQ}) is the local logarithmic contribution to the squashing factor, which one may be tempted to interpret as the local (logarithmic) squashing factor rate. However, it depends on the cross-sectional  ellipse matrix $E$, which in turn depends on the history of how the flux tube of neighboring field lines was deformed up to $\tau$. Thus, this rate is not a truly local quantity. In what follows, we show a way around that issue.

\subsection{Constructing an approximate local squashing factor rate}\label{sec:localQden}

In this section, we manipulate eq.~(\ref{dQ}) to obtain the \textit{squeeze factor} ($\mathcal{Z}$): an estimate of the squashing factor as an integral over purely local quantities ($\rho_{\mathcal{Z}}$):
\be\label{QlocDef}
\ln {Q}\sim\ln \mathcal{Z}\equiv\int d\tau \rho_{\mathcal{Z}}(\tau)\ .
\ee 
We use the symbol $\mathcal{Z}$  to denote that approximation in anticipation of recovering the results of Section~\ref{sec:saddleSq}. 

In order to construct the approximate $\mathcal{Z}$, we explored a variety of ad hoc modifications to eq.~(\ref{dQ}) which result in a purely local right-hand side. To decide which modification is most ``reasonable'', we performed numerical experiments with realistic simulated data, the final results of which are presented in \citep{Tassev2019}, where we show maps of $\mathcal{Z}$ and $Q$ which are morphologically comparable. To keep with the spirit of this paper, however, below we present an analytical motivation for our choice for $\mathcal{Z}$  (but see Section~\ref{sec:sym}  for a discussion of an alternative), developed after the numerical experiments informed our intuition. The final result of this section matches that of  Section~\ref{sec:saddleSq} (eq.~(\ref{rhoQreal1})). However, that result was obtained for constant $m$. Here we derive that result using kinematic arguments for tube-like flux surfaces undergoing deformations by slowly-varying $m$ in the presence of non-zero field-line curvature.

We will treat the following two cases separately: $m$ having two complex conjugate eigenvalues, and $m$ having two real eigenvalues. Let us start with the latter. 

If $m(\tau)$ has two real eigenvalues and is changing only slowly, then the cross-sectional ellipse  will be preferentially stretched in the direction of the (normalized) eigenvector $\hat \v_{+}$ of $m$, which corresponds to its largest eigenvalue $\lambda_{+}$ (such that $m\hat\v_{+}=\lambda_{+}\hat\v_{+}$). Indeed, we are going to assume that the semi-major axis of the cross-sectional ellipse is collinear with $\hat\v_{+}$. The ellipse matrix, $E$, can then be approximated locally as $E_{\mathrm{loc.}}\sim E$:
\begin{eqnarray}\label{Eansatz}
\diamondsuit \quad E_{\mathrm{loc.}} \equiv 
R_{+}
\begin{pmatrix}
\vspace{2mm}
s_a^{-2} &{\hspace{5mm}}& 0\cr
&\vspace{-5mm}\cr
0 &{  }& s_i^{-2}
\vspace{2mm}
\end{pmatrix}
R_{+}^{\mathrm{T}} \ ,\  \mathrm{with}\quad
R_{+}\equiv \begin{pmatrix}
   \vert &{\hspace{5mm}}&\vert \cr
   \hat\v_{+} &{ }& \hat\v_{\perp}\cr
   \vert &{ }& \vert
\end{pmatrix}\ .
\end{eqnarray}
In the above equation, $\hat\v_{\perp}$ is defined to be orthonormal to  $\hat\v_{+}$, and is therefore not necessarily an eigenvector of $m$. With the construction above, $\hat\v_{+}^{\mathrm{T}}E_{\mathrm{loc.}}\hat\v_{+}=1/s_a^2$,   $\hat\v_{\perp}^{\mathrm{T}}E_{\mathrm{loc.}}\hat\v_{\perp}=1/s_i^2$ and  $\hat\v_{+}^{\mathrm{T}}E_{\mathrm{loc.}}\hat\v_{\perp}=0$, which show that indeed the cross-sectional ellipse is oriented as described above. Note that $R_{+}$ is an orthogonal matrix as its columns are the orthonormal vectors $\hat\v_{+}$ and $\hat\v_{\perp}$, and is, therefore, a rotation matrix. 

Next we  substitute $E_{\mathrm{loc.}}$ for $E$ in eq.~(\ref{dQ}). First, we need to calculate:
\be\label{trEm}
\diamondsuit \quad \mathrm{tr}\big(E_{\mathrm{loc.}}m\big)&=&\mathrm{tr}\left[
\begin{pmatrix}
\vspace{2mm}
s_a^{-2} &{\hspace{3mm}}& 0\cr
&\vspace{-5mm}\cr
0 &{  }& s_i^{-2}
\vspace{2mm}
\end{pmatrix}
\begin{pmatrix}
\vspace{2mm}\mbox{--- } &\hat\v_{+}&\mbox{---} \cr
&\vspace{-5mm}\cr
   \mbox{--- }  &\hat\v_{\perp} &\mbox{---} \vspace{2mm}
\end{pmatrix}m\begin{pmatrix}
   \vert &{\hspace{3mm}}&\vert \cr
   \hat\v_{+} &{ }& \hat\v_{\perp}\cr
   \vert &{ }& \vert
\end{pmatrix}\right]\nonumber\\
&=&\mathrm{tr}\left[
\begin{pmatrix}
\vspace{2mm}
s_a^{-2} &{\hspace{3mm}}& 0\cr
&\vspace{-5mm}\cr
0 &{  }& s_i^{-2}
\vspace{2mm}
\end{pmatrix}
\begin{pmatrix}
\vspace{2mm}
\lambda_{+} & {\hspace{3mm}}& \hat\v_{+}^{\mathrm{T}}m\hat\v_{\perp}\cr
&\vspace{-5mm}\cr
0 &{  }& \hat\v_{\perp}^{\mathrm{T}}m\hat\v_{\perp}
\vspace{2mm}
\end{pmatrix}
\right]\ ,
\ee
where we used the cyclic property of the trace.
Note that\footnote{As $R_{+}^{\mathrm{T}}mR_{+}$ equals an upper triangular matrix, as seen in eq.~(\ref{trEm}), it is the real Schur form of $m$. The off-diagonal element can be interpreted as a finite shear due to the non-orthogonal eigenvectors of $m$, which causes a non-zero force-free parameter $\alpha$. See Section~\ref{sec:saddleSq} for further discussion.} $R_{+}^{\mathrm{T}}mR_{+}$, which appears above, is a rotation of $m$, and therefore its eigenvalues must match those of $m$. Thus, we can conclude that $\hat\v_{\perp}^{\mathrm{T}}m\hat\v_{\perp}=\lambda_{-}$, where $\lambda_{-}$ is the smaller of the two eigenvalues of $m$. We can then write:
\be
\diamondsuit \quad \mathrm{tr}\big(E_{\mathrm{loc.}}m\big)=\frac{\lambda_{+}}{s_a^2}+\frac{\lambda_{-}}{s_i^2}\ .
\ee
Substituting in eq.~(\ref{dQ})  and simplifying, we obtain:
\be\label{dQ1}
\diamondsuit \quad \mathrm{tr}\left(m\right)-2\frac{\mathrm{tr}\left(E_{\mathrm{loc.}}m\right)}{\mathrm{tr}\left(E_{\mathrm{loc.}}\right)}=(\lambda_{+}-\lambda_{-})\, \frac{s_a^2-s_i^2}{s_a^2+s_i^2}\ .
\ee
In order for the above expression to be purely local, we need to make it independent of the semi-major and semi-minor axis of the cross-sectional ellipse. A trivial way to do that is to maximize the above expression, which entails taking $s_i\ll s_a$.  Indeed, this inequality holds  for large $\tau$ when $m$ is constant.

Thus, we are assuming that the cross-sectional ellipse is infinitely squeezed, with a semi-major axis oriented along the eigenvector $\v_{+}$ of $m$. With that assumption, we can write the squeeze factor $\mathcal{Z}$ -- our  approximation to the squashing factor -- as:
\be\label{rhoQreal}
\diamondsuit \quad \rho_{\mathcal{Z}}=\frac{d \ln \mathcal{Z}}{d\tau}\equiv \lim\limits_{s_i/s_a\to 0}\left[ \mathrm{tr}\left(m\right)-2\frac{\mathrm{tr}\left(E_{\mathrm{loc.}}m\right)}{\mathrm{tr}\left(E_{\mathrm{loc.}}\right)}\right]= \lambda_{+}-\lambda_{-}  \quad \mbox{for real eigenvalues.}
\ee
This expression is completely local as sought for. Analytically, it is rather neat as it is simply the difference in the logarithmic rates with which the field lines are being pulled along the eigenvectors of the transverse flow set up by $m$ (cf. eq.~(\ref{squeezrate})). 


Note that the eigenvalue difference above depends on the antisymmetric part of $m$. This is due to our assumption that $m$ is slowly varying; and therefore, the squeezed cross-sectional ellipse ends 
up oriented along one of the eigenvectors of $m$ and not, for example, its symmetrized counterpart (see Section~\ref{sec:sym}). Thus, one may worry that the above expression is sensitive to physically irrelevant infinitesimal rotations between planes (1) and (2), resulting from poor choice of basis. Fortunately, $m$ has non-zero eigenvalues matching $m^{\vtd}$ (Section~\ref{sec:inv}), which in turn was written in the minimally rotated bent basis (Section~\ref{sec:uniq}). This guarantees that the components of $\r$ expressed in the normal plane rotate at the same rate as those expressed in the fixed 3D Cartesian basis, i.e. the antisymmetric part of $m$ captures only rotations that are indeed physical. For further discussion, see eq.~(\ref{minimR}) and the discussion around it.

Equation~(\ref{rhoQreal}) is valid for real eigenvalues of $m$. For two complex conjugate eigenvalues, the expression would result in an imaginary contribution to $\ln(Q)$ and is thus invalid. To deal with that case, we can use the fact that the eigenvalues will then have equal real parts. Thus, the transverse flow set up by  $m$ is a stable or unstable spiral (or center), which periodically rotates the cross-sectional ellipse, returning it to the same aspect ratio after a full rotational period (assuming constant $m$). Therefore, from eq.~(\ref{Qaspect}) we can see that $Q$ would on average remain unaffected by such a flow. Thus, a reasonable choice for $\rho_{\mathcal{Z}}$ would be to set it to zero in this case. That corresponds to eq.~(\ref{possibleImRhoZ1}) in Section~\ref{sec:sqIm}.

Another way of motivating setting $\rho_{\mathcal{Z}}$ to zero for complex eigenvalues of $m$ is to note that the initial orientation of the cross-sectional ellipse can be arbitrary as there is no preferred ``stretching'' direction set up by the flow of $m$. Thus, for $E_{\mathrm{loc.}}$ we can again use eq.~(\ref{Eansatz}) with $R_{+}$ this time being a rotation matrix rotating the cross-sectional ellipse at a random angle $\theta$. Plugging that ansatz into (\ref{dQ}) and averaging over the angle $\theta$, after some algebra we again find:
\be\label{rhoQimag}
\diamondsuit \quad \rho_{\mathcal{Z}}=0\quad \mbox{for complex eigenvalues.}
\ee
If the localized periodic squashing of the flux tube is of interest, however, one can use, eq.~(\ref{possibleImRhoZ2}) for complex eigenvalues of $m$ (see Section~\ref{sec:sqIm}).

Equations (\ref{rhoQreal}) and (\ref{rhoQimag}) are the final result of this section. Next, we  write down the eigenvalue difference in eq.~(\ref{rhoQreal}) using the local properties of the magnetic field.

\subsection{The squeeze factor}

In order to evaluate eq.~(\ref{rhoQreal}), we need to calculate the eigenvalues of the $2\times2$ matrix $m$. As we showed in Section~\ref{sec:inv}, the two non-zero eigenvalues of $M$ must match the two eigenvalues of $m$. The difference between those two eigenvalues for the $2\times 2$ matrix $m$ can be written as $\lambda_{+}-\lambda_{-}=\sqrt{2\,\mathrm{tr}(m^2)-\left(\mathrm{tr}(m)\right)^2}$, which can in turn be written as $\lambda_{+}-\lambda_{-}=\sqrt{2\,\mathrm{tr}(M^2)-\left(\mathrm{tr}(M)\right)^2}$ as the third eigenvalue of $M$ is zero. In Appendix~\ref{sec:app} (cf. eq.~(\ref{covM})), 
we show that in curvilinear coordinates, $M_{\  j}^{i}$ is (unsurprisingly) given by the covariant derivative  $\nabla_j \hat B^i$. Thus, in any coordinate system we can finally write equations (\ref{rhoQreal}) and (\ref{rhoQimag}) as\footnote{It may be of interest to point out the relationship between $Q$ and the Lyapunov exponents of an attractor flow specified by some $\hatbB(\tau)$. For a particular attractor, the Lyapunov exponents ($l_i$) are defined through the eigenvalues ($\Lambda_i(\tau)>0$) of the symmetric matrix $J^\mathrm{T}(\tau)J(\tau)$ as: $$l_i =\lim\limits_{\tau\to\infty}\frac{\ln\big(\Lambda_i(\tau)\big)}{2\tau}\ .$$ Assuming that the footpoints are $\tau\to\infty$ apart along the field lines; and that the field is defined over a region, where taking that limit is meaningful; and that the field-line flow asymptotically is described by an attractor -- all assumptions that clearly break down for  magnetic fields in the solar corona -- one can then use eq.~(\ref{QJ}) to find $Q=(\Lambda_1+\Lambda_2)/\sqrt{\Lambda_1\Lambda_2}$. Therefore, in the limit $Q\gg 1$, we obtain $\ln Q(\tau)\sim \tau |\Delta l|$, where $\Delta l$ is the difference between the two Lyapunov exponents of the flow in the normal plane. Comparing with eq.~(\ref{qhoQFinal}), we can see that $|\Delta l|$ can then be approximated by the $\tau$-average of $\Re(\Delta \lambda)$. For small $\tau$, assuming that we still have $Q\gg 1$, one can then identify $\rho_\mathcal{Z}$ as the difference between the \textit{local} Lyapunov exponents \citep{1991JNS.....1..175A}. }:
\be\label{qhoQFinal}
\boxed{\frac{d \ln Q}{d\tau}\sim\frac{d \ln \mathcal{Z}}{d\tau}=\rho_{\mathcal{Z}}=\Re\left\{\lambda_{+}-\lambda_{-}\right\}=\Re\left\{\sqrt{2\nabla_i \hat B^j\nabla_j \hat B^i-\left(\nabla_i \hat B^i\right)^2}\right\}
}\ ,
\ee
with $\Re(x)$ denoting the real part of $x$. To set the minimum value of $\mathcal{Z}$ to that of ${Q}$, we use $\mathcal{Z}(\tau_0)=2$ as an initial condition for eq.~(\ref{qhoQFinal}). Note that for an imaginary eigenvalue difference, if one is interested
in taking into account  the localized periodic squashing of the flux tube then, instead of setting $\rho_{\mathcal{Z}}$ to zero in that case, one can use eq.~(\ref{possibleImRhoZ2}) (see Section~\ref{sec:sqIm}).

Note that the equation above is invariant of the direction of integration along the reference field line as required for $Q$ \citep{Titov02,Titov07}. Moreover, since $\rho_{\mathcal{Z}}\geq 0$, the contributions to the squeeze factor, $\mathcal{Z}$, cannot be undone as one integrates over a field line, especially if one uses  eq.~(\ref{possibleImRhoZ2}) in the case of an imaginary eigenvalue difference. That is in contrast to $Q$, which can remain small in the presence of large localized squashing, as long as the flux tube around the field line in question is squashed and then un-squashed between its footpoints. As $Q$ is growing exponentially large in the vicinity of HFTs and nulls, a large $\rho_{\mathcal{Z}}$ can be used as an indicator for the presence of those features.

The first  equality in eq.~(\ref{qhoQFinal}) is approximate in a sense that can be quantified by comparing equations~(\ref{dQ}) and (\ref{rhoQreal}),  with the difference arising from using the cross-sectional ellipse matrix ($E$) in $Q$, while using its local approximation ($E_{\mathrm{loc.}}$) in $\mathcal{Z}$.  
Thus, one can think of $\rho_{\mathcal{Z}}$ integrated over a stretch of $\tau$ as the extra logarithmic squashing of an already infinitely squeezed ellipse (that is, the quadratic form, corresponding to $E_{\mathrm{loc.}}$) that is pointed along the eigenvectors of $m$. Given the choice of  $E_{\mathrm{loc.}}$, described in Section~\ref{sec:localQden}, that orientation of the cross-sectional ellipse can be assumed to be caused by a slowly evolving $m$, which preferentially squashes flux tubes along its eigenvectors; or can be thought of as being produced by a deliberate choice of footpoint location ($\tau_0$), which produces a tube with that particular cross-sectional ellipse at some $\tau$.

For an alternative justification for approximating the squashing factor with $\mathcal{Z}$, compare the result above with the discussion in Section~\ref{sec:saddleSq}, and eq.~(\ref{rhoQreal1}) in particular. Equation (\ref{qhoQFinal}) is the final result of this section. It is the generalization of eq.~(\ref{rhoQreal1})  for field lines with non-zero curvature. 

\subsection{The symmetrized squeeze rate}\label{sec:sym}

As we discussed previously, there is an inherent ambiguity of how one quantifies the squashing rate using local quantities. In the previous section, we presented a kinematic argument for our choice of $E_{\mathrm{loc.}}$, eq.~(\ref{Eansatz}). We assumed that $m$ varies slowly enough, so as to guarantee that the cross-sectional ellipse is stretched along the eigenvector of $m$ corresponding to the largest eigenvalue. 

One may wonder how our results would change if we picked $R_{+}$ in eq.~(\ref{Eansatz}) to be a rotation matrix with a rotation angle which maximizes the resulting $d\ln \mathcal{Z}/d\tau$. After lengthy algebra, one can confirm that such a choice still recovers equations~(\ref{dQ1}) and (\ref{rhoQreal}), but this time the eigenvalue difference ($\Delta \lambda_s$) that would enter in (\ref{rhoQreal}) would correspond to the symmetrized matrix $m_s\equiv(m+m^\mathrm{T})/2$. Thus, this corresponds to an ellipse aligned with the eigenvectors of $m_s$, and not of $m$. It should be clear, however, that such an orientation for the cross-sectional ellipse can only happen by chance as a transient in a regime when $m$ is changing fast.

In the language of Section~\ref{sec:decompstd}, $\Delta\lambda_s$  equals the  difference between the eigenvalues of the rate-of-shear tensor ($\Sigma$; see eq.~(\ref{stddecomprates})) as a non-zero rate-of-expansion does not affect $\Delta\lambda_s$. Therefore, $\Delta \lambda_s$ gives the relative logarithmic rate of squeeze in the standard rotation-squeeze decomposition (first and second row of Fig.~\ref{fig:decomp}). 
Thus, the symmetrized squeeze rate, $\Delta \lambda_s$, is non-zero  for elliptical flows. One can show that for any $2\times2$ matrix $m$, $\Delta \lambda_s^2=\Delta \lambda^2+\alpha^2$, with the generalized force-free parameter, $\alpha$, being the difference between the two off-diagonal elements of $m$ (e.g. eq.~(\ref{alpha_tmp})). Combining with equations~(\ref{omega0lambdafluids}) and (\ref{finalCoiling1}) we arrive at:
\be
\Delta\lambda_s=|\omega_c|\left(\frac{S_a}{S_i}-\frac{S_i}{S_a}\right)\ ,
\ee
which is clearly different from the physically motivated localized rate of squashing that we obtained earlier for periodic flows:  equations~(\ref{possibleImRhoZtmp}) and (\ref{possibleImRhoZ2}) in Section~\ref{sec:sqIm}.

Similarly, a dependence of $Q$ on the eigenvalues of $m_s$ can be obtained from eq.~(\ref{Q2dflowExact}) for the steady-state 2D flow (discussed in Section~\ref{sec:toy}) for  small $\Delta\tau$, such that $Q$ is not exponentially boosted. Expanding $Q$ in that equation to second order in $\Delta\tau$, one can show that  $Q\approx2+\Delta\tau^2\Delta\lambda_s^2$. So, clearly $\Delta\lambda_s$ can play a role in regions of low $Q$. However, in this paper, we are only interested in $Q\gg1$ which, as we argue in Section~\ref{sec:saddleSq}, depends on the eigenvalues of the unsymmetrized $m$.

Despite these arguments against using the symmetrized eigenvalue difference, one may still consider using $\Delta\lambda_s$ in (\ref{rhoQreal}), especially if interested in regions that do not necessarily contribute to an exponential boosting of $Q$. Therefore, we did modify \qsl  \citep{QSL_Squasher} to allow for the calculation of $\rho_{\mathcal{Z}}$ using $\Delta\lambda_s$ in (\ref{rhoQreal}) as an option.

\section{Coiling number}\label{sec:coiling}

In this section we write down our local approximation for the coiling rate of neighboring field lines in the normal plane. We relate it to the generalized force-free parameter, and we interpret the latter in two different ways: as an angle average of the angular rate of motion of neighboring field lines, and as a geometrically boosted coiling rate. The results of this section parallel those for the simple 2D planar flow presented in Section~\ref{sec:ellSq}.

\subsection{The generalized force-free parameter}

For the purposes of calculating the local twist number (cf. our eq.~(\ref{twistN}), as well as  eq.~(16) of \citep{2006JPhA...39.8321B}) the force-free parameter, $\alpha$, can be generalized for any magnetic field as:
\be
\alpha\equiv \frac{\B\cdot\left(\bm{\nabla}\times \B\right)}{B^2}\ .
\ee
Note that $\alpha$ can also be written as:
\be
\alpha&=& \frac{\hatbB\cdot\left(\bm{\nabla}\times \left(\hatbB B\right)\right)}{B}=\hatbB\cdot\left(\bm{\nabla}\times \hatbB\right)-\frac{\hatbB\cdot\left(\hatbB\times (\bm{\nabla} B)\right)}{B}\ .\nonumber
\ee
The last term  above vanishes because it is proportional to the dot product of two orthogonal vectors.  Therefore, we obtain:
\be
\alpha&=& \hatbB\cdot\left(\bm{\nabla}\times \hatbB\right) \ .\label{alphaFinal}
\ee

Let us now calculate the angular rate $\omega$ at which neighboring field lines are rotating around a reference field line.  The rate is given by (e.g. \cite{2006JPhA...39.8321B}; \cite{2016ApJ...818..148L}; cf. our eq.~(\ref{dphi1})):
\be\label{omegaInit}
\omega (\x(\tau),\r(\tau))=\hatbB(\x(\tau))\cdot \left(\hat{\bm{r}}\times \frac{d\hat{\bm{r}}}{d\tau}\right)\ .
\ee
To simplify the above expression, let us use eq.~(\ref{drhat}) and express $\tilde m$ from (\ref{dr3}). We find that the specific combinations of dot and cross products allow for the elimination of many of the terms. In the end, we obtain (cf. eq.~(18) of \cite{2016ApJ...818..148L}, as well as our eq.~(\ref{dphi2})):
\be\label{omegaFinal}
\omega=\hatbB\cdot \left(\hat \r\times \left((\hat \r\cdot\bm{\nabla})\hatbB\right)\right)\ .
\ee

To compare the expression for $\omega$ with that for $\alpha$, let us pick a basis in which $\hatbB$ has components $(0,0,1)$. In that basis, we can write $\r$ as $(r_x,r_y,0)$. Plugging those into (\ref{alphaFinal}) and (\ref{omegaFinal}), we obtain:
\be
\alpha&=&\nabla_x \hat B_y-\nabla_y \hat B_x\ ,\nonumber\\
\omega&=&\frac{r_x^2\nabla_x \hat B_y-r_y^2 \nabla_y \hat B_x - r_xr_y \left(\nabla_x \hat B_x-\nabla_y \hat B_y\right)}{r_x^2+r_y^2}\ .\label{alpha_tmp}
\ee
Clearly, $\omega$ depends on the location of the neighboring field line, while $\alpha$ does not. So, let us write $r_x=r\cos(\phi)$ and $r_y=r\sin(\phi)$ in polar coordinates and average $\omega$ over $\phi$. We obtain:
\be
\langle \omega \rangle\equiv\frac{1}{2\pi}\int_0^{2\pi} \omega d\phi=\frac{\nabla_x \hat B_y-\nabla_y \hat B_x}{2}\ .
\ee
Comparing with (\ref{alpha_tmp}), we can finally write:
\be\label{omegaAve}
\boxed{
\alpha=2\langle\omega\rangle
}\ .
\ee
The expression above is coordinate independent and represents the final result of this section. It allows us to interpret $\alpha$  as the average angular rate of motion of neighboring field lines around the reference field line passing through the location at which $\alpha$ is calculated. The average is taken over the neighboring field lines, assuming they are initially uniformly distributed in angle. 

The above result  mirrors   the discussion around eq.~(\ref{twist_fluids}) for vorticity (the fluid flow equivalent of $\alpha$) as a measure of the angular rate of rotation of a floating vane uniformly ``sampling'' (in angle)  nearby streamlines. Thus, the angular average in eq.~(\ref{omegaAve}) is non-zero even for sheared saddle transverse flow of neighboring field lines, when none of the field lines make full turns around one another. Thus, $\alpha$ is a poor measure of local magnetic field line twist. To improve on it, we introduce the coiling rate below.

\subsection{The coiling rate}\label{sec:w0alpha}


Below we obtain our expression for the coiling rate. We start with the unambiguously defined rate of rotation of individual neighboring field lines in the normal plane. Then we show how the coiling rate is related to a specific averaging of the rate of rotation of individual field lines in a way which is distinct from the averaging that lead to the generalized force-free parameter (see eq.~(\ref{omegaAve})).

So, let us start by exploring the behavior of the vector $\r=(r_x,r_y)$  in the plane orthogonal to $\hatbB$. That vector obeys a first-order differential equation (DE) (cf. eq.~(\ref{dr})) with $(r_x,r_y)=(0,0)$ being a critical point. Let us modify the right-hand side of eq.~(\ref{dr}) by taking the traceless part of $m$, and denote its solution vector with $r^\Psi$:
\be\label{drA} 
\diamondsuit \quad \frac{dr^\Psi_i(\tau)}{d\tau}=\Psi_{ij}(\tau)r^\Psi_j(\tau) \ ,
\ee
with $\Psi$ defined in eq.~(\ref{stddecomprates}).
 Using equations~(\ref{m3d}) and (\ref{extm}), we can write the 3D extension of the above equation in the bent basis as:
\be\label{A3}
\frac{dr^{\Psi\vtd}_i}{d\tau}=\left( P_\perp M P_\perp - \frac{\mathrm{tr}(M)}{2}P_\perp\right)_{ij}r^{\Psi\vtd}_j\equiv \Psi^\vtd_{ij}r^{\Psi\vtd}_j\ ,
\ee
with  $\Psi^\vtd$ being the traceless-transverse part of $m^\vtd$. Above we used the fact that $\mathrm{tr}(P_\perp)=2$. Note that $\Psi^\vtd$ is then simply the traceless-transverse part of $M$. 

We can check that $\hat \r^\Psi$ rotates at the same rate as $\hat \r$ in the normal plane if they start off collinear. To do that let us initialize the two vectors in 3D such that $\hat \r^{\Psi\vtd}(\tau_0)=\hat \r(\tau_0)$. The angle $\hat \r^\Psi$  sweeps in the normal plane up to $\tau_0+\delta\tau$ is by construction equal to the angle its 3D extension $\hat r^{\Psi\vtd}$ sweeps, which in turn  can be calculated as:
\be\label{phiAvtd}
\delta \phi^\vtd_\Psi&\equiv&\epsilon_{ijk} \hat B_{i}(\tau_0)\hat r^{\Psi}_j(\tau_0)\hat r^{\Psi\vtd}_k(\tau_0+\delta\tau)=\delta\tau\epsilon_{ijk} \hat B_{i}(\tau_0)\hat r^\Psi_j(\tau_0)\Psi_{kl}^\vtd\hat r^\Psi_l(\tau_0)\nonumber\\
&=&\delta\tau\epsilon_{ijk} \hat B_{i}(\tau_0)\hat r_j(\tau_0)\left(m_{kl}^\vtd-\frac{\mathrm{tr}(M)}{2}P_{\perp,kl}\right)\hat r_l(\tau_0)\nonumber\\
&=&\delta\tau\epsilon_{ijk} \hat B_{i}(\tau_0)\hat r_j(\tau_0)m_{kl}^\vtd\hat r_l(\tau_0)\ .
\ee
The last line above equals $\delta\phi^\vtd(\tau_0)$ (cf. eq.~(\ref{phivtd})), and therefore $\delta \phi^\vtd_\Psi$ indeed matches $\delta \phi^\vtd$.

Therefore, in order to study how $\r$ is rotating in the normal plane, it is sufficient to understand the rotation of $\r^\Psi$. So, from now on let us focus on eq.~(\ref{drA}) (and where necessary, its 3D extension, eq.~(\ref{A3})). As we are interested in rotations in the normal plane (and not in saddle-like behavior), let us focus on the case where $\Psi$ has a positive determinant. For constant $\Psi$, the DE in eq.~(\ref{drA})  has a solution which is a center at $\r^\Psi=0$. Then, the phase portrait of $\r^\Psi$ is a set of concentric ellipses. Let the semi-major axis of one of them is given by $cS_a$, and the semi-minor axis -- by $cS_i$ (with $c$ being a constant). We can then pick $c$ and its units such that $S_iS_a=\omega_c$, with $\omega_c$ being the angular rate of rotation of $\r(\tau)$ which, in anticipation of recovering the results of Section~\ref{sec:ellSq}, we will call the \textit{coiling rate}. In the basis aligned with the semi-axes of the ellipse, the DE can then be written as:
\be\label{ellipseODE}
\diamondsuit \quad \begin{pmatrix}
dr_x/d\tau\cr
dr_y/d\tau
\end{pmatrix}=
\begin{pmatrix}
0 & \ -S_a^2\cr
S_i^2 & \ 0
\end{pmatrix}
\begin{pmatrix}
 r_x\cr
r_y
\end{pmatrix}
= \Psi_{ij}r_j\ ,
\ee
where the last equality allows us to read off the $2\times 2$ matrix $\Psi$ in this particular basis. As a check, let us write the general solution to the system above: $r_x(\tau)=c S_a \cos\left[\omega_o\tau+\varphi\right]$, $r_y(\tau)=c S_i \sin\left[\omega_o\tau+\varphi\right]$, with $c$ and $\varphi$ being constants of integration. This confirms that eq.~(\ref{ellipseODE}) has as a solution the ellipse described above. 

Note that the ellipse with semi-axes specified by  $s_a$ and $s_i$ in eq.~(\ref{Mai}) is the cross-sectional  ellipse formed by the intersection of a flux tube of neighboring field lines with a plane orthogonal to the reference field line. In contrast, the ellipse with semi-axes specified by  $S_a$ and $S_i$ above is the ellipse which \textit{each one} of those field lines would describe   in the normal plane around the reference field line under the constant flow specified by eq.~(\ref{ellipseODE}).

From eq.~(\ref{ellipseODE}), we can write the following relationships between $S_i$, $S_a$, $\omega_c$ and the invariants of the matrix $\Psi$:
\be
\diamondsuit \quad\omega_c^2&=&\mathrm{det}(\Psi)=\frac{\big(\mathrm{tr}(\Psi)\big)^2-\mathrm{tr}\big(\Psi^2\big)}{2}=-
\frac{\mathrm{tr}\big(\Psi^2\big)}{2}\ ,\nonumber\\
\diamondsuit \quad\left(\frac{S_i}{S_a}\right)^2+\left(\frac{S_a}{S_i}\right)^2&=&\frac{\mathrm{tr}\left(\Psi\,\Psi^{\mathrm{T}}\right)}{\mathrm{det}(\Psi)}=2\frac{\mathrm{tr}\left(\Psi\,\Psi^{\mathrm{T}}\right)}{\big(\mathrm{tr}(\Psi)\big)^2-\mathrm{tr}\big(\Psi^2\big)}=-2
\frac{\mathrm{tr}\left(\Psi\,\Psi^{\mathrm{T}}\right)}{\mathrm{tr}\big(\Psi^2\big)}\label{localEllipse}\ ,
\ee
where we used the fact that $\Psi$ is traceless.

Now, let us allow for a varying $\hatbB(\tau)$. We can still define $\omega_c$, $S_i$ and $S_a$ at each instant through the equations above. We can then substitute the 3D extension $\Psi^\vtd$ for $\Psi$ from eq.~(\ref{A3}) in eq.~(\ref{localEllipse}) after noting that $\Psi^\vtd$ has one zero eigenvalue (for the eigenvector equal to $\hatbB$). Thus, we can write the instantaneous coiling rate as:
\be\label{tmp3}
\omega_c^2=-\frac{\mathrm{tr}\left(\left(\Psi^\vtd\right)^2\right)}{2}=-\frac{1}{4}\left[2\,\mathrm{tr}\left(M^2\right)-\left(\mathrm{tr}\left(M\right)\right)^2\right]\ ,
\ee
where we used (\ref{usefulEq}). Note that this can be written as (cf. eq.~(\ref{qhoQFinal})):
\be\label{omega0Final}
\boxed{
\omega_c=\frac{1}{2}|\Im\left\{\lambda_{+}-\lambda_{-}\right\}|\times \mathrm{sign}(\alpha)=\frac{1}{2}\left|\Im\left\{\sqrt{2\nabla_i \hat B^j\nabla_j \hat B^i-\left(\nabla_i \hat B^i\right)^2}\right\}\right|\times \mathrm{sign}(\alpha)
} \ ,
\ee
with $\Im(x)$ giving  the imaginary part of $x$. Thus, up to a sign, $\omega_c$ is the imaginary part of one of the two complex conjugate eigenvalues of $M$. The sign of $\omega_c$ is picked to be the same as the sign of $\alpha$ (or equivalently, the average $\langle \omega\rangle$): positive for right-handed twist, and negative for left-handed.

Following the same procedure we used to obtain eq.~(\ref{tmp3}), using eq.~(\ref{localEllipse}) we can calculate the following combination for non-zero $\omega_c$:
\be\label{tmp2}
\left[\left(\frac{S_a}{S_i}\right)^2+\left(\frac{S_i}{S_a}\right)^2+2\right]\omega_c^2=\mathrm{tr}\left[MP_\perp\left(M^{\mathrm{T}}-M\right)\right]\ .
\ee
One can show that the right hand side above equals $\alpha^2$ from eq.~(\ref{alphaFinal}). The easiest way to see that is to write the equation in a basis in which $\hatbB(\tau)$ has components (0,0,1). Thus, we can conclude that:
\be\label{finalCoiling}
\boxed{
\alpha=\left(\frac{S_i}{S_a}+\frac{S_a}{S_i}\right)\omega_c \  \ \mbox{for complex $\lambda_{+}$; $\omega_c=0$ otherwise.}
}
\ee
Therefore, the above equation is the generalization of eq.~(\ref{omegazcenter})  for field lines with non-zero curvature. Similar to our results of Section~\ref{sec:ellSq}, we again see that the generalized force-free parameter, which is the magnetic field equivalent of  vorticity of planar fluid flows, is boosted  relative to the actual coiling rate by geometric factors related to the finite squeeze deformation needed to produce an elliptical transverse flow of field lines (see Section~\ref{sec:ellSq} for further discussion).

Equations~(\ref{omegaAve}) and (\ref{finalCoiling}) lead to the following inequality:
\be
|\alpha|=2\,|\langle\omega\rangle|\geq 2\,|\omega_c|\ .
\ee
Thus, we can conclude that the magnitude of the generalized force-free parameter, $\alpha$, can be enhanced relative to twice the coiling rate, $2\omega_c$, for two different reasons. (1) When the orthogonal flow produced by $M$ is not a spiral or center, i.e. when it is a saddle or a node, $\alpha$ generally can still assign a sense of rotation to that flow as long as the transverse part of $M$ is not symmetric (cf. right-hand side of eq.~(\ref{tmp2})). (2) When the orthogonal flow caused by $\Psi$ is a center (i.e. $M$ has a pair of complex conjugate eigenvalues), but the center produces an elliptical flow pattern, $\alpha$ is still enhanced relative to $2\omega_c$. The reason is that for elliptical flows (esp. with large aspect ratios), field lines passing close to the reference field line can sweep large angles even when $\omega_c$ is low, thus enhancing the average $\langle\omega\rangle$, which results in an enhanced $\alpha$ (see also the discussion after eq.~(\ref{omegaAve})).

\subsection{The coiling number}
The local twist number $\mathrm{N_t}$ is defined as the integral over a field line $L$  \citep[see][eq.~(16)]{2006JPhA...39.8321B}:
\be\label{twistN}
\mathrm{N_t}=\frac{1}{2\pi}\int_L d\tau \frac{\alpha(\tau)}{2}=\frac{1}{2\pi}\int_L d\tau \langle\omega(\tau)\rangle\ ,
\ee
where the last equality comes from eq.~(\ref{omegaAve}). 
However, the above average includes contributions from saddle-like behavior of the field-lines in the transverse direction to $\hatbB$. Thus, one can envision constraining the integral only to those parts of the reference field line for which the phase portrait of the transverse flow is a center/spiral (i.e. $\Im(\lambda_{+})\neq 0$). Thus, one can define an alternative local twist number:
\be
\mathrm{N_{t,im}}\equiv\frac{1}{2\pi}\int\limits_{L\cap \Im(\lambda_{+})\neq 0} d\tau \frac{\alpha(\tau)}{2}=\frac{1}{2\pi}\int\limits_{L\cap \Im(\lambda_{+})\neq 0} d\tau \langle\omega(\tau)\rangle\  .
\ee
The difference between $\mathrm{N_t}$ and $\mathrm{N_{t,im}}$ can be used to quantify the saddle-flow contributions to the local twist number.

We define the coiling number as:
\be\label{coilingFinal}
\boxed{
\mathrm{N_{c}}\equiv\frac{1}{2\pi}\int_L d\tau\  \omega_c(\tau)
}\ .
\ee
The coiling number, $\mathrm{N_c}$, gives an unbiased estimate of the number of turns neighboring field lines make around one another. It is unbiased in the following strict sense: for steady-state transverse flows (implying constant $\omega_c$) and integer $\mathrm{N_c}$, $\mathrm{N_c}$ matches the actual number of turns neighboring field lines make around one another (see Sections~\ref{sec:ellSq} and \ref{sec:w0alpha}). In contrast, for elliptical transverse flows, the local twist number is biased by a factor depending to the aspect-ratio of the flow. The difference between $\mathrm{N_c}$ and $\mathrm{N_{t,im}}$ can be used to quantify that bias; see eq.~({\ref{finalCoiling}}).

\section{Axially-symmetric, force-free flux ropes}\label{sec:simpleFR}

In this Section, we are going to study the properties of generic, axially-symmetric, force-free flux ropes. We are first going to write down the general equations governing the magnetic field and current in such flux ropes, as well as the coiling and local twist rates.  Then we are going to apply them to several specific flux-rope examples, including ones not discussed previously in the literature.  We are going to collect our results in Fig.~\ref{fig:fr} along the way.

\begin{figure}[t!]\epsscale{1.}
	\plotone{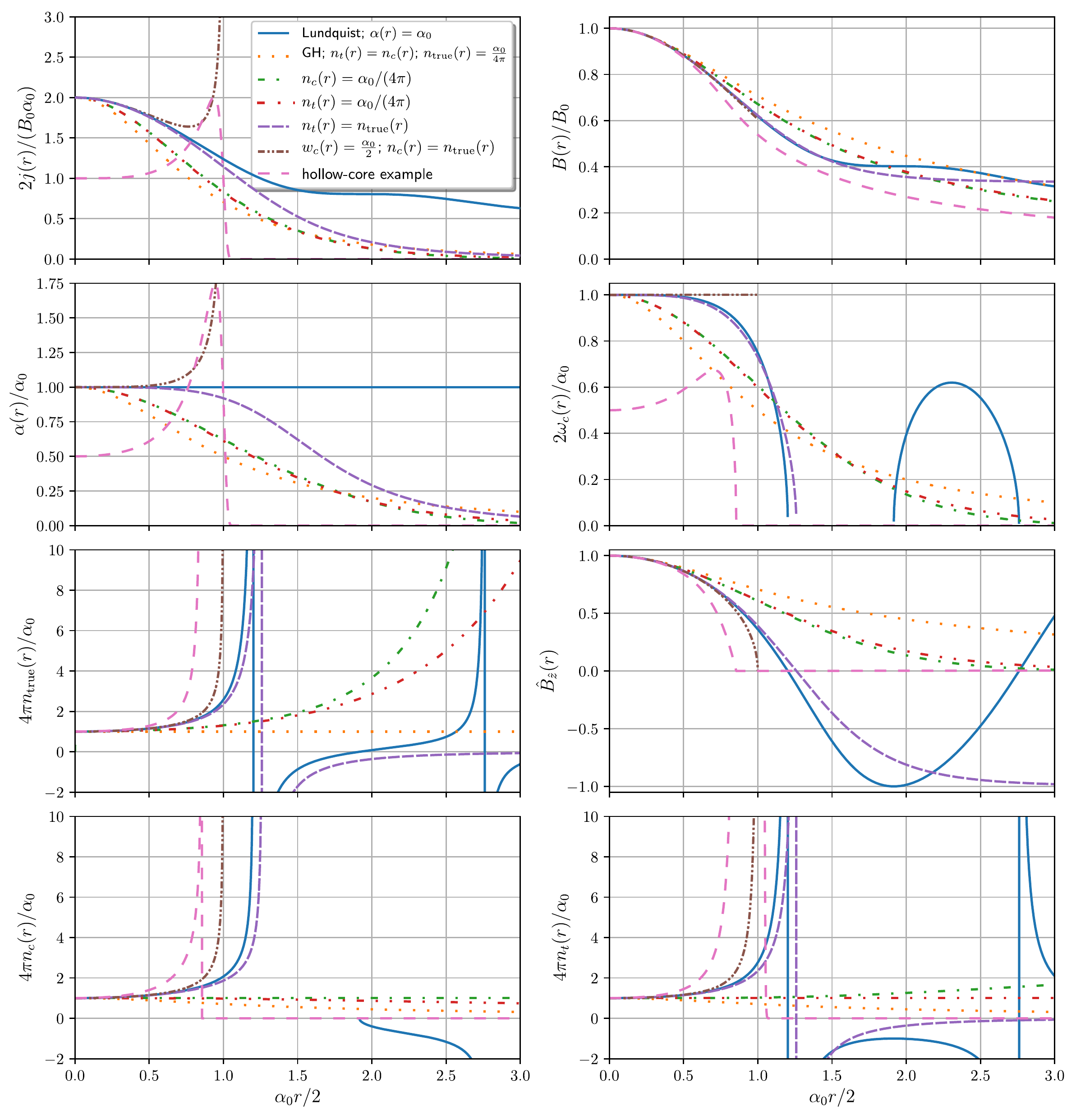}\caption{Here we summarize the results of Section~\ref{sec:simpleFR}. We show the current density ($j$), magnetic field magnitude ($B$), force-free parameter ($\alpha$), coiling rate ($\omega_c$), true non-local twist number per unit flux-rope length ($n_{\mathrm{true}}$), component of the normalized magnetic field along the flux-rope axis ($\hat B_{\hat z}$), coiling  number per unit flux-rope length ($n_c$), and local twist  number per unit flux-rope length ($n_t$). These quantities are plotted as a function of radial distance ($r$) from the axis of the flux rope for seven different axially-symmetric, force-free flux rope models: (1) the Lundquist model, featuring a constant $\alpha(r)$; (2) the Gold-Hoyle (GH) model, for which the coiling rate matches the local twist rate, and the non-local twist number per unit flux-rope length is fixed: $n_{\mathrm{true}}(r)=\alpha_0/(4\pi)$; (3) a flux rope with a fixed coiling number per unit flux-rope length: $n_c(r)=\alpha_0/(4\pi)$; (4)  a flux rope with a fixed local twist number per unit flux-rope length: $n_t(r)=\alpha_0/(4\pi)$; (5)  a flux rope for which the twist number equals the true non-local twist number ($n_t(r)=n_{\mathrm{true}}(r)$); (6)  a flux rope with a constant coiling rate ($\omega_c=\alpha_0/2$), for which we show that the coiling number equals the true non-local twist number: $n_c(r)=n_{\mathrm{true}}(r)$. That flux rope model is defined up to $\alpha_0r/2<1$. (7) A hollow-core flux rope example. We scaled $\alpha(r)$, $j(r)$ and $\omega_c(r)$ for the last model by dividing those quantities by 2. All models are parametrized using $\alpha_0\equiv\alpha(r=0)$. See the text for discussion. 
	}\label{fig:fr}
\end{figure}

\subsection{Generic, axially-symmetric, force-free flux ropes\label{app:gen}}
Let us calculate the local coiling rate for flux ropes with cylindrical symmetry. In the orthonormal cylindrical basis $(\bm{\hat r},\, \bm{\hat \phi},\, \bm{\hat z} )$, we impose the following  conditions for the non-normalized magnetic field components:
\be\label{conditionsBcyl}
B_{\hat r}=0, \quad \lim\limits_{r\to0}B_{\hat\phi}(r)=0+\mathcal{O}(\alpha_0r)\ , 
\ee
with non-zero $B_{\hat\phi}(r)$ and $B_{\hat z}(r)$, which are functions solely of the cylindrical radial coordinate $r$. In the above equation, $\alpha_0$ is a constant to be specified below. The limit in the  second equality above is well-defined for non-diverging current distributions that close to the flux-rope axis are linear in $r$ (cf. eq.~(\ref{bc_all})). 
Given the above conditions, from equations~(\ref{alphaFinal}), (\ref{omega0Final}), as well as eq.~(\ref{cylcov}) from Appendix~\ref{sec:app} for the gradient of the magnetic field in cylindrical coordinates, we can express the coiling rate and the \textit{generalized} force-free parameters as:
\be\label{cylSym}
\omega_c^2&=&\frac{1}{2r}\frac{\partial}{\partial r}\left( \frac{B_{\hat \phi}^2(r)}{B^2(r)}\right)=\frac{1}{2r}\frac{\partial}{\partial r}\left( \hat B_{\hat \phi}^2(r)\right)\ ,\nonumber\\
\alpha &=& \frac{1}{r}\left(\frac{B_{\hat z}}{B}\right)^2\frac{\partial}{\partial r}\left(r\, \frac{B_{\hat \phi}(r)}{B_{\hat z}(r)}\right)=\frac{\hat B_{\hat z}^2}{r}\,\frac{\partial}{\partial r}\left(r\, \frac{\hat B_{\hat \phi}(r)}{\hat B_{\hat z}(r)}\right)\ .
\ee
From the first equation above, we can see that the coiling rate is defined for increasing $\hat B_\phi^2$, and vanishes when $\hat B_\phi$ has an extremum, or when $\hat B_\phi$ vanishes. The first condition is always satisfied near the axis of the flux ropes we consider with finite current distribution as can be seen from eq.~(\ref{conditionsBcyl}).  This behavior can be seen by comparing the right panels on the second and third rows of Fig.~\ref{fig:fr} for the sample of flux ropes discussed below.

Let us also write down the coiling number, local twist number, and the true non-local twist number, all evaluated per unit flux-rope length measured along the rope axis ($L_z$). We are going to denote those quantities by $n_c$, $n_t$ and $n_{\mathrm{true}}$, respectively. Using eq.~(\ref{coilingFinal}), one can calculate $n_c$ as:
\be\label{twist_linden}
n_c\equiv\mathrm{N_c}/L_z=(2\pi L_z)^{-1}\int_L \omega_c d\tau=(2\pi L_z)^{-1}\int\limits_0^{L_z} \frac{\omega_c}{\hat B_{\hat z}} d\tau_z=\frac{\omega_c}{2\pi\hat B_{\hat z} }\ ,
\ee
where $\tau_z$ is the field-line length parameter along the axis of the flux rope, and the integrals are taken over a fixed flux-rope length, $L_z$. Similarly, we can write \citep[see also][]{2016ApJ...818..148L}:
\be\label{rates_n}
n_t\equiv\mathrm{N_t}/L_z=\frac{\alpha}{4\pi\hat B_{\hat z} }\ , \quad n_{\mathrm{true}}=\frac{\hat B_{\hat \phi}}{2\pi r\hat B_{\hat z} } \ .
\ee
One can eliminate the radial derivative of $\hat B_{\hat \phi}$ from eq.~(\ref{cylSym}), and combine with equations~(\ref{twist_linden}) and (\ref{rates_n}) to write a concise relationship between the rates above:
\be
n_c^2=\big(2 n_t -n_{\mathrm{true}}\hat B_{\hat z}\big)n_{\mathrm{true}}\hat B_{\hat z}  \ .
\ee
To obtain the  equation above, we used the normalization condition:
\be\label{gen_fr_norm}
\hat B_{\hat \phi}^2+\hat B_{\hat z}^2=1\ .
\ee

For generic flux ropes with cylindrical symmetry, one can Taylor expand $B_{\hat\phi}(r)$ and $B_{\hat z}(r)$ and then substitute those series in eq.~(\ref{cylSym}). As long as the conditions in eq.~(\ref{conditionsBcyl}) are satisfied, one can then explicitly check that:
\be\label{coretwist}
\lim\limits_{r\to0}\omega_c=\lim\limits_{r\to0}\left(\frac{\alpha}{2}\right)=\lim\limits_{r\to0}\left(\frac{B_{\hat\phi}}{rB_{\hat z}}\right)
\ee
at the flux rope axis, which is confirmed in all examples in this section (compare the panels in the second row of Fig.~\ref{fig:fr} together with the left panel on the third row of the same figure). The last quantity is the true non-local twist rate (equal to $2\pi n_{\mathrm{true}}$; cf. eq.~(\ref{rates_n})) of the axially symmetric rope, and therefore, $\omega_c$ and $\alpha/2$ are equally good measures of twist at the core of such flux ropes\footnote{The fact that $\alpha/2$ gives the correct coiling rate at the core of axially-symmetric flux ropes with smooth current density, was shown by \cite{2016ApJ...818..148L}. In Section~\ref{sec:simpleFR}, we show that that requirement can be relaxed to include non-diverging current distributions, which close the flux-rope axis become linear functions of radius.}. As one moves away from the axis of the rope, both $\alpha/2$ and $\omega_c$ diverge from the true non-local rate of twisting of field lines around the axis of the rope (see Fig.~\ref{fig:fr}). One can quantify that deviation by using the $c_3$ parameter introduced by \cite{2016ApJ...818..148L}. However, predicting that \textit{non-local} twisting rate from purely local properties of the field is not possible for generic 3D flux ropes, which is why in this paper we focused only on predicting as accurately as possible the \textit{local} rate of coiling of infinitesimally separated field lines.

To apply our analysis to simple flux ropes, first let us construct a generic, force-free, axially-symmetric flux rope using the normalized components of the magnetic field, $\hat B_{\hat z}(r)$ and $\hat B_{\hat \phi}(r)$. We do that by imposing the force-free condition $\bm j=\alpha \B$, where we use units for current, such that the current density is given by $\bm j=\bm{\nabla}\times\B$. The force-free condition implies that $j_{\hat\phi}/B_{\hat \phi}=j_{\hat z}/B_{\hat z}$, in which we can  substitute the components of the normalized magnetic field, multiplied by a radially-dependent magnitude of the magnetic field $B(r)$, to find:
\be\label{ff_fr}
B(r)=B_0\exp\left[-\int\limits_{0}^r \frac{\hat B_{\hat \phi}^2(r')}{r'}dr'\right]\mbox{\quad for force-free flux ropes,}
\ee
where $B_0\equiv B(r=0)$ is a constant, and we applied eq.~(\ref{gen_fr_norm}). From the above equation, we can see that $B(r)$ is a monotonically decreasing function (top-right panel of Fig.~\ref{fig:fr}). The integral above converges as long as we require that $\hat B_{\hat \phi}$ decays at the rope axis sufficiently fast, which holds when the boundary condition, eq.~(\ref{conditionsBcyl}), is satisfied. 

One can use equations~(\ref{gen_fr_norm}) and (\ref{ff_fr}), combined with eq.~(\ref{cylSym}), to show that the current density for a generic, axially-symmetric, force-free flux rope is given by:
\be\label{gen_fr_j}
j^2(r)=\alpha^2(r) B^2(r)=-\frac{\bigg[\big((r B)'B\big)'\bigg]^2}{4 r (rB)'B'}\mbox{ \quad for force-free flux ropes,}
\ee
where a prime denotes a derivative in $r$. 

If one is solving eq.~(\ref{gen_fr_j}) for $B(r)$, one needs to solve it for $B''(r)$ first. The equation has two roots, only one of them being physical. We write it down here for completeness:
\be\label{b_from_j}
B''(r)+\frac{B'(r)}{r B(r)}\left[rB'(r)+3B(r)-2r j(r)\sqrt{-\left(1+\frac{B(r)}{rB'(r)}\right)}\right]=0\ .
\ee
As can be seen from eq.~(\ref{ff_fr}), the quantity under the square root above is guaranteed to be non-negative by the requirement $0\leq\hat B_{\hat \phi}^2\leq1$, which serves as a constraint on the allowed $j(r)$ profiles. Similarly, $\omega_c$ can be expressed solely using $B(r)$:
\be\label{gen_fr_wc}
\omega_c^2(r)=-\frac{\big(\ln(B)'r\big)'}{2r}\mbox{ \quad for force-free flux ropes.}
\ee

Equations~(\ref{cylSym}, \ref{gen_fr_norm}, \ref{ff_fr}, \ref{gen_fr_j}, \ref{b_from_j}, \ref{gen_fr_wc}), combined with the boundary condition, eq.~(\ref{conditionsBcyl}), allow one to construct a generic, force-free, axially-symmetric flux rope, as long as only \textit{one} of the functions $\hat B_{\hat \phi}(r)$, $\hat B_{\hat z}(r)$, $\alpha(r)$, $\omega_c(r)$, $j(r)$ or $B(r)$ is specified. Using those equations, one can explicitly check (by Taylor expansion) that the boundary condition, eq.~(\ref{conditionsBcyl}), translates to the following behavior with $r\to0^+$:
\be\label{bc_all}
\alpha(r)&\approx& \alpha_0+\mathcal{O}(\alpha_0r)\nonumber\\
\omega_c(r)&\approx& \frac{\alpha_0}{2}+\mathcal{O}(\alpha_0r)\nonumber\\
2\pi n_{\mathrm{true}}(r)&\approx& \frac{\alpha_0}{2}+\mathcal{O}(\alpha_0r)\nonumber\\
B(r)&\approx&B_0\left[1- \frac{\alpha_0^2r^2}{8}+\mathcal{O}(\alpha_0^3r^3)\right]\nonumber\\
j(r)&\approx&\alpha_0 B_0+\mathcal{O}(\alpha_0r)\nonumber\\
\hat B_{\hat \phi}(r)&\approx&\frac{\alpha_0r}{2}+\mathcal{O}(\alpha_0^2r^2)
\nonumber\\
\hat B_{\hat z}(r)&\approx&1-\frac{\alpha_0^2r^2}{8}+\mathcal{O}(\alpha_0^3r^3)
\ ,
\ee
where $\alpha_0\equiv \alpha(r=0)$.  The above set of equations can be used to set the boundary condition at $r=0$ for equations~(\ref{cylSym}, \ref{gen_fr_norm}, \ref{ff_fr},  \ref{gen_fr_j}, \ref{b_from_j}, \ref{gen_fr_wc}). So, for example, from eq.~(\ref{bc_all}), we can see that at $r=0$ we have $B'(r=0)=0$ and $\hat B_{\hat z}'(r=0)=0$.
Next, we are going to use those expressions to study specific examples of  force-free, axially-symmetric flux-ropes\footnote{\label{ft:kappa} For completeness, we give the expressions for the curvature ($\kappa$) and torsion ($\tau_{r}$) of field lines in axially-symmetric flux ropes. Using the Frenet-Serret formulas from differential geometry, as well as eq.~(\ref{cylcov}), we obtain: 
\be
\kappa=\frac{\hat B_{\hat \phi}^2}{r}\ , \ \ \tau_r=\frac{\hat B_{\hat \phi}\hat B_{\hat z}}{r}\ .
\ee
At the flux rope axis, those quantities approach:
\be
\kappa\approx \frac{\alpha_0^2 r}{4}+\mathcal{O}(\alpha^3_0 r^2)\ ,\ \ \tau_r\approx \frac{\alpha_0}{2}+\mathcal{O}(\alpha_0 r)\ ,
\ee
where we used eq.~(\ref{bc_all}). 
Note that similar to the coiling rate, the torsion approaches the true non-local twist rate at the flux rope axis (cf. eq.~(\ref{coretwist})).}.

\newpage
\subsection{The Gold-Hoyle and Lundquist flux ropes}
In the Gold-Hoyle model of a flux rope \citep[GH;][]{1960MNRAS.120...89G}, we have $B_{\hat \phi} = B_0 br /( 1 + b^2 r^2 )$ and $B_{\hat z} = B_0/( 1 + b^2 r^2 )$ (with $b$ and $B_0$ being constant), thus satisfying eq.~(\ref{conditionsBcyl}). The GH model has a constant non-local twist rate given by $n_{\mathrm{true}}(r)=b/(2\pi)$ (see eq.~(\ref{rates_n})). In Fig.~\ref{fig:fr}, we show the results of Section~\ref{app:gen} applied to the GH model.

Equation~(\ref{cylSym}), applied to the GH model, gives:
\be
\alpha=2\omega_c=\frac{2b}{1+b^2r^2}\ .
\ee
Comparing the first equality above with eq.~(\ref{finalCoiling}), we can see that, independent of $r$, the transverse field-line flow for this model corresponds to a center with purely circular trajectories around it. Thus, for this flux rope model, $\alpha/2$ does not overestimate the local field-line coiling rate. 

However, we can immediately point to a counterexample: the Lundquist model \citep{lundquist}, which again satisfies eq.~(\ref{conditionsBcyl}). In the Lundquist model, $B_{\hat \phi}=B_0J_1(\alpha r)$ and $B_{\hat z}=B_0J_0(\alpha r)$, where $B_0$ and $\alpha(r)=\alpha_0$ are constants, the latter matching the force-free parameter; and $J_n$ are the Bessel functions of the first kind. From eq.~(\ref{cylSym}), we obtain:
\be
2\omega_c(r)\approx\alpha\left[1 -\frac{\alpha^4 r^4}{128} + \mathcal{O}(\alpha^6 r^6)\right]\ .
\ee
Clearly, the coiling rate approaches $\alpha/2$ at the flux rope core; however, away from the rope axis, the two diverge, indicating that the transverse flow becomes elliptical (cf. eq.~(\ref{finalCoiling})). 

We demonstrate the properties of the Lundquist model in Fig.~\ref{fig:fr}. One can see that the true non-local twist rate becomes infinite when $\hat B_{\hat z}(r)=0$, which happens at the zeros of $J_0(\alpha_0 r)$: $\alpha_0r/2\approx 1.20$ and $2.76$, and so on  (third row, right panel of Fig.~\ref{fig:fr}). Beyond the first zero at $\alpha_0r/2\approx 1.20$, the coiling rate becomes imaginary for a range of radii, implying that locally the field lines are no longer twisting around one another, and instead they form a saddle flow in the normal plane. Thus, one can argue that the first zero of $\hat B_{\hat z}(r)$ delineates the innermost boundary of the Lundquist flux rope. Indeed, beyond that radius, the sense of winding of the field lines around the flux-rope axis reverses (third row, left panel of Fig.~\ref{fig:fr}).

\subsection{Further flux-rope examples}
\subsubsection{Flux rope with constant coiling rate}
One may be curious whether there are closed-form expressions for axially-symmetric flux ropes with constant $\omega_c$. Combining eq.~(\ref{cylSym}) with eq.~(\ref{conditionsBcyl}) one can show that for $\omega_{c}=$constant, we have $\hat B_{\hat \phi}=\omega_c r$, with $\hat B_{\hat z}=\sqrt{1-\hat B_{\hat \phi}^2}$; and therefore, such a flux-rope is defined only within $r\leq |\omega_c|^{-1}$.  Knowing the normalized components of the magnetic field, one can construct a force-free flux rope using eq.~(\ref{ff_fr}):
\be
\left[B_{\hat r},\ B_{\hat \phi},\ B_{\hat z}\right]=\left[0,\  B_0e^{-(\omega_c r)^2/2}\omega_c r,\   B_0 e^{-(\omega_c r)^2/2} \sqrt{1-(\omega_c r)^2}\right],\quad \mbox{for } |r\omega_c|\leq 1,
\ee
where $B_0$ and $\omega_c$ are constants. Combining the above equation with equations~(\ref{twist_linden}) and (\ref{rates_n}), one can easily show that $n_c(r)=n_{\mathrm{true}}(r)$ for this flux rope, i.e. the coiling number per unit flux-rope length matches the true non-local twist number.

The corresponding $\alpha$ is given by:
\be
\alpha=2\omega_c\left(\frac{1-\frac{(\omega_cr)^2}{2}}{\sqrt{1-(\omega_c r)^2}}\right)\approx2\omega_c\left[1 +\frac{(\omega_cr)^4}{8} + \mathcal{O}(\omega_c^6 r^6)\right]
\ .
\ee
The current distribution for this flux rope diverges at $r\to|\omega_c|^{-1}$, has a non-zero minimum at $r=|\omega_c|^{-1}\sqrt{2-\sqrt{2}}\approx 0.765 |\omega_c|^{-1}$ and a finite maximum at $r=0$. See Fig.~\ref{fig:fr} in which we demonstrate the properties of this flux rope with a constant $\omega_c(r)=\alpha_0/2$. This  exhibits a core in which the current density is smaller than  at flux-rope periphery. Indeed, one is free to use the equations of  Section~\ref{app:gen} to construct hollow-core flux ropes as well (see Fig.~\ref{fig:fr}, where we include such an example); see below for a discussion.

\subsubsection{Flux rope with constant coiling number per unit flux-rope length}
Another analytic example is  a flux rope with a constant  coiling number per unit flux-rope length: $n_c=$const., with $n_c$ given by eq.~(\ref{twist_linden}). Combining that with eq.~(\ref{cylSym}), 	 one can show that in this case $\hat B_{\hat z}(r)=\exp\left[-(2\pi r n_c)^2/2\right]$, with  $\hat B_{\hat \phi}=\pm\sqrt{1-\hat B_{\hat z}^2}$. Once again, one can construct a force-free flux rope by a suitable choice of $B(r)$. That flux-rope is given by:
\be\label{b_fr}
\left[B_{\hat r},\ B_{\hat \phi},\ B_{\hat z}\right]&=&\left[0,\  \pm B(r)\sqrt{1-e^{-(2\pi r n_c)^2}},\  B(r)e^{-\frac{(2\pi r n_c)^2}{2}}\right]\ ,
\ee
with $B(r)$ given by eq.~(\ref{ff_fr}) with $\hat B_{\hat \phi}=B_{\hat\phi}/B$, and $B_{\hat\phi}$ given by eq.~(\ref{b_fr}).
The corresponding current distribution is peaked around the $r=0$ axis, similar to the majority of the models shown in Fig.~\ref{fig:fr}. For completeness, we give the corresponding analytical coiling rate and $\alpha$ below:
\be
\omega_c(r)&=&2\pi n_c e^{-\frac{(2\pi r n_c)^2}{2}}\ ,\nonumber\\
\frac{\alpha(r)}{2\omega_c(r)}&=&\frac{1}{2}\left[\frac{\hat B_{\hat \phi}}{2\pi r n_c}+\frac{2\pi r n_c}{\hat B_{\hat \phi}}\right]\approx 1+\frac{(\pi r n_c)^4}{2}+\mathcal{O}(n_c^6r^6)\ .
\ee

\subsection{Discussion}

One can similarly construct a flux rope with a constant local twist number per unit flux-rope length ($n_t(r)$=const.), as well as a flux rope for which the local twist number per unit  flux-rope length matches the true non-local twist number ($n_t(r)=n_{\mathrm{true}}(r)$). We do not provide analytical expressions for those two flux-rope configurations. Instead, in Fig.~\ref{fig:fr} we show our results for them by numerically integrating the equations derived in Section~\ref{app:gen}.

From Fig.~\ref{fig:fr}, one can see that the flux ropes with constant coiling, local twist, and non-local twist rates per flux-rope length (i.e. those in the figure denoted with $n_c(r)=\alpha_0/(4\pi)$, $n_t(r)=\alpha_0/(4\pi)$, and GH, respectively), and especially the first two, show quite similar properties in terms of current density, magnetic field, $\alpha$, and $\omega_c$. However, the true non-local twist rate becomes drastically different beyond $\alpha_0r/2\sim 1$: compare the left panel on the third row with both panels on the last row of Fig.~\ref{fig:fr}. 

From this set of examples, one can see that as long as $\alpha_0r/2\lesssim 1$, the non-local twist number is well approximated by both the coiling and local twist numbers to within several tens of percent. This is not surprising in light of equations~(\ref{coretwist}) and (\ref{bc_all}). This statement is valid independent of whether we are looking at hollow-core or centrally-peaked current distributions. Indeed, in Fig.~\ref{fig:fr} we include a hollow-core flux rope example (with a current density given by the more or less random $j(r)=\alpha_0 B_0\exp\left[(r\alpha_0/2)^4-(r\alpha_0/2)^{40}\right]$, which produces such a current profile). 

However, the numerical results quoted above show that any study relying on determining an accurate non-local twist number far from the flux-rope axis, cannot rely solely on the local twist or coiling numbers. Instead, one would have to resort to finding the flux-rope axis to calculate the non-local twist rate \citep[e.g.][]{2016ApJ...818..148L}. In turn, for complex magnetic field configurations, especially those found in global models of the solar corona, the first step to finding the flux-rope axis is finding the flux ropes themselves. As argued in Section~\ref{sec:coiling} and as shown in a follow-up paper \citep{Tassev2019},  using the natural $\mathrm{N_c}\gtrsim 1$ threshold as a flux-rope detection threshold is superior to using $\mathrm{N_t}\gtrsim 1$ due to the latter giving many false-positives. As  the results discussed above (and shown in Fig.~\ref{fig:fr}) indicate, once the presence of a flux rope has already been established, both the coiling and local twist number are equally reliable estimates of the non-local twist sufficiently close to the flux rope axis.

\section{Summary}\label{sec:summary}

We study the properties of the local transverse deviations of magnetic field lines. Those deviation vectors follow a planar flow in a plane normal to one of the field lines (the ``reference'' field line, which can be any field line of interest: e.g. the one passing through the voxel for which we are calculating the coiling and squeezing rates) as the  plane is moved along the length of that field line. As long as we focus only on field lines with infinitesimal separations, the resulting planar field-line flow is linear with a critical point at the location where the reference field line intersects the normal plane. Therefore, the types of transverse field-line flows  are given by the  standard equilibrium solutions of linear systems, such as a saddle, cycle, spiral or node. The type of solution is
unique for the normal plane that is  \textit{minimally rotated} around its normal when moved along the reference field line (see Section~\ref{sec:uniq}). We show that that solution is determined by the two non-zero eigenvalues of the gradient of the normalized magnetic field. 

We argue that the eigenvalue difference, $\Delta\lambda$, can be used to quantify the degree of squeezing or coiling of neighboring field lines. Thus, when $\Delta\lambda$ is real, we define it as the squeezing rate ($\rho_{\mathcal{Z}}$) which, when integrated over a field line, gives $\ln(\mathcal{Z})$, with $\mathcal{Z}$ defined as the squeeze factor. We demonstrate that  $\mathcal{Z}$ can be approximated by the squashing factor, $Q$.  Thus, the squeeze factor offers an alternative, with which one can check the robustness of a QSL detection.

QSL maps have  become largely synonymous with maps\footnote{However, calculating maps of the  squashing factor are computationally expensive, which lead \cite{QSL_Squasher} to propose an alternative, computationally cheap, crude method for locating QSLs using so-called FLEDGE maps.} of the squashing factor, $Q$, with QSLs identified as the locations of exponentially large $Q$  \citep[e.g.][]{Pariat12}. Yet, $Q$ has the following drawback: it can miss QSLs corresponding to regions of large local field-line squeezing, when that squeezing is undone before the field lines reach the footpoints between which $Q$ is calculated. As the squashing factor is an inherently globally-defined quantity, we show that one cannot construct an exact local squashing factor rate, which could account for such localized squeezing. However, we show that $\rho_{\mathcal{Z}}$ can be treated as a local approximation to the rate of squashing in a certain strict sense, described in the paper. As by definition, $\rho_{\mathcal{Z}}\geq 0$, the squeeze factor, $\mathcal{Z}$, does not suffer from the problem of missing unsqueezed regions described above. Thus, in that regard, $\mathcal{Z}$ is a better proxy for QSLs than the squashing factor.   Moreover, given by just an integral over a field line, the squeeze factor is computationally cheaper than the squashing factor, which in addition, depends on computing the relative deviations of neighboring field lines \citep[e.g.][]{QSL_Squasher}.

As  $\mathcal{Z}$ grows exponentially large in the vicinity of HFTs and null points, one can use a threshold in the local $\rho_{\mathcal{Z}}$ to find the locations of nulls, HFTs and current sheets (see \cite{Tassev2019}). This is in contrast to the squashing factor, for which an equivalent, purely local squashing rate does not exist; and thus, one has to run a full non-local squashing factor calculation to obtain a similar HFT detection \citep[e.g.][]{Savcheva16a}. One may argue that when calculated on a grid, $\rho_{\mathcal{Z}}$, lacks discriminatory power to distinguish a null point from a short HFT, for example. However, finding a null point from a discrete sampling of the magnetic field is inherently ambiguous as it depends on the interpolation scheme used to supersample the magnetic field inside each grid cell\footnote{Whether a null is present inside  a grid cell depends on whether the three surfaces defined by $B_i=0$ intersect at a point within the cell or not. As the shape of those surfaces is determined by the interpolation scheme used to interpolate $\B$ within the cell, so does the presence of a null point.  For an algorithm for finding null points using trilinear interpolation, see \citep{2007PhPl...14h2107H}.}.

When $\Delta\lambda$ is imaginary, we define it as twice the coiling rate ($\omega_c$) which, when integrated over a field line, gives the coiling number, $\mathrm{N_c}$. We show that unlike the standard local twist number \citep[e.g.][eq. 16]{2006JPhA...39.8321B}, the coiling number gives an unbiased estimate of the local number of twists neighboring field lines make around one another. The local twist number is defined as a field-line integral over what-we-call the generalized force-free parameter ($\alpha$), which is proportional to the component of the current parallel to the magnetic field. As $\alpha$ is the equivalent of vorticity of the planar transverse field-line flow, it is non-zero for saddle flows which exhibit shear. Yet, field lines do not revolve around one another in such flows. Therefore, $\alpha$ clearly gives a biased estimate of local twist, resulting in $\mathrm{N_t}\geq 1$ even in regions exhibiting shear, well outside of flux ropes.  We find that $\alpha$ is biased even for elliptical/spiral transverse field-line flows. We offer a geometric interpretation of that bias: $\alpha$ is enhanced relative to the average rate of  rotation of neighboring  field lines with factors depending on the aspect ratio of the transverse field-line flow. We show that $\omega_c$ is unbiased in that regard, as for steady-state flows, $2\pi/\omega_c$ exactly equals the period of rotation of infinitesimally-separated field lines around one another. We also show that for non-diverging current distribution, $\omega_c$ approaches $\alpha/2$ at the core of axially-symmetric flux ropes, and therefore, is an  equally good measure of twist at the core of such flux ropes (Section~\ref{sec:simpleFR}). The coiling number can be useful in the study of flux rope instabilities, such as the kink instability \citep[e.g.][]{2016ApJ...818..148L}. Moreover, one can use the natural inequality $\mathrm{N_c}\gtrsim 1$ for detecting flux ropes in complicated magnetic field configurations. We demonstrate that in \citep{Tassev2019}.

We give closed-form expressions for $\rho_{\mathcal{Z}}$ and $\omega_c$ in terms of the gradient of the normalized magnetic field in both Cartesian and curvilinear coordinates. We implemented the calculation of the squeeze factor and coiling number, as well as their respective local rates, in the public code\footnote{The code is available here: \url{https://bitbucket.org/tassev/qsl_squasher/}. The code allows one to output the squeeze rate as calculated using either of the two options proposed for handling elliptical flows: equations~(\ref{possibleImRhoZ1}) or (\ref{possibleImRhoZ2}); or using the squeeze rate corresponding to the symmetrized $m$ (Section~\ref{sec:sym}). The code also outputs the local type of critical point (e.g. spiral, node, saddle) -- thus, offering another type of partitioning of magnetic fields -- as well as the trace of the gradient of the normalized magnetic field. The code can be easily modified to output the aspect ratio of elliptical flows using eq.~(\ref{finalCoiling}); the orientation of the eigenvectors of the flow (or for elliptical flows -- the orientation of the semi-axes; see footnote~\ref{ft:eig}); as well as non-transverse quantities such as the curvature of the field lines using eq.~(\ref{curv}). We are going to explore those field-line properties elsewhere.} \qsl introduced by \cite{QSL_Squasher}. In a follow-up paper, we show numerical calculations of those quantities for realistic magnetic fields \citep{Tassev2019}.

\begin{acknowledgements}
We would like to thank Edward DeLuca for useful discussions and comments on the manuscript.    
\end{acknowledgements}

\appendix
\section{Results for curvilinear coordinates}\label{sec:app}

In this section, we  obtain the expressions for the coiling and squeezing  rates in curvilinear coordinates.
Below we  first write our results for any coordinate basis, and then we  focus on spherical and cylindrical coordinates. We  also address an approximation introduced in the squashing factor calculation in spherical coordinates in \citep{QSL_Squasher}.

\subsection{The field-line deviation equation}\label{sec:fldevcurve}

For two infinitesimally separated points in space at positions $\x$ and $\x+d\x$, the  difference between their radius-vectors can  be written as $d\x=dx^i\e_i(\x)$ only in  \textit{coordinate} basis $\e_i(\x)$ (indeed, this can serve as a definition of a coordinate basis). The equation for the field-line radius vector $\x(\tau)$ (eq.~(\ref{fl})) can then be written in any coordinate basis as:
\be\label{flcoo}
\frac{d\x(d\tau)}{\delta\tau}=\frac{dx^i(\tau)}{d\tau}\e_i(\x(\tau))=\hat B^i( \x(\tau))\e_i(\x(\tau))=\hatbB(\x(\tau))\ ,
\ee
and similarly for a neighboring field line with radius vector $\y(\tau)$. 

The field line deviation equation (eq.~(\ref{dev})) for $\delta \x(\tau)=\y(\tau)-\x(\tau)$ then takes the form: 
\be\label{rhsdev}
\frac{d\delta \x(\tau)}{d\tau}&=&\frac{d\y(\tau)}{d\tau}-\frac{d\x(\tau)}{d\tau}\nonumber\\
&=&\hat B^i( \y(\tau))\e_i(\y(\tau))-\hat B^i( \x(\tau))\e_i(\x(\tau))\nonumber\\
&=&\delta x^k\hat B^i_{\ ,k}(\x(\tau))\e_i(\x(\tau))+\hat B^i(\x(\tau))\delta x^k \partial_k\e_i(\x(\tau))\nonumber\\
&=&\delta x^k\hat B^i_{\ ,k}(\x(\tau))\e_i(\x(\tau))+ \delta x^k B^j(\x(\tau))\Gamma^i_{\ jk}\e_i(\x(\tau))\nonumber\\
&=&\delta x^k \hat B^i_{\ ;k}\e_i(\x(\tau))\ ,
\ee
where we linearized in $\delta \x$ and  used the fact that the Christoffel symbols, $\Gamma^{k}_{\ ij}$, give the spatial derivative of the basis vectors: $\partial_j\e_i=\Gamma^{k}_{\ ij}\e_{k}$. In writing the above equation, we used the standard coma notation for  partial derivatives, and the semicolon notation for covariant derivatives. 

The equality between the first and the last lines in eq.~(\ref{rhsdev}) is hardly surprising. From that equation, one can read off the $\tau$ derivative of the components of $\delta \x$ as $ \hat B^i_{\ ;k} \delta x^k$. That is correct if the components of \textit{both} $\delta \x(\tau+\delta\tau)$ and $\delta \x(\tau)$ (entering in $d\delta\x(\tau)$ in the finite difference sense, with $\delta \tau$ being infinitesimal) are written in the  common basis $\e_i(\x(\tau))$. This is exactly what we want for the generalization of $M_{ij}$ in eq.~(\ref{ddx}) to curvilinear coordinates. Thus, we can conclude that in any  coordinate basis, $\e_{i}(\x(\tau))$,  (being careful about index placement):
\be\label{covM}
M^{i}_{\ j}=\hat B^i_{\ ;j}\ ,
\ee
which is a rather unsurprising result (but see Section~\ref{sec:stdQcurv}). We used that result to promote the derivatives, entering in the  squeezing  and coiling rates (see equations~(\ref{qhoQFinal}) and (\ref{omega0Final})), to covariant derivatives. Note that those rates are scalars, and therefore the covariant derivative entering in them can be written in non-coordinate basis as well.

In Section~\ref{sec:Msph}, we  write  the above equation explicitly in the spherical and cylindrical orthonormal bases. However, let us first clear up a possible confusion, which may arise from eq.~(\ref{covM}), about how the standard squashing factor in curvilinear coordinates is calculated.

\subsection{Calculating the standard squashing factor}\label{sec:stdQcurv}
Let us take a brief detour and write the equations for the field line deviation components in a form suitable for calculating the standard (non-local) squashing factor in spherical coordinates, as done in eq.~(6) of \citep{QSL_Squasher}. To obtain eq.~(\ref{covM}) for $M^i_{\ j}$, we had to write all vectors and operators entering in eq.~(\ref{rhsdev}) in the same common basis $\e_i(\x(\tau))$, including both $\delta\x(\tau+\delta\tau)$ and $\delta\x(\tau)$, which enter in $d\delta\x(\tau)$ in the finite difference sense (with $\delta\tau$ being infinitesimal). However, in the standard calculation of $Q$, we need to evaluate the components of $\delta \x(\tau+\delta\tau)$ in its corresponding basis $\e_i(\x(\tau+\delta\tau))$, and \textit{not} in the basis $\e_i(\x(\tau))$. The reason is that one then wants to project out the components of $\delta \x(\tau+\delta\tau)$ that are parallel to $\hatbB(\tau+\delta\tau)$, which in numerical codes is   given in the basis $\e_i(\x(\tau+\delta\tau))$ and \textit{not} in the basis $\e_i(\x(\tau))$. Thus, we write the left-hand side of eq.~(\ref{rhsdev}) as:
\be\label{lhsdev}
\frac{d\delta \x(\tau)}{d\tau}=\frac{\delta\x(\tau+\delta\tau)-\delta\x(\tau)}{\delta\tau}&=&\frac{\delta x^i(\tau+\delta\tau)\e_i(\x(\tau+\delta\tau))-\delta x^i(\tau)\e_i(\x(\tau))}{\delta\tau}\nonumber\\
&=&\frac{\delta x^i(\tau+\delta\tau)-\delta x^i(\tau)}{\delta\tau}\e_i(\x(\tau)) +\delta x^i(\tau) \frac{dx^j(\tau)}{d\tau}\partial_j\e_i(\x(\tau)) \nonumber\\
&=&\frac{d\delta x^i(\tau)}{d\tau}\e_i(\x(\tau))+ \delta x^k(\tau)\hat B^j(\x(\tau))\Gamma^{i}_{\ kj}\e_{i}(\x(\tau))\ ,
\ee
where we used eq.~(\ref{flcoo}) and linearized in $\delta\tau$. Note again (first line above), that $\delta x^i(\tau+\delta\tau)$ and $\delta x^i(\tau)$ are written in \textit{different} coordinate bases.

Setting equations~(\ref{rhsdev}) and (\ref{lhsdev}) equal, we can see that the terms containing the Christoffel symbols cancel out. To do that, we note that the Christoffel symbols are symmetric in their lower two indices, which is true for any coordinate basis. The end result is:
\be\label{flDevQSLSQ}
\frac{d\delta x^i(\tau)}{d\tau}=\hat B^i_{\ ,k}(\x(\tau))\delta x^k\ ,
\ee
with the partial derivative entering above, and \textit{not} the covariant derivative which enters in eq.~(\ref{covM}). The above equation is valid in any coordinate basis, and therefore can be used to write the field line deviation equation in cylindrical or other coordinates, as needed. Below we focus our attention to spherical coordinates.

We  use the right-handed coordinate basis $\bm{e}_\phi,\ \bm{e}_\theta,\  \e_r$, which is related to the right-handed non-coordinate orthonormal basis $\hat{\bm\phi},\ \hat{\bm\theta}, \ \hat{\bm r}$ through:
\be\label{basis}
\left[\hat{\bm\phi},\ \hat{\bm\theta}, \ \hat{\bm r}\right]=\left[\frac{1}{r\cos(\theta)}\bm{e}_\phi,\ \frac{1}{r}\bm{e}_\theta,\  \e_r\right]\ .
\ee
Note that we use the convention used in \citep{QSL_Squasher}, where $\theta$ stands for latitude, and not co-latitude, so one needs to be careful when comparing with results in the right-handed orthonormal basis [radius, co-latitude, longitude]. 

In numerical datasets, the magnetic field in spherical coordinates is usually given in the non-coordinate orthonormal basis given above. We  denote the components of $\hatbB$ in that basis with hats over the indices, i.e. $\hat B^{\hat i}=[\hat B^{\hat \phi},\hat  B^{\hat \theta},\hat  B^{\hat r}]$. The components of $\hatbB$ in the coordinate spherical basis, we  denote by dropping the hats over the indices: $\hat B^{i}=[\hat B^{\phi},\hat  B^{\theta},\hat  B^{r}]$. Since $\hatbB=\hat B^{i}\e_i$ holds in any basis, then using eq.~(\ref{basis}) we can write:
\be\label{componentsB}
\left[\hat B^{\hat \phi},\hat  B^{\hat \theta},\hat  B^{\hat r}\right]=\left[r\cos(\theta)B^\phi,\ r B^\theta,\  B^r\right]\ .
\ee

Noting that the components of $\delta \x$ in the coordinate spherical basis are given by $\delta x^i=[\delta \phi,\delta\theta,\delta r]$, and using eq.~(\ref{componentsB}) to write eq.~(\ref{flDevQSLSQ})  using the components of $\hatbB$ in the non-coordinate spherical orthonormal basis, after a bit of algebra we find:
\be\label{dev_sph}
r\cos(\theta )\frac{d \delta \phi(\tau)}{d\tau}&=&\left[\delta\phi\frac{\partial}{\partial \phi}+\delta\theta\frac{\partial}{\partial \theta}+\delta r\frac{\partial}{\partial r}\right] \hat B^{\hat \phi} -\left[\frac{\delta r}{r}-\delta \theta \tan(\theta)\right]B^{\hat\phi}\ ,
\nonumber\\
r\frac{d\delta \theta(\tau)}{d\tau}&=&\left[\delta\phi\frac{\partial}{\partial \phi}+\delta\theta\frac{\partial}{\partial \theta}+\delta r\frac{\partial}{\partial r}\right] \hat B^{\hat \theta}-\frac{\delta r}{r}\hat B^{\hat \theta}\ ,
\nonumber\\
\frac{d\delta r(\tau)}{d\tau}&=&\left[\delta\phi\frac{\partial}{\partial \phi}+\delta\theta\frac{\partial}{\partial \theta}+\delta r\frac{\partial}{\partial r}\right] \hat B^{\hat r}\ ,
\ee
with the components of $\delta \x$ in the non-coordinate orthonormal spherical basis given by:
\be
\delta x^{\hat i}=\left[r \cos(\theta)\delta\phi,\ r\delta\theta, \ \delta r\right]\ ,
\ee
similar to eq.~(\ref{componentsB}) above, as well as to eq.~(8) of \citep{QSL_Squasher}. 

Comparing eq.~(\ref{dev_sph}) above with eq.~(6) of \citep{QSL_Squasher}, we can see that the terms neglected in eq.~(6) of that paper correspond to the terms $\left[\delta r/r-\delta \theta \tan(\theta)\right]B^{\hat\phi}$ and $(\delta r/r)\hat  B^{\hat \theta}$ which are correspondingly subtracted on the right-hand sides of the first two equations in (\ref{dev_sph}) above. In the above equations, those terms are suppressed relative to the rest by the ratio of the typical scale over which $\hatbB$ varies and the radius of the Sun. We implemented those terms in the open-source \qsl code, introduced by \cite{QSL_Squasher}. After performing numerical experiments with realistic data, we find that the additional terms affect the calculation of $Q$ only negligibly, given the overall qualitative nature of $Q$. Yet, we recommend one to implement those terms as they have the potential to be important in spherical coordinates for magnetic fields with small gradients.

\subsection{The magnetic field gradient}\label{sec:Msph}

The local  squeezing   ($\rho_{\mathcal{Z}}$) and coiling rates ($w_0$) are given by equations~(\ref{qhoQFinal}) and (\ref{omega0Final}) which, as argued in Section~\ref{sec:fldevcurve}, are valid in any  basis.  Here, we would like to write those equations explicitly in spherical and cylindrical coordinates.

To implement those equations numerically, it is convenient to write the covariant derivative of $\hatbB$ in the non-coordinate orthonormal basis given by eq.~(\ref{basis}). Calculating that derivative is a standard exercise, but we include it here for the sake of completeness. To do that, we need to start with the metric for spherical coordinates and calculate the Christoffel symbols, which allow us to find the covariant derivative $\hat B^i_{\ ;j}$ in the coordinate spherical basis. Then we substitute in the components of $\hatbB$ in the orthonormal basis (eq.~(\ref{componentsB})). Finally, we write the tensor components in the non-coordinate orthonormal basis (eq.~(\ref{basis})). We perform those steps with the open-source mathematics software, SageMath\footnote{\url{https://www.sagemath.org/}}. The final result is:
\allowdisplaybreaks[4]\be\label{BiCovJ}
\hat B^{\hat \phi}_{\ ;\hat \phi}&=&\frac{1}{r}\hat B^{\hat r}-\frac{\tan(\theta)}{r}\hat B^{\hat \theta}+\frac{1}{r\cos(\theta)}\frac{\partial\hat B^{\hat \phi}}{\partial\phi}\ ,\nonumber\\
\hat B^{\hat \theta}_{\ ;\hat \theta}&=&\frac{1}{r}\hat B^{\hat r}+\frac{1}{r}\frac{\partial\hat B^{\hat \theta}}{\partial\theta}\ ,\nonumber\\ 
\hat B^{\hat r}_{\ ;\hat r}&=&\frac{\partial \hat B^{\hat r}}{\partial r}\ ,\nonumber\\
\hat B^{\hat r}_{\ ;\hat \theta}&=&-\frac{1}{r}\hat B^{\hat \theta}+\frac{1}{r}\frac{\partial \hat B^{\hat r}}{\partial \theta}\ ,
\nonumber\\
\hat B^{\hat \theta}_{\ ;\hat r}&=&\frac{\partial \hat B^{\hat \theta}}{\partial r}\ ,\nonumber\\
\hat B^{\hat \phi}_{\ ;\hat r}&=&\frac{\partial \hat B^{\hat \phi}}{\partial r}\ ,\nonumber\\
\hat B^{\hat r}_{\ ;\hat \phi}&=&-\frac{1}{r}\hat B^{\hat \phi}+\frac{1}{r\cos(\theta)}\frac{\partial \hat B^{\hat r}}{\partial \phi}\ ,\nonumber\\
\hat B^{\hat \theta}_{\ ;\hat \phi}&=&\frac{\tan(\theta)}{r}\hat B^{\hat \phi}+
\frac{1}{r\cos(\theta)}\frac{\partial \hat B^{\hat \theta}}{\partial\phi}\ ,\nonumber\\
\hat B^{\hat \phi}_{\ ;\hat \theta}&=&\frac{1}{r}\frac{\partial \hat B^{\hat \phi}}{\partial \theta}\ .
\ee
Given that the above expressions are written in an orthonormal basis without using the fact that $\hatbB$ was normalized, the first three lines add up to the divergence  (as their sum is simply the trace  $\hat B^{\hat i}_{\ ;\hat i}$) of any vector field $\hatbB$ expressed in spherical coordinates; line 4 minus line 5 (one of the three non-zero  antisymmetric parts of  $\hat B^{\hat i}_{\ ;\hat j}$) gives the longitudinal component of the curl of that vector field; line 6 minus line 7 gives the latitudinal component of the curl; line 8 minus line 9 gives the radial component of the curl. It is straightforward to check that these results match standard textbook results for the divergence and curl of a vector field in spherical coordinates with a few differences due to the fact that our $\theta$ stands for latitude, and not the usual co-latitude. Those differences are as follows: in the right-handed non-coordinate basis [radius, co-latitude, longitude], if $\theta$ above is to stand for the standard co-latitude, then all one has to do to find $\hat B^{\hat i}_{\ ;\hat j}$ in that basis is to replace $\cos(\theta)$ with $\sin(\theta)$, as well as $\tan(\theta)$ with $(-\cot(\theta))$, in eq.~(\ref{BiCovJ}).

One can follow the same procedure we outlined above to express the covariant derivative of $\hatbB$ in the cylindrical orthonormal basis $(\hat \r,\hat{\bm\phi},\hat {\bm z})$, with $r$ being the cylindrical radial coordinate:
\be\label{cylcov}
\hat B^{\hat i}_{\ ;\hat j}&=&\hat B^{\hat i}_{\ ,\hat j} \mbox{  for $\hat j\neq \hat\phi$; and:}\nonumber\\
\hat B^{\hat r}_{\ ;\hat \phi}&=&-\frac{1}{r} \hat B^{\hat \phi}+\frac{1}{r}\frac{\partial \hat B^{\hat r}}{\partial \phi}\ , \quad 
\hat B^{\hat \phi}_{\ ;\hat \phi}=\frac{1}{r} \hat B^{\hat r}+\frac{1}{r}\frac{\partial \hat B^{\hat \phi}}{\partial \phi}\ , \quad 
\hat B^{\hat z}_{\ ;\hat \phi}=\frac{1}{r}\frac{\partial \hat B^{\hat z}}{\partial \phi}\ .
\ee

Since we are using an orthonormal basis in writing out  equations~(\ref{BiCovJ}) and (\ref{cylcov}), index placement is unimportant, and the generalized force-free parameter $\alpha$ (eq.~(\ref{alphaFinal})) can be written as:
\be\label{alphaCov}
\alpha=\epsilon_{\hat i\hat j\hat k} \hat B^{\hat i}\hat B^{\hat k}_{\ ;\hat j}\ .
\ee
Equations (\ref{qhoQFinal}), (\ref{omega0Final}), (\ref{BiCovJ}) and (\ref{alphaCov})  give the final result  in spherical coordinates for the local  squeezing  and coiling rates. These equations were implemented in the public code \qsl and were used in the calculations presented in the follow-up paper \citep{Tassev2019}.

\bibliographystyle{apj}  
\bibliography{Antonia_sig_topo_v3}       
\IfFileExists{\jobname.bbl}{}  
{ 
\typeout{} 
\typeout{****************************************************} 
\typeout{****************************************************} 
\typeout{** Please run "bibtex \jobname" to obtain}  
\typeout{**the bibliography and then re-run LaTeX}  
\typeout{** twice to fix the references!} 
\typeout{****************************************************} 
\typeout{****************************************************} 
\typeout{} 
 }

\end{document}